\documentclass{article}

\usepackage{arxiv}

\usepackage[utf8]{inputenc} 
\usepackage[T1]{fontenc}    
\usepackage{url}            
\usepackage{booktabs}       
\usepackage{amsfonts}       
\usepackage{nicefrac}       
\usepackage{microtype}      
\usepackage{lipsum}		
\usepackage{graphicx}
\usepackage{natbib}
\usepackage{doi}
\usepackage{a4wide}
\usepackage{amssymb}
\usepackage{amsmath}
\usepackage{mathrsfs}
\newtheorem{theorem}{Problem}
\usepackage{float}
\usepackage{caption}
\usepackage{subcaption}
\usepackage{paralist}
\usepackage{dirtytalk}
\usepackage{tikz}
\usetikzlibrary{fit, positioning}
\usetikzlibrary{arrows}
\usetikzlibrary{backgrounds}
\usetikzlibrary{er}
\usetikzlibrary{decorations}
\usetikzlibrary{topaths}
\usetikzlibrary{fadings}
\usetikzlibrary{external}
\usetikzlibrary{angles, quotes, intersections}
\definecolor{lblue}{RGB}{166, 206, 227} 
\definecolor{blue}{RGB}{31 , 120, 180} 
\definecolor{lgreen}{RGB}{178, 223, 138} 
\definecolor{green}{RGB}{51 , 160, 44 } 
\definecolor{lred}{RGB}{251, 154, 153} 
\definecolor{red}{RGB}{227, 26 , 28 } 
\definecolor{lor}{RGB}{253, 191, 111} 
\definecolor{or}{RGB}{255, 127, 0  } 
\definecolor{lpurple}{RGB}{202, 178, 214} 
\definecolor{purple}{RGB}{106, 61 , 154} 
\definecolor{lbrown}{RGB}{255, 255, 153} 
\definecolor{brown}{RGB}{177, 89 , 40 } 

\definecolor{lblue}{RGB}{166, 206, 227} 
\definecolor{blue}{RGB}{31 , 120, 180} 
\definecolor{lgreen}{RGB}{178, 223, 138} 
\definecolor{green}{RGB}{51 , 160, 44 } 
\definecolor{lred}{RGB}{251, 154, 153} 
\definecolor{red}{RGB}{227, 26 , 28 } 
\definecolor{lor}{RGB}{253, 191, 111} 
\definecolor{or}{RGB}{255, 127, 0  } 
\definecolor{lpurple}{RGB}{202, 178, 214} 
\definecolor{purple}{RGB}{106, 61 , 154} 
\definecolor{lbrown}{RGB}{255, 255, 153} 
\definecolor{brown}{RGB}{177, 89 , 40 } 

\definecolor{mblack}{RGB}{45, 42, 46}
\definecolor{mpurple}{RGB}{241,78,78}
\definecolor{msea}{RGB}{171, 157, 242}
\definecolor{mpink}{RGB}{255, 97, 136}
\definecolor{morgane}{RGB}{252,152,103}

\title{An Overset Algorithm for Multiphase Flows using 3D Multiblock Polyhedral Meshes}

\author{ \href{https://orcid.org/0009-0008-3853-6085}{\includegraphics[scale=0.06]{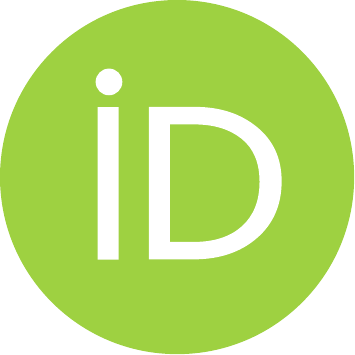}\hspace{1mm}Spiros Zafeiris} \\
	School of Naval Architecture \& Marine Engineering \\
	Zografos, Athens, PA 15780, Attiki, Greece \\
	  \\
	\And
	\href{https://orcid.org/0000-0002-2742-5258}{\includegraphics[scale=0.06]{orcid.pdf}\hspace{1mm}George Papadakis} \\
	School of Naval Architecture \& Marine Engineering \\
	Zografos, Athens, PA 15780, Attiki, Greece\\
	\\
}

\renewcommand{\shorttitle}{ }
\hypersetup{
pdftitle={A template for the arxiv style},
pdfsubject={q-bio.NC, q-bio.QM},
pdfauthor={David S.~Hippocampus, Elias D.~Striatum},
pdfkeywords={First keyword, Second keyword, More},
}

\begin{document}
\maketitle

\begin{abstract}
	In this study, we present a parallel topology algorithm with a suitable interpolation method for chimera simulations in CFD. The implementation is done in the unstructured Finite Volume (FV) framework and special attention is given to the numerical algorithm. The aim of the proposed algorithm is to approximate fields with discontinuities with application to two-phase incompressible flows. First, the overset topology problem in partitioned polyhedral meshes is addressed, and then a new interpolation algorithm for generally discontinuous fields is introduced. We describe how the properties of FV are used in favor of the interpolation algorithm and how this intuitive process helps to achieve high-resolution results. The performance of the proposed algorithm is quantified and tested in various test cases, together with a comparison with an already existing interpolation scheme. Finally, scalability tests are presented to prove computational efficiency. The method suggests to be highly accurate in propagation cases and performs well in unsteady two-phase problems executed in parallel architectures.
\end{abstract}


\section{Introduction}

Nowadays, CFD is used to model and recreate many categories of physical phenomena, and as computing power tends to increase, more complex applications are added to the scope of investigation. Many applications that involve moving solid (or deformable) boundaries have lead to the development of the overset method (chimera) which was introduced by Volkov \cite{volkov1968method} and then developed and refined by Steger \cite{steger1987use}, Starius \cite{starius1977composite}, Benek \cite{benek1983flexible}.

Overset grids are composed of multiple overlapping grids, each of which covers a specific portion of the simulation domain. The grids are typically structured or unstructured, and are usually of different resolutions. The overlapping regions between the grids are where the solution is interpolated from one grid to another using various interpolation techniques.

The overset grid technique has several advantages over other methods. First, it enables the simulation of complex geometries without the need for complex grid generation. Second, it allows for local grid refinement, which can increase the accuracy of the solution in areas of interest while minimizing computational resources. Finally, the method can handle moving boundaries or objects, which is crucial in many practical applications such as aeroelasticity and fluid-structure interaction (FSI) .

Even though the overset grid method comes with several benefits, there are also some limitations and challenges that must be addressed. One of the drawbacks is the complexity and computational cost of the interpolation process, which can be especially challenging. The interpolation itself is also a challenging task, especially when strong discontinuities are present. Without the implementation of a proper interpolation algorithm, the quality of the solution may be compromised in the overlapping regions due to interpolation errors. This may happen, because field variables are (partially) interpolated from one grid to another, which is another source of numerical diffusion. 

Another challenge is the \say{hole cutting} task, the need to maintain grid connectivity and consistency during the simulation. This involves ensuring that the grids overlap correctly, the holes are properly defined, and the solution is accurately interpolated at the interfaces between the grids. This can be particularly difficult when dealing with moving boundaries or complex geometries. Specifying the relative position of the grids and identifying the interpolation points at each timestep of the numerical algorithm can be a complex and computationally expensive process which can have deteriorating effects in massively parallel frameworks. Several efficient \say{hole cutting}  algorithms have been developed in the literature such as \cite{guerrero2006overset}, \cite{koblitz2017direct}, \cite{ha2018development}, \cite{chandar2019overset}, \cite{horne2019massively}, \cite{duan2020high}. A powerful algorithm that integrates parallelism and proper node identification, that inspired this work is presented by J. Sitaraman et. al. \cite{roget2014robust}. 

Regarding the interpolation techniques, they differ when considering structured or unstructured meshes. The explicit structured grid implementation of an overset grid algorithm can also be found in the work of Chesshire \cite{chesshire1990composite} in which the information from structured node indices is used to calculate a bi-linear, bi-quadratic or a Lagrange interpolant. Alabi et. al. \cite{alabi2004parallel} presented a framework that uses a search algorithm for grid topology definition and high order stencil-based interpolation for solving the compressible Navier-Stokes with the WENO reconstruction procedure, addressing also the importance of near wall node search. 

When considering unstructured grids, high-order interpolation methods were developed and assessed by Sharma \cite{sharma2021overset}. Grid cell values are interpolated from one grid to the other using high-order scattered data interpolation schemes such as the Radial Basis Functions (and their modified counterpart) to achieve minimal energy loss and retain the conservation properties of the underlying discretization. 

In steady or unsteady problems, the effects of interpolation errors are observed through the inability of the system to conserve its energy (internal and mechanical) since composite grids tend to weaken the accuracy and the convergence rate. Conservativity and preservation of accuracy on composite meshes has been a topic addressed by Chesshire \cite{chesshire1990composite}, Hadzic \cite{hadzic2006development}, Sharma \cite{sharma2021overset} and Crabill \cite{crabill2016high}. The majority of the literature is either focusing on the performance of grid data interpolation or the coupling with the solver and the corrector steps for fluxes. In \cite{hadzic2006development}, the mass loss is balanced by correcting the mass fluxes with an iterative procedure. Other flux correction techniques were implemented from Chandar \cite{chandar2019overset} aside from interpolation schemes. The coupling with the solver can be implemented in many ways and would probably depend on the current solution methodology (e.g., the pressure-coupling procedure). However, there is an additional source of computing cost that belongs in this iterative convergence process on top of the processing cost of interpolating and solving the partial differential equations. 

When considering flows with strong discontinuities the challenges associated with overset grids become even more pronounced. High-order interpolation techniques designed for smooth fields fail to yield a total variance diminishing (TVD) solution due to the occurrence of oscillations in regions with a field jump. This types of problems are easily found in nonlinear conservation laws in cases of shocks (compressible flows) or free-surface flows with a gravity field (gravity-induced water waves).
The problem of overset interpolation in discontinuity regions is addressed by Lee \cite{lee2011high} for compressible Euler solutions in structured grids. In \cite{lee2011high} it is shown that such interpolation problems may be overcome by implementing TVD limiters on high resolution interpolation (or namely reconstruction) schemes. The most important criterion in this procedure is the use of a stencil for buffer values, which can only be achieved (at least explicitly) in structured meshes. Lastly, an approximation method that serves both purposes (locally scattered interpolating points \& discontinuous fields) is the Pad\'e approximation. An examination of this method is thoroughly examined in \cite{hesthaven2006pade}, \cite{guillaume1998generalized} and applied in multidimensional problems in \cite{chantrasmi2009pade}. The multidimensional extension is non-trivial and has to be implemented together with certain geometric requirements (one of them is the avoidance of extrapolation). 


In this work, we will focus on the interpolation schemes for overset unstructured grids in a multi-phase flow framework. The goal is to present a scheme that respects the discontinuity of a field, can be applied to a general unstructured mesh, and is computationally efficient. The presented algorithm is developed on top off \textit{MaPFlow}, a massively parallel Finite Volume (FV) solver developed in NTUA \cite{diakakis2019assessment,ntouras2020coupled} which handles multi-block polyhedral meshes. \textit{MaPFlow} is able to handle both compressible and incompressible flow simulations. For this reason, it is noted that the work presented here in applicable to both compressible and incompressible flows without any modifications. However, we will only focus in the incompressible part with the use of volume fraction to model multiple phases \cite{ntouras2020coupled}.


This paper is structured as follows: In \autoref{section:overset} the overset algorithm is described, \autoref{section:interpolation} explains various existing interpolation algorithms and elaborates on the new proposed scheme. In \autoref{section:results} results are shown consisting of a numerical test on an analytical solution, a 2D propagation problem, 3D rigid body dynamics problem and lastly a scalability test. In the last \autoref{section:conclusions} we summarize the current study and evaluate the results.

\section{The Overset Approach}\label{section:overset}

In this section we briefly describe the hole-cutting and donor searching algorithms implemented in \textit{MaPFlow}. The concept emanates from the Topology Independent Overset Grid Assembly (TIOGA) library \cite{roget2014robust}. This library is an open-source tool, and its interface can be used to perform chimera simulations in modern Fortran with the MPI protocol. Even though the method presented here is inspired by TIOGA, it is substantially modified to handle polyhedral meshes.

During the implementation of the overset algorithm in \textit{MaPFlow}, we want to achieve computational robustness and minimal memory manipulations. The reason being that, unfortunately, not all processes (CPUs) will participate in the overset algorithm, and that is because some grid blocks are far from the overset region (in terms of Euclidean distance). This implies that calculations and memory allocations won't scale at the same rate as the number of solver processes increases, regardless of how computing-intensive the overset task is. Taking this into account, it is obvious that we have to be more careful with extra computing cost in the identification of the topology. In the following the \say{hole} cutting algorithm will be briefly described. 

\subsection{Boundary Cutting}

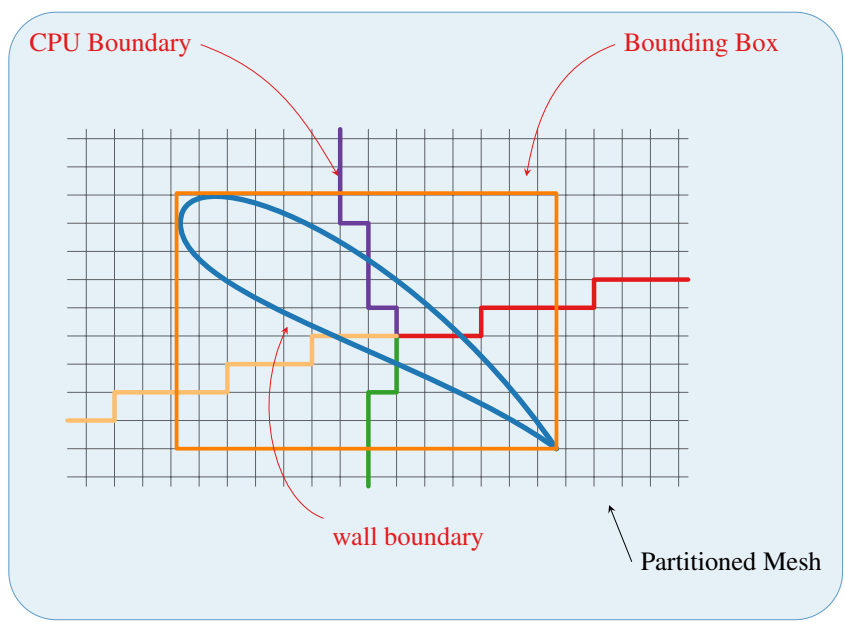
\begin{figure}[h]
    \centering
    \begin{tikzpicture}[line cap=round,line join=round,>=triangle 45, scale=2.5,background rectangle/.style=
{draw=blue!70,fill=blue!11,rounded corners=4ex}, show background rectangle]

\draw[xshift=-40mm, step=0.15, black,line width=0.05mm, opacity=0.6] (2,1) grid (5.3,2.9);
\draw[xshift=-40mm, red,   line width=0.6mm] (5.3,2.1) -- (4.8,2.1) -- (4.8,1.95) -- (4.2,1.95) -- (4.2,1.8) -- (3.75,1.8);
\draw[xshift=-40mm, green, line width=0.6mm] (3.75,1.8) -- (3.75, 1.5) -- (3.6,1.5) -- (3.6,1.0);
\draw[xshift=-40mm, purple,line width=0.6mm] (3.75,1.8) -- (3.75,1.95) -- (3.6,1.95) -- (3.6,2.4) -- (3.45,2.4) -- (3.45,2.9);
\draw[xshift=-40mm, lor,   line width=0.6mm] (3.75,1.8) -- (3.3,1.8) -- (3.3,1.65) -- (2.85,1.65) -- (2.85,1.5) -- (2.25,1.5) -- (2.25,1.35) -- (2.,1.35);

\draw[xshift=-40mm, line width=0.7mm, color=blue] (4.6, 1.2) .. controls (3.94, 1.68) and (2.6, 2) .. (2.6, 2.4);
\draw[xshift=-40mm, line width=0.7mm, color=blue]  (2.6, 2.4).. controls   (2.6, 2.8) and (3.81, 2.35) .. (4.6, 1.2);
\draw[xshift=-40mm, line width=0.5mm, color=orange] (4.6, 1.2) -- (2.58, 1.2) -- (2.58, 2.56) -- (4.6, 2.56) -- (4.6, 1.2);
\draw[xshift=-40mm, line width=0.1mm, color=red,  -stealth] (3.36, 0.83) node[below right] {wall boundary}.. controls (3.09, 0.93) and  (2.98, 1.46) .. (3.17, 1.85);
\draw[xshift=-30mm, style=curve to, line width=0.1mm, color=red, -stealth] (3.91, 3.35) node[right] {Bounding Box} to [out=200, in=80] (3.44,2.65);

\draw[xshift=-30mm, style=curve to, line width=0.1mm, color=red, -stealth] (1.71, 3.35) node[left] {CPU Boundary} to [out=-20, in=120] (2.44,2.65);

\draw[xshift=-30mm, line width=0.1mm,-stealth] (4, 0.6) node [right] {Partitioned Mesh} -- (3.88, 0.9);

\draw[xshift=-22mm] (0,0.4) node {  };

\end{tikzpicture}
    \caption{Representation of hole-cutting of wall boundary while viewing the background grid}
    \label{fig:hole_cut_tikz}
\end{figure}

The goal is to mark the bulk size of the total cells that will be unused. That way, loop statements in the following task will be accelerated by reducing the size of the cell count. The identification will be done with a masking function. The first type of cells that is marked are the hole cells with an indicator value of $0$. The task can be seen in \autoref{fig:hole_cut_tikz}.

We can distinguish the wrapping of the boundary by a cartesian box (the body-fitted grid is not shown). This cartesian box (BB : bounding box) is constructed as a superset of those from processes with wall boundaries. Each box is gathered into a single process and then broadcast. After the broadcast, all processes can mark cells inside of the BB as wall cells (they are useless from now on) and the following layers (identified by searching the connectivity of the faces) are marked as receptors (indicator value of $-1$). It is now obvious that the same work has to be done for overset faces, and the resulting superset of bounding boxes will allow for marking cells with a logical mask as \say{contained} (close to the overset region).

An important note is that, in general polyhedral meshes the type of connectivity search is done using faces. More precisely, in \textit{MaPFlow}, faces carry their topology identity, e.g. wall boundary, symmetry, cpu-to-cpu boundary.

As long as a strip of cells is marked with the identification of $0$, we can flood through 3D-space to mark the neighboring cells as \say{value receptors} by using face Left \& Right connectivity.

On the other hand, we have achieved a second, not so obvious, task. As every process checks for \say{wall} or \say{contained} cells we can categorize processes with respect to their participation in the overset algorithm. This way processes can participate by  
\begin{inparaenum}[\itshape a\upshape)]
    \item having cells in the body-fitted grid,
    \item in the background grid and close to the overset boundary and, lastly
    \item to not participate at all by having only cells in the background and far from the overset region.
\end{inparaenum} We split these processes in the first 2 families, and from now on inter-CPU communication is devoted between these 2 groups.

\subsection{Donor-searching}

We should keep in mind that overset boundary conditions through interpolation must be set for both grids, so donor searching is performed both ways (unlike boundary cutting which usually does not refer to wrapping a boundary of the background grid). By setting the total number of overlapping grids at any given position to be at most $2$, we can simply repeat the procedure for every mesh-pair.

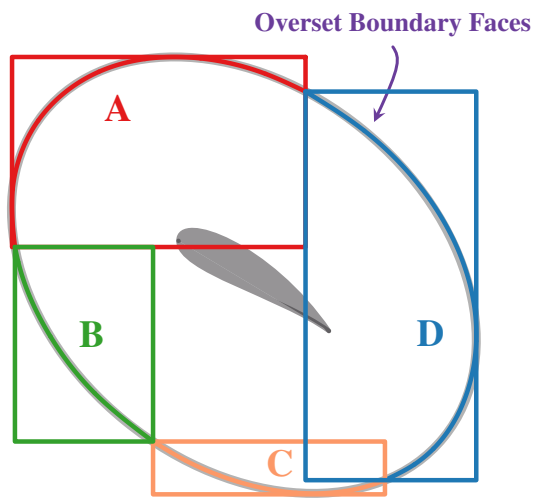
\begin{figure}[H]
    \centering    

\begin{tikzpicture}[line cap=round,line join=round,>=triangle 45, background rectangle/.style=
 {draw=blue!0,fill=blue!0,rounded corners=5ex}, show background rectangle, scale=0.5]
\draw [opacity=0.3, rotate around={-39.1088:(1.53799,0.9259)},line width=1.2mm] (1.53799,0.92591) ellipse (6.8466cm and 5.00273cm);
\draw[fill=white, opacity=0.5, line width=0.6mm, color=mblack, xshift=18mm, yshift=10mm, scale=2] (1, -0.78) .. controls (0.34, -0.3) and (-1, 0.02) .. (-1, 0.42);
\draw[fill=white, opacity=0.5, line width=0.6mm, color=mblack, xshift=18mm, yshift=10mm, scale=2] (-1, 0.42) .. controls (-1, 0.82) and (0.21, 0.37)..((1, -0.78);

   \draw[line width=0.4mm,color=red] (-1.8, 6.0) node[below] {\Large \textbf{A}};
\draw[line width=0.6mm,color=red, smooth,samples=100,domain=0:0.103519] plot[parametric] function{-11.32481*t**(3)-10.15846*t+3.177871,-27.451979*t**(3)+4.860538*t+5.810934};
\draw[line width=0.6mm,color=red, smooth,samples=100,domain=0.103519:0.26095] plot[parametric] function{5.40203*t**(3)-5.194640*t**(2)-9.62072*t+3.159316,-1.200402*t**(3)-8.1526*t**(2)+5.70448*t+5.78181};
\draw[line width=0.6mm,color=red, smooth,samples=100,domain=0.260959:0.39306] plot[parametric] function{6.1920343*t**(3)-5.81311*t**(2)-9.45932*t+3.1452767,-7.0594788*t**(3)-3.56566*t**(2)+4.507478*t+5.885937};
\draw[line width=0.6mm,color=red, smooth,samples=100,domain=0.393067:0.50592] plot[parametric] function{12.20617*t**(3)-12.90501*t**(2)-6.671731*t+2.7800,-8.57989*t**(3)-1.7727*t**(2)+3.80275*t+5.978271};
\draw[line width=0.6mm,color=red, smooth,samples=100,domain=0.505927:0.68973] plot[parametric] function{16.57959*t**(3)-19.54290*t**(2)-3.31343*t+2.21368,7.309934*t**(3)-25.89*t**(2)+16.004*t+3.9205};
\draw[line width=0.6mm,color=red, smooth,samples=100,domain=0.689739:0.86417] plot[parametric] function{4.15941*t**(3)+6.15714*t**(2)-21.0397*t+6.28920,19.96406*t**(3)-52.07423*t**(2)+34.0645*t-0.23172};
\draw[line width=0.6mm,color=red, smooth,samples=100,domain=0.864174:1] plot[parametric] function{-41.5743*t**(3)+124.7229*t**(2)-123.5013*t+35.8041,0.777739*t**(3)-2.333*t**(2)-8.92033*t+12.150444};

   \draw[line width=0.4mm,color=green] (-2.5, -.02) node[below] {\Large \textbf{B}};
\draw[line width=0.6mm,color=green, smooth,samples=100,domain=0:0.0622] plot[parametric] function{19.452744*t**(3)+1.32160*t-4.54860,4.807063*t**(3)-6.362221*t+1.67462};
\draw[line width=0.6mm,color=green, smooth,samples=100,domain=0.062255:0.39451] plot[parametric] function{-1.6669*t**(3)+3.9444*t**(2)+1.076*t-4.54351,0.388238*t**(3)+0.8252*t**(2)-6.4136*t+1.675694};
\draw[line width=0.6mm,color=green, smooth,samples=100,domain=0.394518:0.68988] plot[parametric] function{0.041939*t**(3)+1.921913*t**(2)+1.8739*t-4.64844,0.924348*t**(3)+0.1907717*t**(2)-6.16327*t+1.64277};
\draw[line width=0.6mm,color=green, smooth,samples=100,domain=0.689889:1] plot[parametric] function{-2.15913*t**(3)+6.477*t**(2)-1.2688*t-3.9257,-2.26141*t**(3)+6.784235*t**(2)-10.7120*t+2.6888};

   \draw[line width=0.4mm,color=morgane] (2.5, -3.45) node[below] {\Large \textbf{C}};
\draw[line width=0.6mm,color=morgane, smooth,samples=100,domain=0:0.1560] plot[parametric] function{2.99603*t**(3)+5.5130*t-0.87626,5.55412*t**(3)-3.3985*t-3.50038};
\draw[line width=0.6mm,color=morgane, smooth,samples=100,domain=0.156001:0.31011] plot[parametric] function{-1.183625*t**(3)+1.956098*t**(2)+5.20790*t-0.8604,-0.71432*t**(3)+2.9336*t**(2)-3.8562*t-3.4765};
\draw[line width=0.6mm,color=morgane, smooth,samples=100,domain=0.310110:0.62991] plot[parametric] function{-0.39496*t**(3)+1.2223*t**(2)+5.4354*t-0.88392,0.994442*t**(3)+1.343*t**(2)-3.3632*t-3.5275};
\draw[line width=0.6mm,color=morgane, smooth,samples=100,domain=0.629916:0.95550] plot[parametric] function{-2.18323*t**(3)+4.60175*t**(2)+3.3067*t-0.4369,1.5128*t**(3)+0.3643*t**(2)-2.7462*t-3.6571};
\draw[line width=0.6mm,color=morgane, smooth,samples=100,domain=0.955501:1] plot[parametric] function{12.4084*t**(3)-37.225*t**(2)+43.2725*t-13.1660,-35.213*t**(3)+105.63928*t**(2)-103.33651*t+28.38093};

   \draw[line width=0.4mm,color=blue] (6.5, 0.) node[below] {\Large \textbf{D}};
\draw[line width=0.6mm,color=blue, smooth,samples=100,domain=0:0.0693] plot[parametric] function{-66.35132*t**(3)+11.63757*t+5.28957,108.15031*t**(3)+5.5515*t-4.5293};
\draw[line width=0.6mm,color=blue, smooth,samples=100,domain=0.069358:0.12033] plot[parametric] function{-28.12819*t**(3)-7.9532*t**(2)+12.18920*t+5.27682,-30.8620*t**(3)+28.9249*t**(2)+3.54538*t-4.483};
\draw[line width=0.6mm,color=blue, smooth,samples=100,domain=0.120336:0.36029] plot[parametric] function{0.57779*t**(3)-18.31639*t**(2)+13.43626*t+5.22679,-25.98685*t**(3)+27.1649*t**(2)+3.75717*t-4.49150};
\draw[line width=0.6mm,color=blue, smooth,samples=100,domain=0.360296:0.85395] plot[parametric] function{9.18402*t**(3)-27.6187*t**(2)+16.7878*t+4.8242,-5.81233*t**(3)+5.35853*t**(2)+11.6139*t-5.43509};
\draw[line width=0.6mm,color=blue, smooth,samples=100,domain=0.853950:1] plot[parametric] function{9.33628*t**(3)-28.0088*t**(2)+17.12097*t+4.7294,21.75476*t**(3)-65.26429*t**(2)+71.9223*t-22.6018};
\draw[line width=0.4mm, color=purple, -stealth] (5.5, 7.) node[above] {\textbf{Overset Boundary Faces}} .. controls (5.7, 6.7) and (5.7,6.5) .. (5, 5);

\draw [line width=0.6mm,color=morgane] (-0.87626,-3.50038)-- (-0.876269,-4.906229)-- (+5.28957,-4.90622)-- (+5.28957,-3.50038)-- (-0.876269,-3.500385);
\draw [line width=0.6mm,color=red] (+3.17787,+6.73157)-- (+3.177871,+1.674628)-- (-4.62541,+1.67462)-- (-4.62541,+6.73157)-- (3.177871,6.73157);
\draw [line width=0.6mm,color=green] (-4.54860,+1.67462)-- (-0.876269,+1.674628)-- (-0.87626,-3.50038)-- (-4.54860,-3.50038)-- (-4.54860,1.674628);
\draw [line width=0.6mm,color=blue] (+3.17787,+5.81093)-- (+7.717144,+5.810934)-- (+7.71714,-4.52938)-- (+3.17787,-4.52938)-- (3.17787,5.81093);
\end{tikzpicture}
    \caption{Representation of the receptor bounding boxes for every colored process}
    \label{fig:ident}
\end{figure}

The practicality behind the use of bounding boxes for polyhedral cell containment in partitioned grids is also followed in the \emph{Donor-searching} part. The reason for their use is the minimal data needed to fully represent them and the robust point-to-box containment criterion. More precisely, we simply have to check if the point coordinates are between minimum and maximum values of the box edges (BB criterion). In contrast, a minimum volume convex hull criterion (the user is referred to \cite{ha2018development}) involves the calculation of a series of dot products. The ownership of the overset BB from processes for the case of the body-fitted mesh is described visually in  \autoref{fig:ident}.

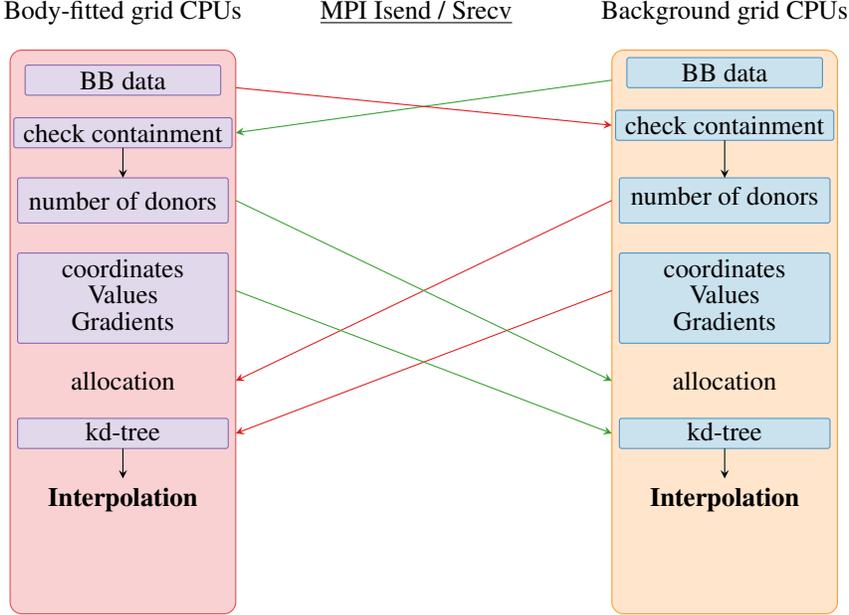
\begin{figure}[H]
    \centering
    \begin{tikzpicture}
    \draw[rounded corners=1ex, line cap=round, draw=red!80, fill=red!20] (0, -2) rectangle (3, 5.5);
    \draw[rounded corners=1ex, line cap=round, draw=or!80, fill=or!20]   (8, -2) rectangle (11,5.5);
    \draw (1.5,6) node {Body-fitted grid CPUs};
    \draw (9.5,6) node {Background grid CPUs};

    \draw[rounded corners=0.2ex, draw=purple!80, fill=purple!20] (0.2,4.9)
    rectangle (2.8,5.3); \draw (1.5,5.1) node {BB data};
    \draw[rounded corners=0.2ex, draw=purple!80, fill=purple!20] (0.05,4.2)
    rectangle (2.95,4.6); \draw (1.5,4.4) node {check containment};
    \draw[-stealth] (1.5,4.2) -- (1.5,3.8);
    \draw[rounded corners=0.2ex, draw=purple!80, fill=purple!20] (0.1,2.2+1) 
    rectangle (2.9,2.8+1);\draw (1.5,2.5+1) node {number of donors};
    \draw (1.5, 1.1) node {allocation};
    \draw[rounded corners=0.2ex, draw=purple!80, fill=purple!20] (0.1,0.6+1.0) 
    rectangle (2.9,1.8+1.0);
    \draw (1.5,1.6+1.0) node {coordinates};
    \draw (1.5,1.25+1.0) node {Values};
    \draw (1.5,0.9+1.0) node {Gradients};

    \draw[rounded corners=0.2ex, draw=purple!80, fill=purple!20] (0.1,0.2) rectangle (2.9,0.6);
    \draw (1.5,0.42) node {kd-tree};

    \draw[-stealth] (1.5,0.2) -- (1.5,-0.2) node[below] {\textbf{Interpolation}};

    \draw[rounded corners=0.2ex, draw=blue!80, fill=lblue!60] 
    (8.2 ,5.0) rectangle (10.8,5.4) ;\draw (9.5, 5.2) node {BB data};
    \draw[rounded corners=0.2ex, draw=blue!80, fill=lblue!60] 
    (8.05,4.3) rectangle (10.95,4.7);\draw (9.5, 4.5) node {check containment};
    \draw[rounded corners=0.2ex, draw=blue!80, fill=lblue!60] (8.1,2.2+1) rectangle (10.9,2.8+1);
    \draw[-stealth] (9.5,3.3+1) -- (9.5,2.8+1) node[below] {number of donors};

    \draw[rounded corners=0.2ex, draw=blue!80, fill=lblue!60] (8.1,0.6+1.0) 
    rectangle (10.9,1.8+1.0);
    \draw (9.5,1.6+1.0) node {coordinates};
    \draw (9.5,1.25+1.0) node {Values};
    \draw (9.5,0.9+1.0) node {Gradients};

    \draw (9.5,1.1) node {allocation};

    \draw[rounded corners=0.2ex, draw=blue!80, fill=lblue!60] (8.1,0.2) rectangle (10.9,0.6);
    \draw (9.5,0.42) node {kd-tree};
    \draw[-stealth] (9.5,0.2) -- (9.5,-0.2) node[below] {\textbf{Interpolation}};
    \draw (5.4,6.0) node {\underline{MPI Isend / Srecv}};
    \draw[-stealth, color=green] (8,5.1) -- (3.,4.4);
    \draw[-stealth, color=red] (8,3.5) -- (3,1.1);
    \draw[-stealth, color=red] (8,3.5-1.2) -- (3,2.4-2);

    \draw[-stealth, color=red] (3,5.0) -- (8,4.5);
    \draw[-stealth, color=green] (3,3.5) -- (8,1.1);
    \draw[-stealth, color=green] (3,3.5-1.2) -- (8,2.4-2);
\end{tikzpicture}
    \caption{Data transfer schematic between groups of CPUs}
    \label{fig:cpus}
\end{figure}

In our case, the following steps are shown in \autoref{fig:cpus}.
\begin{enumerate}
    \item every process with receptor cells constructs a bounding box surrounding them, and then it is sent to processes from the other group (we call it the source process).
    \item The search is done from the opposite group of processes (destination) by searching regular cells (that do not undergo interpolation themselves, i.e. an indicator value of $1$) and using the BB criterion.
    \item The number of donors inside this box is sent back by the destination process for allocation (it might be zero).
    \item Lastly, donor data (coordinates, values and gradients) are also sent to the source process.
\end{enumerate}

These 4 steps are implemented with asynchronous send - synchronous receive calls from the MPI. The result is a number of scattered cell-centered values in the vicinity of the receptor cells of a single process. We still have to find, for every receptor cell, the closest donors. This is done using the kd-tree algorithm. With this algorithm we can achieve an $\mathcal{O}\left( N\log N\right)$ scaling. The implementation of the kd-tree is taken from Matthew B. Kennel \cite{kennel2004kdtree}.

That way, we are left with a group of the closest donor cells, and therefore the topology problem is tackled.

\section{Interpolation Schemes}\label{section:interpolation}
After the establishment of the overset topology, flow information must be interchanged. This is a critical step in overset grid simulations, where solution variables need to be accurately computed in the overlapping regions between the grids. A well-performing interpolation scheme may have a significant impact on the accuracy, stability, and efficiency of the simulation. Various interpolation schemes have been developed over the years, such as in \cite{sharma2021overset}. Each interpolation scheme has its own advantages and limitations, and the choice of scheme depends on the specific requirements of the simulation. The current choice of the interpolation scheme encapsulates the demand for using scattered nodes as interpolating degrees of freedom of a generally discontinuous function, e.g. the volume fraction field.


\subsection{Preliminaries} \label{sub:green_gauss}


One important feature of several finite volume algorithms is the ability to approximate field gradients at the cell center without many calculations on large stencils. The usual way of a gradient calculation is through the Green-Gauss theorem
\begin{align}\label{eq:green_grad}
    & \nabla Q\Bigg|_{el} = \frac{1}{V}\int_V \nabla Q \cdot dV \approx 
    \sum_f Q_f \cdot S_f \cdot \Vec{n}_f
\end{align}

where $Q$ is a primitive variable, $V$ is the volume of the cell under investigation, $f$ is one of the cell's faces and $\Vec{n}_f$ is the normal vector of the face $f$ which points outside of the cell. The summation involves the collection of the cell's faces. The left side of \autoref{eq:green_grad} represents the approximated gradient on the cell center which is the desired product. The values $Q_f$ which are currently unknown, are approximated by the following interpolation procedure (PLR : piecewise linear reconstruction), which among others, can be found at \cite{queutey2007interface}.

\subsection{Definition}
The general interpolation problem can be stated as follows:
Consider a set of scattered nodes $\mathcal{X}_N = \left\{\mathbf{x_1}, \mathbf{x_2}, \dotsc , \mathbf{x_N}\right\} \subseteq \Omega $, which are strictly distinct, i.e. $\mathbf{x_i} \neq \mathbf{x_j}, \iff i\neq j,\ 1\leq i,j\leq N$. Generally $\Omega$ is either $\mathbb{R}^2$ or $\mathbb{R}^3$, that is, every element of $\mathcal{X}$ is a vector of coordinates.

Additionally, let $\mathcal{Y} = \left\{y_1, y_2, \dotsc, y_N\right\} \subseteq  \mathbb{R}$, $\mathcal{U} = \left\{u_1, u_2, \dotsc, u_N\right\} \subseteq \mathbb{R}$ and $\mathcal{V} = \left\{v_1, v_2, \dotsc, v_N\right\} \subseteq \mathbb{R}$ be three sets of corresponding real values of the nodes comprising $\mathcal{X}$, that is every node $\mathbf{x}_i,\ i=1,2,3,\dotsc,N$ carries the values $y_i,\ u_i$ and $v_i$.

The interpolation problem is defined by the following constraints.

\begin{theorem}
    Find $f: \Omega \to \mathbb{R}$  that satisfies
    \begin{align}\label{eq:inter_f}
        & f\left(\mathbf{x}_i\right) = y_i,\ i=1,2,3,\dotsc,N
    \end{align} where $y_i,\ i\in\{1,2,3,\dotsc,N\}$ are known data.

    Furthermore, assume the existence of the first derivative of $f\left(\mathbf{x}\right)$ and therefore it follows that 
    \begin{align}\label{eq:inter_grad}
        & \nabla f \left(\mathbf{x}_i\right) = \left(u_i, v_i \right),\ i=1,2,3,\dotsc,N
    \end{align}
    where $u_i,v_i$ are also known data.
\end{theorem}

Of course, the existence of a derivative in broken functions is artificial (or existent only in the weak sense). However, the definition takes place given that known data $ \mathcal{Y}, \mathcal{U}, \mathcal{V}$ emerge as approximation data of the finite volume discretization (calculated with the help the Green-Gauss relation shown in \autoref{sub:green_gauss}).

The interpolation function will be denoted as $s\left(\mathbf{x}\right)$ and the Euclidean distance between points $\mathbf{x}, \mathbf{y}$ is denoted as $d(\mathbf{x},\mathbf{y}): = || \mathbf{x}-\mathbf{y}||_2$.


\subsection{Non-linear Interpolation on smooth regions}

We seek to employ a high-order explicit interpolation procedure that fits values in interpolation problems with very smooth regions. The choice is the inverse distance weighting method introduced by Shepard \cite{shepard1968two}. This method reads

\begin{align}\label{eq:idw}
    & s(\mathbf{x}) = \dfrac{\sum_{i=1}^N w_i\left(\mathbf{x} \right)\cdot y_i}{\sum_{i=1}^N w_i\left(\mathbf{x} \right)} && w_i\left(\mathbf{x} \right) = \frac{1}{d\left(\mathbf{x},\mathbf{x}_i \right)^p},\quad p>0
\end{align}

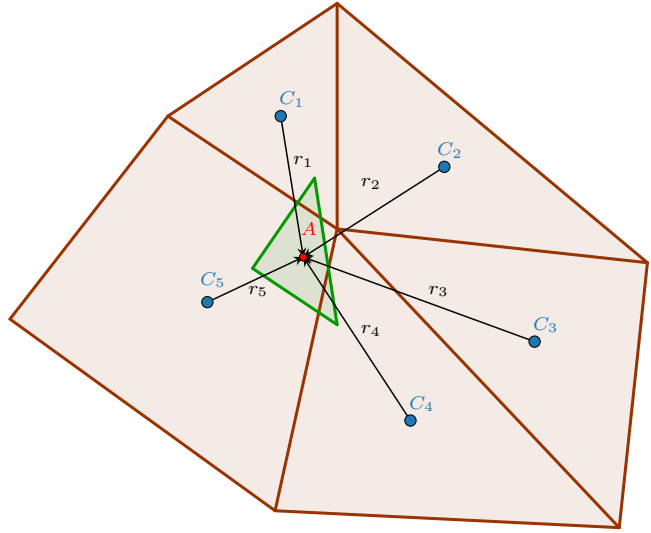
\begin{figure}[H]
    \centering
    \definecolor{ffqqqq}{rgb}{1.,0.,0.}
\definecolor{qqzzqq}{rgb}{0.,0.6,0.}
\definecolor{rfxybu}{rgb}{0.12,0.47,0.70}
\definecolor{zzttqq}{rgb}{0.6,0.2,0.}
\begin{tikzpicture}[line cap=round,line join=round,>=triangle 45, scale=7.5]
\fill[line width=0.4mm,color=zzttqq,fill=zzttqq,fill opacity=0.100] (-1.2,1.2) -- (-1.5,1.) -- (-1.2,0.8) -- cycle;
\fill[line width=0.4mm,color=zzttqq,fill=zzttqq,fill opacity=0.100] (-1.5,1.) -- (-1.78,0.64) -- (-1.31,0.30) -- (-1.2,0.8) -- cycle;
\fill[line width=0.4mm,color=zzttqq,fill=zzttqq,fill opacity=0.100] (-1.31,0.30) -- (-0.70,0.27) -- (-1.2,0.8) -- cycle;
\fill[line width=0.4mm,color=zzttqq,fill=zzttqq,fill opacity=0.100] (-0.70,0.27) -- (-0.65,0.74) -- (-1.2,0.8) -- cycle;
\fill[line width=0.4mm,color=zzttqq,fill=zzttqq,fill opacity=0.100] (-0.65,0.74) -- (-1.2,1.2) -- (-1.2,0.8) -- cycle;
\fill[line width=0.4mm,color=qqzzqq,fill=qqzzqq,fill opacity=0.100] (-1.24,0.89) -- (-1.35,0.73) -- (-1.20,0.63) -- cycle;
\draw [line width=0.4mm,color=zzttqq] (-1.2,1.2)-- (-1.5,1.);
\draw [line width=0.4mm,color=zzttqq] (-1.2,0.8)-- (-1.2,1.2);
\draw [line width=0.4mm,color=zzttqq] (-1.5,1.)-- (-1.78,0.64);
\draw [line width=0.4mm,color=zzttqq] (-1.78,0.64)-- (-1.31,0.30);
\draw [line width=0.4mm,color=zzttqq] (-1.31,0.30)-- (-1.2,0.8);
\draw [line width=0.4mm,color=zzttqq] (-1.2,0.8)-- (-1.5,1.);
\draw [line width=0.4mm,color=zzttqq] (-1.31,0.30)-- (-0.70,0.27);
\draw [line width=0.4mm,color=zzttqq] (-0.70,0.27)-- (-0.65,0.74);
\draw [line width=0.4mm,color=zzttqq] (-0.65,0.74)-- (-1.2,0.8);
\draw [line width=0.4mm,color=zzttqq] (-1.2,0.8)-- (-0.70,0.27);
\draw [line width=0.4mm,color=zzttqq] (-0.65,0.74)-- (-1.2,1.2);
\draw [line width=0.4mm,color=qqzzqq] (-1.24,0.89)-- (-1.35,0.73);
\draw [line width=0.4mm,color=qqzzqq] (-1.35,0.73)-- (-1.20,0.63);
\draw [line width=0.4mm,color=qqzzqq] (-1.20,0.63)-- (-1.24,0.89);
\draw [-stealth,line width=0.2mm] (-1.3,1.) -- (-1.26,0.75);
\draw [-stealth,line width=0.2mm] (-1.01,0.91) -- (-1.26,0.75);
\draw [-stealth,line width=0.2mm] (-0.85,0.60) -- (-1.26,0.75);
\draw [-stealth,line width=0.2mm] (-1.07,0.46) -- (-1.26,0.75);
\draw [-stealth,line width=0.2mm] (-1.43,0.67) -- (-1.26,0.75);
\begin{scriptsize}
\draw [fill=rfxybu] (-1.3,1.) circle (0.1mm);
\draw[color=rfxybu] (-1.28,1.03) node {$C_1$};
\draw [fill=rfxybu] (-1.01,0.91) circle (0.1mm);
\draw[color=rfxybu] (-1.00,0.94) node {$C_2$};
\draw [fill=rfxybu] (-0.85,0.60) circle (0.1mm);
\draw[color=rfxybu] (-0.83,0.63) node {$C_3$};
\draw [fill=rfxybu] (-1.07,0.46) circle (0.1mm);
\draw[color=rfxybu] (-1.05,0.49) node {$C_4$};
\draw [fill=rfxybu] (-1.43,0.67) circle (0.1mm);
\draw[color=rfxybu] (-1.42,0.71) node {$C_5$};
\draw [fill=ffqqqq] (-1.26,0.75) circle (0.07mm);
\draw[color=ffqqqq] (-1.25,0.80) node {$A$};
\draw[color=black] (-1.26,0.92) node {$r_1$};
\draw[color=black] (-1.14,0.88) node {$r_2$};
\draw[color=black] (-1.02,0.69) node {$r_3$};
\draw[color=black] (-1.14,0.62) node {$r_4$};
\draw[color=black] (-1.34,0.69) node {$r_5$};
\end{scriptsize}
\end{tikzpicture}
    \caption{Inverse distance weighting schematic}
    \label{fig:idw}
\end{figure}

where $p$ is a user-fixed parameter concerning the range of contributions of every interpolating node. This non-linear averaging process can be efficient even with large numbers of interpolation points due to its explicit expression. The exponent can take any value from the number of dimensionality and above : $p\ge 2$ for 2D problems and $p\ge 3$ for 3D problems. However, this user-defined parameter depends on the density of nodes in 3D space and therefore, optimal results require an optimization of parameter $p$.



\subsection{Nearest-Neighbor Value}

Regions that are not smooth require a different manipulation of the available scattered data. The least intricate way to deal with such a region is the simple method of the nearest-neighbor value. Using this method, the value of a calculation point comes directly from its closest distant donor.

Formally, the interpolant $s\left(\mathbf{x}\right)$ in this case takes the following expression.

\begin{align}
    & \varrho = \min_{i=1,N} \left\{ d\left(\mathbf{x}_c,\mathbf{x}_i\right)\right\} \\
    & \left\{l=1,2,3,\dotsc,N\ :\ d\left(\mathbf{x}_c,\mathbf{x}_{l}\right) = \varrho\right\} \\
    & s(\mathbf{x}) = y_{l}\cdot H_{\mathcal{B}(\mathbf{x}_c,\varrho)}(\mathbf{x})
\end{align}

With $H_{\mathcal{B}(\mathbf{x},\varrho)}$ we express the Heaviside function that is activated inside the ball $\mathcal{B}(\mathbf{x},\varrho)$. Tests with this procedure can be found in two phase flows with the PLIC-VOF formulation in the work of Chang \cite{chang2020comparative}.

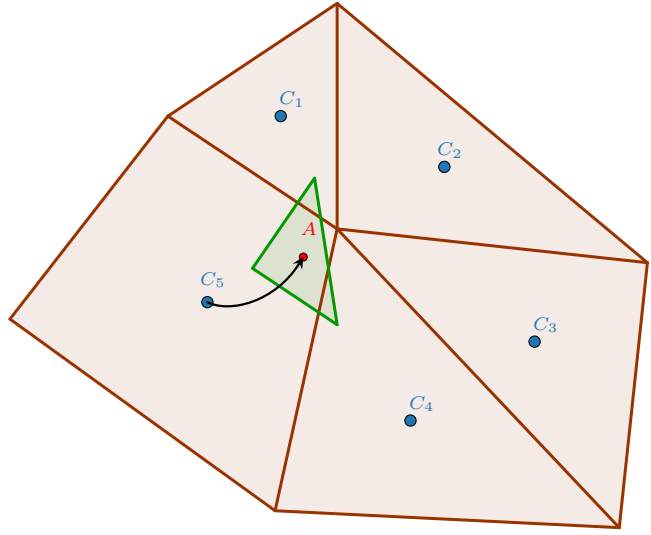
\begin{figure}[H]
    \centering
    \definecolor{ffqqqq}{rgb}{1.,0.,0.}
\definecolor{qqzzqq}{rgb}{0.,0.6,0.}
\definecolor{rfxybu}{rgb}{0.12,0.47,0.70}
\definecolor{zzttqq}{rgb}{0.6,0.2,0.}
\begin{tikzpicture}[line cap=round,line join=round,>=triangle 45, scale=7.5]
\fill[line width=0.4mm,color=zzttqq,fill=zzttqq,fill opacity=0.100] (-1.2,1.2) -- (-1.5,1.) -- (-1.2,0.8) -- cycle;
\fill[line width=0.4mm,color=zzttqq,fill=zzttqq,fill opacity=0.100] (-1.5,1.) -- (-1.78,0.64) -- (-1.31,0.30) -- (-1.2,0.8) -- cycle;
\fill[line width=0.4mm,color=zzttqq,fill=zzttqq,fill opacity=0.100] (-1.31,0.30) -- (-0.70,0.27) -- (-1.2,0.8) -- cycle;
\fill[line width=0.4mm,color=zzttqq,fill=zzttqq,fill opacity=0.100] (-0.70,0.27) -- (-0.65,0.74) -- (-1.2,0.8) -- cycle;
\fill[line width=0.4mm,color=zzttqq,fill=zzttqq,fill opacity=0.100] (-0.65,0.74) -- (-1.2,1.2) -- (-1.2,0.8) -- cycle;
\fill[line width=0.4mm,color=qqzzqq,fill=qqzzqq,fill opacity=0.100] (-1.24,0.89) -- (-1.35,0.73) -- (-1.20,0.63) -- cycle;
\draw [line width=0.4mm,color=zzttqq] (-1.2,1.2)-- (-1.5,1.);
\draw [line width=0.4mm,color=zzttqq] (-1.2,0.8)-- (-1.2,1.2);
\draw [line width=0.4mm,color=zzttqq] (-1.5,1.)-- (-1.78,0.64);
\draw [line width=0.4mm,color=zzttqq] (-1.78,0.64)-- (-1.31,0.30);
\draw [line width=0.4mm,color=zzttqq] (-1.31,0.30)-- (-1.2,0.8);
\draw [line width=0.4mm,color=zzttqq] (-1.2,0.8)-- (-1.5,1.);
\draw [line width=0.4mm,color=zzttqq] (-1.31,0.30)-- (-0.70,0.27);
\draw [line width=0.4mm,color=zzttqq] (-0.70,0.27)-- (-0.65,0.74);
\draw [line width=0.4mm,color=zzttqq] (-0.65,0.74)-- (-1.2,0.8);
\draw [line width=0.4mm,color=zzttqq] (-1.2,0.8)-- (-0.70,0.27);
\draw [line width=0.4mm,color=zzttqq] (-0.65,0.74)-- (-1.2,1.2);
\draw [line width=0.4mm,color=qqzzqq] (-1.24,0.89)-- (-1.35,0.73);
\draw [line width=0.4mm,color=qqzzqq] (-1.35,0.73)-- (-1.20,0.63);
\draw [line width=0.4mm,color=qqzzqq] (-1.20,0.63)-- (-1.24,0.89);
\begin{scriptsize}
\draw [fill=rfxybu] (-1.3,1.) circle (0.1mm);
\draw[color=rfxybu] (-1.28,1.03) node {$C_1$};
\draw [fill=rfxybu] (-1.01,0.91) circle (0.1mm);
\draw[color=rfxybu] (-1.00,0.94) node {$C_2$};
\draw [fill=rfxybu] (-0.85,0.60) circle (0.1mm);
\draw[color=rfxybu] (-0.83,0.63) node {$C_3$};
\draw [fill=rfxybu] (-1.07,0.46) circle (0.1mm);
\draw[color=rfxybu] (-1.05,0.49) node {$C_4$};
\draw [fill=rfxybu] (-1.43,0.67) circle (0.1mm);
\draw[color=rfxybu] (-1.42,0.71) node {$C_5$};
\draw [fill=ffqqqq] (-1.26,0.75) circle (0.07mm);
\draw[color=ffqqqq] (-1.25,0.80) node {$A$};
\draw[color=black, line width=0.3mm, -stealth] (-1.43, 0.67) .. controls (-1.39, 0.65) and (-1.31, 0.67) .. (-1.26, 0.75);
\end{scriptsize}
\end{tikzpicture}
    \caption{Nearest neighbor schematic}
    \label{fig:NN}
\end{figure}

\subsection{Gradient-Aware Interpolation}

Taking into account the aforementioned interpolation algorithms and the features of unstructured FV methods we introduce the following procedure.

From the set of $N$ scattered nodes consider the $N_d < N$ points that are closest to the calculation point $\mathbf{x}_c$, that is:
\begin{align}
    & \mathcal{X}_d = \left\{\mathbf{x}_1, \mathbf{x}_2, \dotsc , \mathbf{x}_{N_d}\right\} : \max_{i=1,N_d}\left[ d(\mathbf{x}_c,\mathbf{x}_i)\right] < d(\mathbf{x}_c,\mathbf{x}_j),\ j\in \left\{N_d+1,\dotsc,N\right\}
\end{align}

Data sets of real values ($\mathcal{Y},\mathcal{U},\mathcal{V}$) with the subscript $d$ denote respective values to points $\mathcal{X}_d$, coming from distance sorting with respect to point $\mathbf{x}_c$. 
Next, a vector interpolant $\mathbf{p} \left(\mathbf{x}\right)$ is introduced to satisfy \autoref{eq:inter_grad} for the set $\mathcal{X}_d$, that is:
\begin{align}
    & \mathbf{p} \left(\mathbf{x}_i\right) = \left(u_i,v_i\right),\ \mathbf{x}_i\in \mathcal{X}_d,\ u_i \in \mathcal{U}_d,\ v_i \in \mathcal{V}_d
\end{align}

\begin{figure}[h]
    \centering
    \begin{tikzpicture}[scale=0.5]
    \draw [line width=0.5mm] (-5.46923,3.6461)-- (0,0) coordinate(O) -- (4.34769,-2.89846) coordinate(Oright);
    \draw [->,line width=0.5mm, color=blue!70] (0.,-3.) -- (2.22,-1.94);
    \draw [->,line width=0.5mm, color=blue!70] (4.,2.) -- (6.,5.);
    \draw [->,line width=0.5mm, color=blue!70] (1.,5.) -- (1.78,7.64);
    \draw [->,line width=0.5mm, color=blue!70] (-5.,-2.) -- (-4.18,-0.5);
    \draw [->,line width=0.5mm, color=red] (0.,0.) -- (1.22,1.86) coordinate(n);
    
    \draw [color=red, line width=0.5mm] (0.7,1.2) node[left] {$\overrightarrow{n}$};
    \draw [fill=orange] (0.,0.) circle  (1.3mm) node[below left] {$A$};
    
    \pic[ draw,<->,>=stealth,red!60!black, "\small{$90^{\circ}$}"{fill=white},inner sep=0.5pt, circle, angle eccentricity=1.1, angle radius = 7.mm] {angle = Oright--O--n};

    \draw [fill=purple] (0.,-3.) circle (1.3mm) node[below left] {$D_1$};
    \draw [fill=purple] (-5.,-2.) circle(1.3mm) node[below left] {$D_2$};
    \draw [fill=purple] (4.,2.) circle  (1.3mm) node[below left] {$D_3$};
    \draw [fill=purple] (1.,5.) circle  (1.3mm) node[below left] {$D_4$};
    \end{tikzpicture}
    \caption{Gradient approximation schematic}
    \label{fig:axis}
\end{figure}
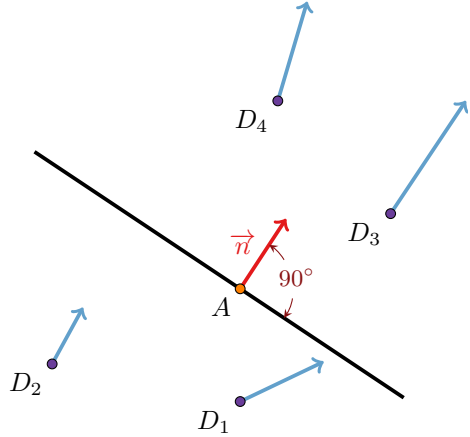

The vector function which interpolates $\mathbf{p}(\mathbf{x})$ is chosen to be the inverse distance averaging interpolant expressed in \autoref{eq:idw} applied in every component of the vector individually. A schematic for this procedure can be seen in \autoref{fig:axis}. Conventional linear interpolation methods such as barycentric interpolation or trilinear mappings are not used due the disorder of donors in space when unstructureds grids are involved. For the trilinear case, the disability to reform to a structured orientation is obvious on purely unstructured meshes (at least without an intermediate mapping), and for the barycentric case, we face several algorithmic obstacles that will force us to use triangulation algorithms (which, in contrast to kd-tree searching, are computationally heavy).

Proceeding with further manipulations and adding further notation
\begin{align}
    & \Vec{n} = \frac{\Vec{p}\left(\mathbf{x}_c\right)}{||\Vec{p}\left(\mathbf{x}_c\right)||} \\
    & A = \mathbf{x}_c && D_i = \mathbf{x}_i,\quad i=1,2,3,\dotsc,N_d 
\end{align}

a parametric curve is constructed (a line in 2D or 3D space) that reads

\begin{align}
    & \mathbf{x} = A + t \cdot \overrightarrow{n}\ , && t\in \mathbb{R}
\end{align}

Using the nomenclature from \autoref{fig:axis} we perform the following dot product.

\begin{align}
    & \overrightarrow{AD_i}\cdot \overrightarrow{n} = t_i && \overrightarrow{AD_i} = \overrightarrow{OA} - \overrightarrow{OD_i} \\
    & \mathbf{H}_i = \mathbf{A} + t_i \cdot \overrightarrow{n} && i=1,2,3,\dotsc,N_d
\end{align}

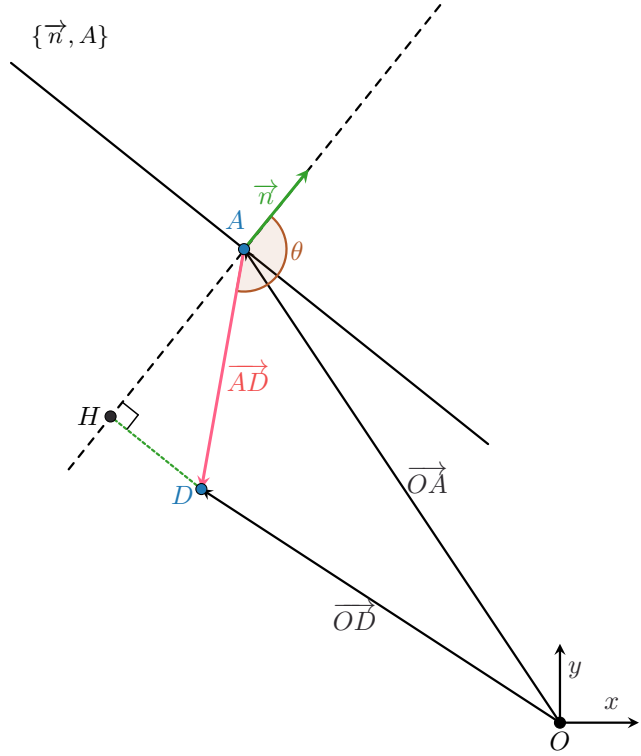
\begin{figure}[h]
    \centering

\begin{tikzpicture}[line cap=round,line join=round,>=triangle 45,x=1.cm,y=1.cm, scale=1.05]
   \draw [line width=0.3mm] (-6.95,8.36)-- (-0.91,3.53);
   \draw [-stealth,line width=0.3mm] (0.,0.) -- (-4.,6.);
   \draw [-stealth,line width=0.3mm] (0.,0.) -- (-4.54,2.96);
   \draw [-stealth,line width=0.3mm] (0.,0.) -- (1.,0.);
   \draw [-stealth,line width=0.3mm] (0.,0.) -- (0.,1.);
   \draw [line width=0.3mm,dash pattern=on 0.4mm off 0.4mm,color=green] (-4.54,2.96)-- (-5.69,3.88);
   \draw [line width=0.3mm,dash pattern=on 3pt off 3pt] (-6.22,3.21)-- (-1.47,9.15);
   \draw [shift={(-4.,6.)},line width=0.3mm,color=brown,fill=brown,fill opacity=0.10]  (0,0) --  plot[domain=-1.749:0.8961,variable=\t]({1.*0.5379*cos(\t r)+0.*0.537*sin(\t r)},{0.*0.5379*cos(\t r)+1.*0.5379*sin(\t r)}) -- cycle ;
   \draw [-stealth,line width=0.4mm,color=mpink] (-4.,6.) -- (-4.54,2.96);
   \draw [line width=0.2mm] (-5.54,4.06)-- (-5.34,3.90);
   \draw [line width=0.2mm] (-5.34,3.90)-- (-5.48,3.72);
   \draw [-stealth,line width=0.4mm,color=green] (-4.,6.) -- (-3.18,7.01);
   \draw [fill=blue] (-4.,6.) circle (0.7mm);
   \draw[color=blue] (-4.115,6.37) node {$A$};
   \draw [fill=blue] (-4.54,2.96) circle (0.7mm);
   \draw[color=blue] (-4.78,2.89) node {$D$} ;
   \draw [fill=black] (0.,0.) node[below] {$O$}circle (0.7mm);
   \draw[color=green] (-3.72,6.70) node {$\overrightarrow{n}$};
   \draw[color=mpurple] (-3.95,4.38) node {$\overrightarrow{AD}$};
   \draw[color=mblack] (-1.68,3.09) node {$\overrightarrow{OA}$};
   \draw[color=mblack] (-2.6,1.35)  node {$\overrightarrow{OD}$};
   \draw[color=mblack] (0.6550,0.25) node {$x$};
   \draw[color=mblack] (0.1932,0.67) node {$y$};
   \draw [fill=mblack] (-5.69,3.88) node[left] {$H$}circle (0.7mm);
   \draw[color=brown] (-3.33,5.98) node {$\theta$};
   \draw (-6.2,8.7) node {\small $\{\overrightarrow{n}, A\}$};
\end{tikzpicture}

    \caption{Projection of donor points}
    \label{fig:AA}
\end{figure}

The result of this projection gives an abscissa with respect to $A$ and a corresponding projected point $\mathbf{H}_i$. This means that when $t_i = 0$, $\mathbf{H}_i$ coincides with $A$, otherwise positive and negative values indicate orientation with respect to the direction of normal vector $\overrightarrow{n}$. A more detailed geometric description of the projection of donors is found at \autoref{fig:AA}.

The donor gradients are then reused to approximate values along the line in previously projected points using first order Taylor expansion:
\begin{align}
    & c_{i} = f(\mathbf{x}_i) + \nabla f(\mathbf{x}_i) \cdot \left(\mathbf{H}_{i} - \mathbf{x}_{i} \right)
\end{align}

from which the following is noted. A simple Taylor expansion from the nearest donor (before projection) to the desired point $\mathbf{A}$ would give unbounded results. In order to validate this claim, we tested the application of direct first order Taylor and the two-phase simulations diverged. The explanation is that, parallel to the direction of the field jump, the linear Taylor reconstruction method lacks boundness without the use of a limiter.


By the current use of a first order Taylor truncation, we ensure that the expansion is performed along a plane nearly perpendicular to the donor gradient direction. This indicates that the dot product from the expansion is calculated between two vectors with a large respective angle (close to $90^{\circ}$) and therefore along the vector $\overrightarrow{DH}_i$, the solution field is smooth.

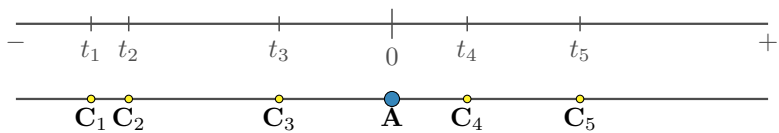
\begin{figure}[h]
    \centering
    \begin{tikzpicture}
        \draw[-,line width=0.3mm, color=black!70] (-5,1) node[below] {$-$} -- (5,1) node[below] {$+$};
        \draw[-,line width=0.2mm, color=black!70] (-4,0.9)  node[below] {$t_1$}--(-4,1.1);
        \draw[-,line width=0.2mm, color=black!70] (-3.5,0.9)node[below]{$t_2$}--(-3.5,1.1);
        \draw[-,line width=0.2mm, color=black!70] (-1.5,0.9)node[below] {$t_3$}--(-1.5,1.1);
        \draw[-,line width=0.2mm, color=black!70] (1.0,0.9) node[below] {$t_4$}--(1.0,1.1);
        \draw[-,line width=0.2mm, color=black!70] (2.5,0.9) node[below] {$t_5$}--(2.5,1.1);
        \draw[-,line width=0.2mm, color=black!70] (0.0,0.8) node[below] {$0$}--(0.0,1.2);
        \draw[-,line width=0.3mm, color=black!70] (-5,0) -- (5,0);
        \draw[fill=yellow!90] (-4.0,0) circle (0.5mm) node[below] {$\mathbf{C}_1$};
        \draw[fill=yellow!90] (-3.5,0) circle (0.5mm) node[below] {$\mathbf{C}_2$};
        \draw[fill=yellow!90] (-1.5,0) circle (0.5mm) node[below] {$\mathbf{C}_3$};
        \draw[fill=yellow!90] (+1.0,0) circle (0.5mm) node[below] {$\mathbf{C}_4$};
        \draw[fill=yellow!90] (+2.5,0) circle (0.5mm) node[below] {$\mathbf{C}_5$};
        \draw[fill=blue!90]  (+0.0,0)  circle (1.0mm) node[below] {$\mathbf{A}$};
    \end{tikzpicture}
    \caption{1D projected points along the axis}
    \label{fig:1d}
\end{figure}

Revising the above, all the points and their values are projected on an axis and the interpolation problem is reduced to a 1D problem of irregularly spaced abscissa (as shown in \autoref{fig:1d}) containing possibly irregular values. With $\mathcal{T} = \left\{t_1, t_2, \dotsc, t_{N_d} \right\} \subseteq \mathbb{R}$ and $\mathcal{C} = \left\{c_1, c_2, \dotsc, c_{N_d}\right\} \subseteq \mathbb{R}$ the abscissa and projected values are expressed respectively. Therefore, this problem is expressed below

\begin{align}\label{eq:inter_red}
    & g(t_i) = c_i, & t_i\in \mathcal{T},\ c_i \in \mathcal{C},\ i \in[1,N_d]
\end{align}

It is now obvious that an interpolation problem expressed from \autoref{eq:inter_f} and \autoref{eq:inter_grad} is now reduced to a more convenient problem (\autoref{eq:inter_red}).

Assuming that the last step involves interpolating on a steep direction, we are lead to choose low order schemes. To achieve a bounded result we work as follows:

\begin{itemize}
    \item \textbf{upwind}   : assuming all nodes lie on the negative side of point $A$
    \item \textbf{linear}   : assuming the existence of nodes on either side
    \item \textbf{downwind} : assuming all nodes lie on the positive side of point $A$
\end{itemize}

%
%
%

With the terms \textit{upwind} and \textit{downwind} we are not referring to the classical definition involving the dot product of the velocity vector with the face normal. However the definition in this case incorporates the direction of the approximated gradient on the calculation point and the orientation of the abscissa of $\mathcal{T}$.

A special case of the whole procedure is when donor gradients represent vectors with zero lengths (or below a given tolerance). This brings us to the conclusion that the field in the vicinity of the calculation point has no sharp changes and interpolation values have very small differences among the rest. In this case, the explicit inverse distance weighting in used once again, this time for the data $\mathcal{Y}$. That way, highly accurate results are obtained in very smooth regions, whereas the gradient-aware algorithm is enabled for stable results in steep regions.

Also, by sorting a small number $N_d$ of the total donor number, we have produced an interpolation scheme with a local nature, therefore ensuring that the computing weight of the interpolation task for all the receptors will scale linearly, i.e. $\mathcal{O}\left(N\right)$. 

\subsection{Correction for hydrostatic pressure}

When considering multi-phase flows under gravitational force, special attention has to be given on the hydrostatic portion of total pressure. The reason being that, in this family of problems, pressure varies linearly in the direction of gravity with a rate equal to $\rho_m g$ ($\rho_m$ and $g$ are the density of mixture and gravitational acceleration respectively as presented in appendix). Additionally, pressure is continuous along the interface of fluids, assuming the absence of surface tension, with a discontinuous gradient due to the density jump. The jump condition is $\left[\frac{\nabla p}{\rho_m}\right] = 0$. 
Further work on the treatment of reconstructions that cohere with these jump conditions can be found in \cite{ntouras2020coupled}. The rate of hydrostatic change becomes extreme when working with high-density fluids (such as water, in our case). Assuming  knowledge of the behavior of hydrostatic pressure, we proceed to add the exact hydrostatic pressure difference during its interpolation.



\subsubsection{Nearest-Neighbor Value}
With the use of the nearest-neighbor value interpolation scheme between two points, the hydrostatic difference is omitted. The addition is done in the following way.

It is assumed that, without any loss of generality, the gravitational force is applied in the $z$-direction. Indices $d$ and $r$ express donor and receptor values respectively.

\begin{align}
    & \rho_m = \alpha_l \rho_w + \left(1-\alpha_l \right) \rho_a \label{eq:hydro_dp}\\
    & \left|p(\mathbf{x}_r) - p(\mathbf{x}_d)\right| = 
    \rho_m\Big|_{a=a(\mathbf{x}_d)} g\delta z + 
    \mathcal{O}\left(\left|\left|\Delta \mathbf{x}\right|\right|_2^{n}\right),\quad n\geq 1,\quad \Delta \mathbf{x} = \mathbf{x}_r - \mathbf{x}_d\\
    & \delta z = z_d - z_r
\end{align}

In this case, a hydrostatic addition step is performed.

\begin{align}
    & p\left( \mathbf{x}_r\right) = p\left(\mathbf{x}_d\right) +\rho_m\Big|_{a_l=a_l(\mathbf{x}_d)} g\delta z
    \label{eq:press_corr}
\end{align}

\subsubsection{Gradient-Aware Method}

The pressure corrector step in \autoref{eq:press_corr} is not straightforward in the GA algorithm. More precisely, there could be a case where the volume fraction will be interpolated as a convex linear combination (it lies between a left and right donor) and the pressure will have a direct assignment (upwind or downwind). This may occur because, for every variable, a different line direction is constructed and the donors are not always both left and right of the calculation point.

This may seem to be a minor problem; however, errors due to hydrostatic inaccuracies in this case can lead to spurious velocities near the free surface because of the large density difference between the two fluids. This phenomenon is amplified when compressive flux reconstruction schemes are used, which means that the density difference between every donor and the receptor is larger. A remedy is to first interpolate the volume fraction (the last component of the solution vector $\overrightarrow{Q}$ defined in appendix) and then the pressure and velocity components. In that way, the hydrostatic term can be calculated with respect to the receptor value of volume fraction (\autoref{eq:press_corr_ga}), and then pressure addition may be done safely.

\begin{align}
    & \Delta p= \rho_m\Big|_{a_l=a_l(\mathbf{x}_r)} g\delta z
    \label{eq:press_corr_ga}
\end{align}



\section{Numerical Results}\label{section:results}

In this section, the performance of the proposed algorithm is evaluated through several test cases. The first test case involves an analytical solution, which provides a baseline for the accuracy and efficiency of the algorithm. The analytical solution can be used to validate the accuracy of the algorithm and to identify any sources of error or instability. The second test case involves a 2D water wave, which serves as a baseline case for free surface flows. The algorithm is evaluated based on its ability to accurately capture the wave propagation and to handle the sharp gradients in the solution variables. The third test case involves a heave decay test of a sphere, where comparison is made with experimental measurements. Finally, a strong scalability test is performed on the third test case in order to extract algorithmic performance information.
\subsection{Two-dimensional Interpolation}

Starting with the simplest case, we focus on a pure interpolation problem for a function of our choice. The purpose of this numerical experiment is twofold. On one hand, we have to ensure that the proposed algorithm gives strictly bounded results, which is the most important feature in such problems. Secondly, we test the performance of the interpolant in a smooth counterpart of the analytical function. In this context, to predict the gradient direction of $\nabla f$, central differences are used (extended for ghost nodes with analytical values) from the interpolation points.

The test function, is borrowed from F. Arandiga \cite{arandiga2020reconstruction} which was used for validation, although extended in two dimensions in our case. For this problem, structured points on a grid with known values taken from the following discontinuous function will be interpolated.

\begin{align}
    f(x,y) = \left\{\begin{array}{lr}
        e^{x+y-1/2} &, \left[0,0.5\right]\times \left[0,0.5\right] \\
        1+e^{x+y-1/2} &, \left[0,0.5\right]\times \left(0.5,1\right]\cup \left(0.5,1\right]\times\left[0,0.5\right]  \\
        2 + e^{x+y-1/2} &, \left[0.5,1\right]\times \left[0.5,1\right]
    \end{array}\right.
\end{align}

\begin{figure}[H]
    \centering
    \begin{subfigure}{0.48\textwidth}
        \includegraphics[width=1.\textwidth]{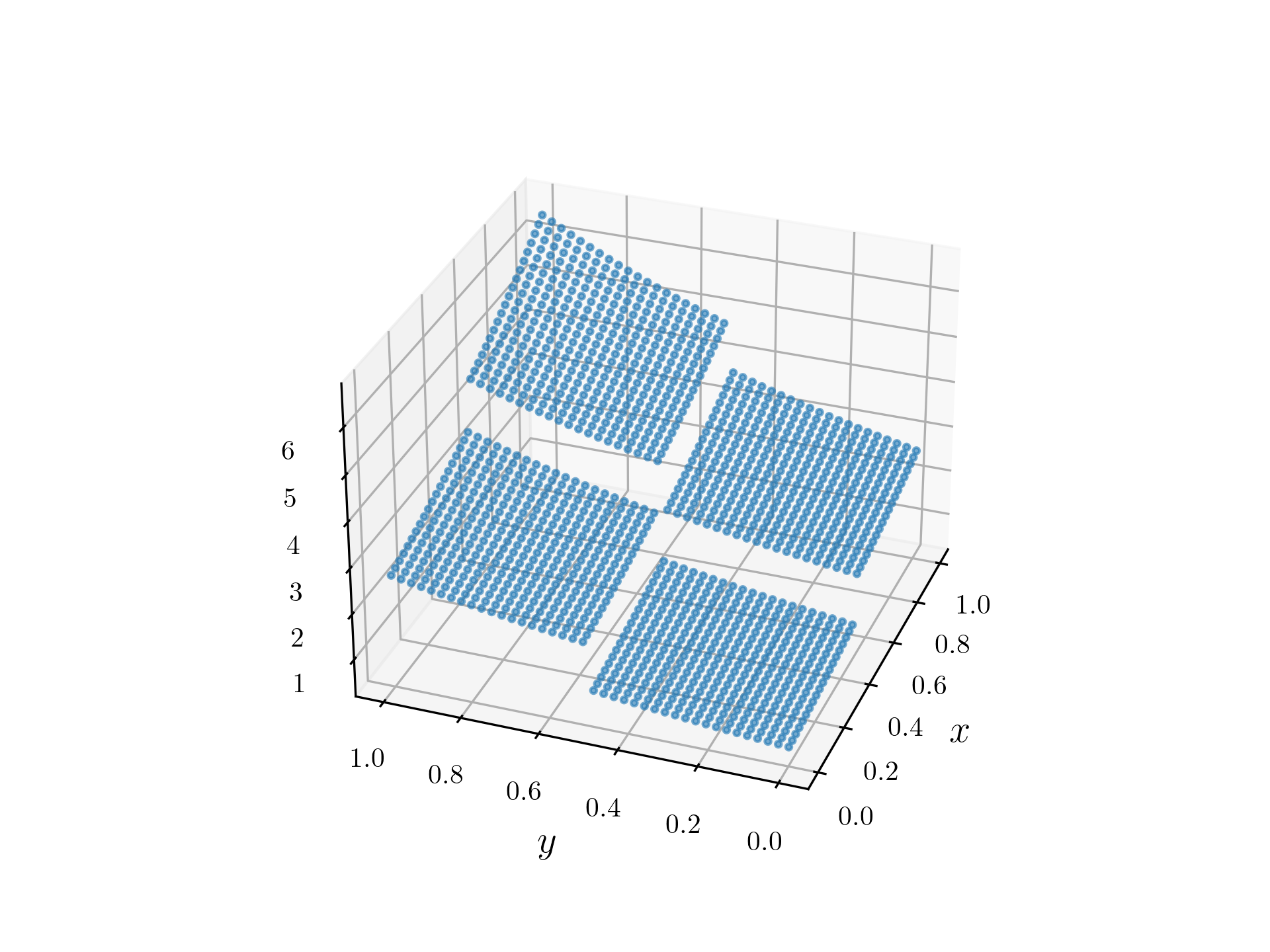}
    \end{subfigure}
    \begin{subfigure}{0.48\textwidth}
        \includegraphics[width=1.\textwidth]{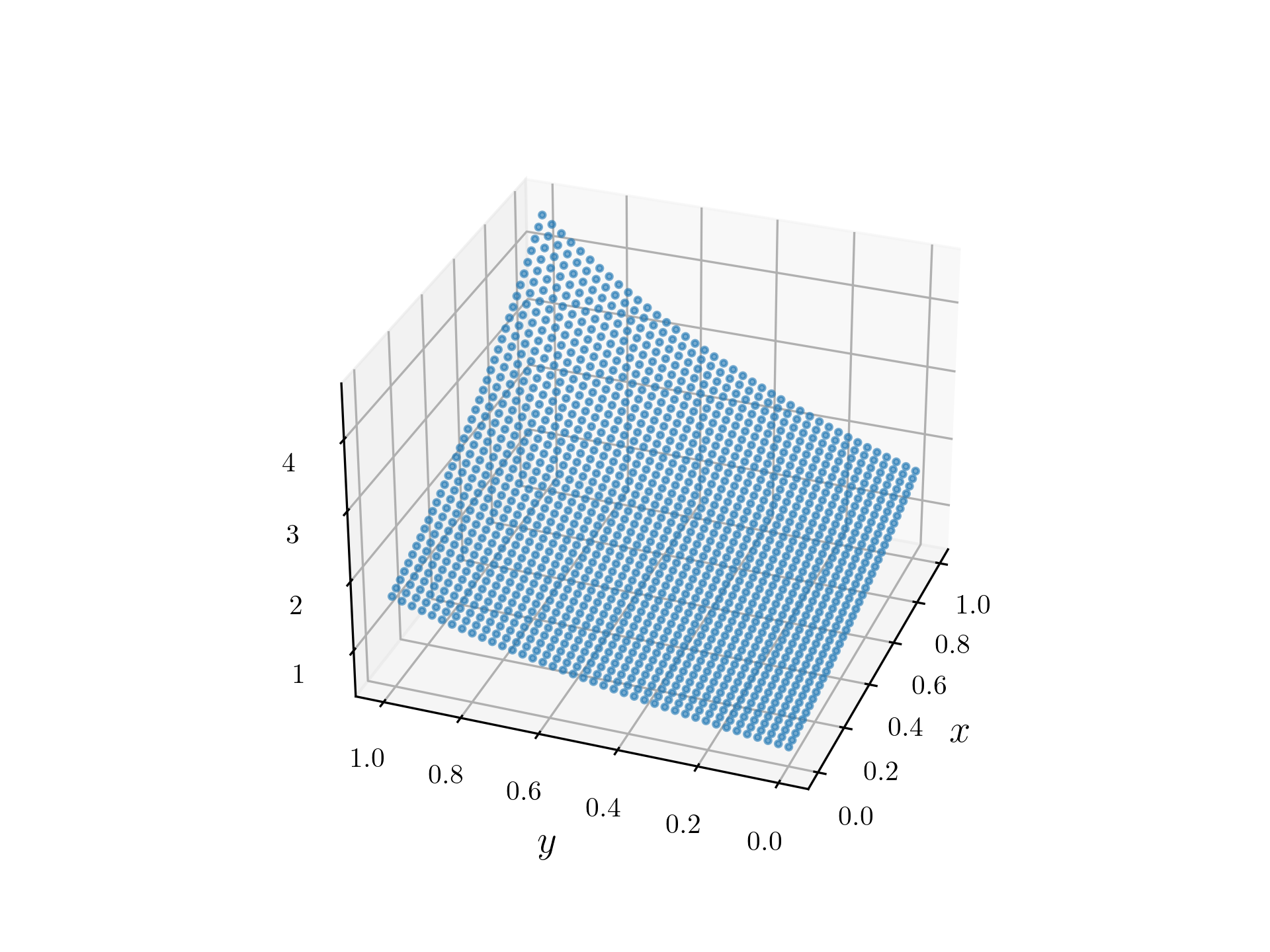}
    \end{subfigure}
    \caption{3D scatter representation of $f$ (left graph) and $g$ (right graph)}
    \label{fig:f_and_g}
\end{figure}

The continuous analogue to $f(x,y)$ is chosen to be $g(x,y) = e^{x+y-0.5}$. The two functions are visualized below in \autoref{fig:f_and_g}. In these two figures, the vertical axis of the graphs is the value of the function.

The calculation of error under $||\cdot||_{2},||\cdot||_{\infty}$ is as follows.

\begin{align}\label{eq:interp_function}
    & E\left(\mathbf{x}\right) = f\left(\mathbf{x} \right) - s\left(\mathbf{x} \right),\quad \mathbf{x} = \left(x,y\right) \\
    & \left|\left|E\left(\mathbf{x}\right) \right|\right|_2 = \left( \int_{\Omega} 
      \left|E\left(\mathbf{x}\right) \right|^2\, dx \right)^{1/2} \approx \left|\Omega \right|\left( \sum_{i=1}^{N_c} |E(\mathbf{x}_c)|^2\right)^{1/2} \\
    & \left|\left|E\left(\mathbf{x}\right) \right|\right|_{\infty} = \sup_{x \in \Omega} \left|E\left(\mathbf{x}\right) \right|
\end{align}

The domain of definition $\Omega = [0,1]\times[0,1]$ has a unit volume (area of a 3D surface). For the experiment, two sets of points are defined :
\begin{itemize}
    \item Interpolation points : $\mathcal{X}_I$
    \item Calculation points : $\mathcal{X}_C$
\end{itemize}
Both are slightly shifted with respect to the discontinuity curves so as to calculate $f$ with stability and $\mathcal{X}_C$ are 4 times as many as $\mathcal{X}_I$ per direction.

\begin{figure}[H]
    \centering
    \begin{subfigure}{0.48\textwidth}
        \includegraphics[width=\textwidth]{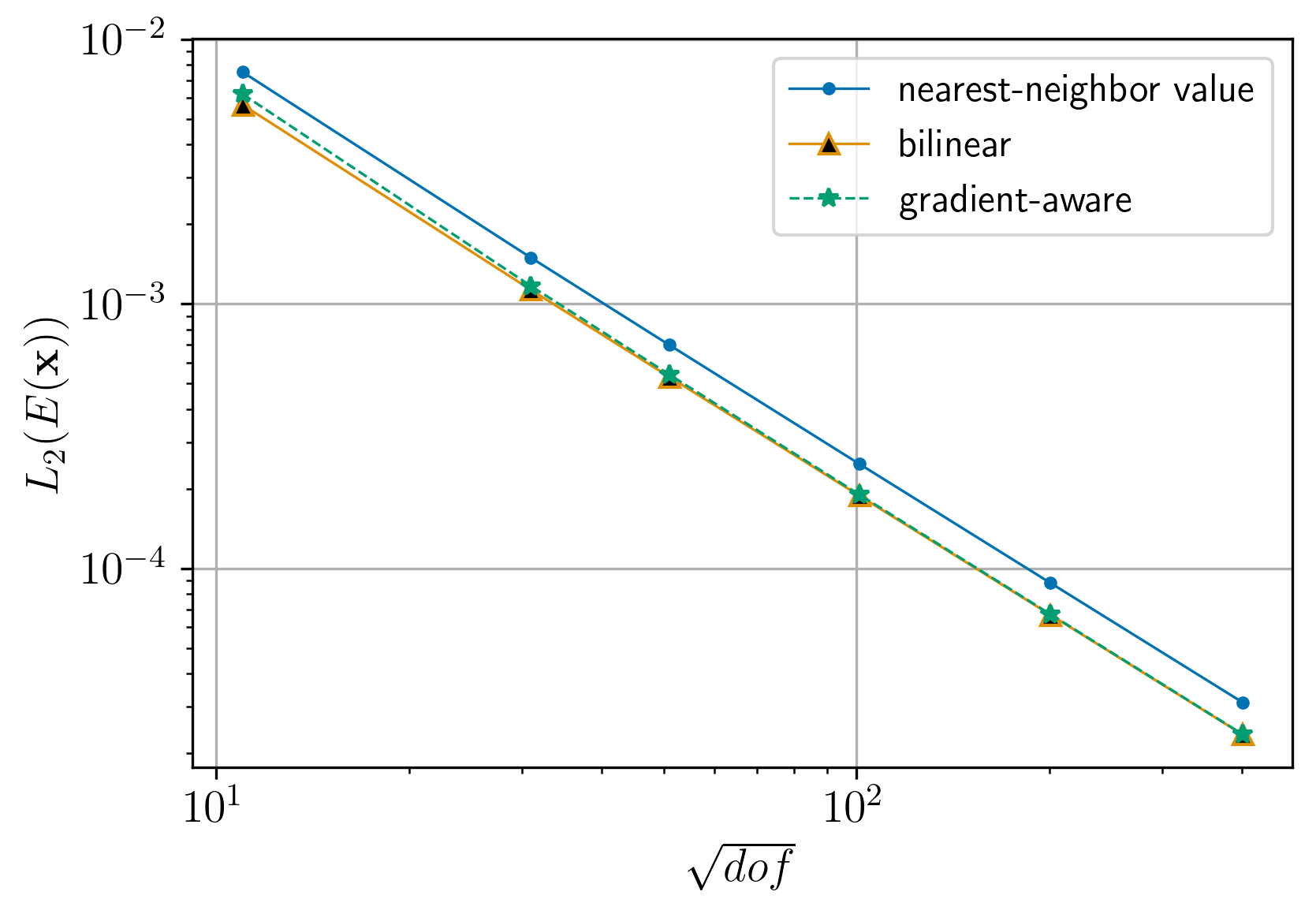}
    \end{subfigure}
    \begin{subfigure}{0.48\textwidth}
        \includegraphics[width=\textwidth]{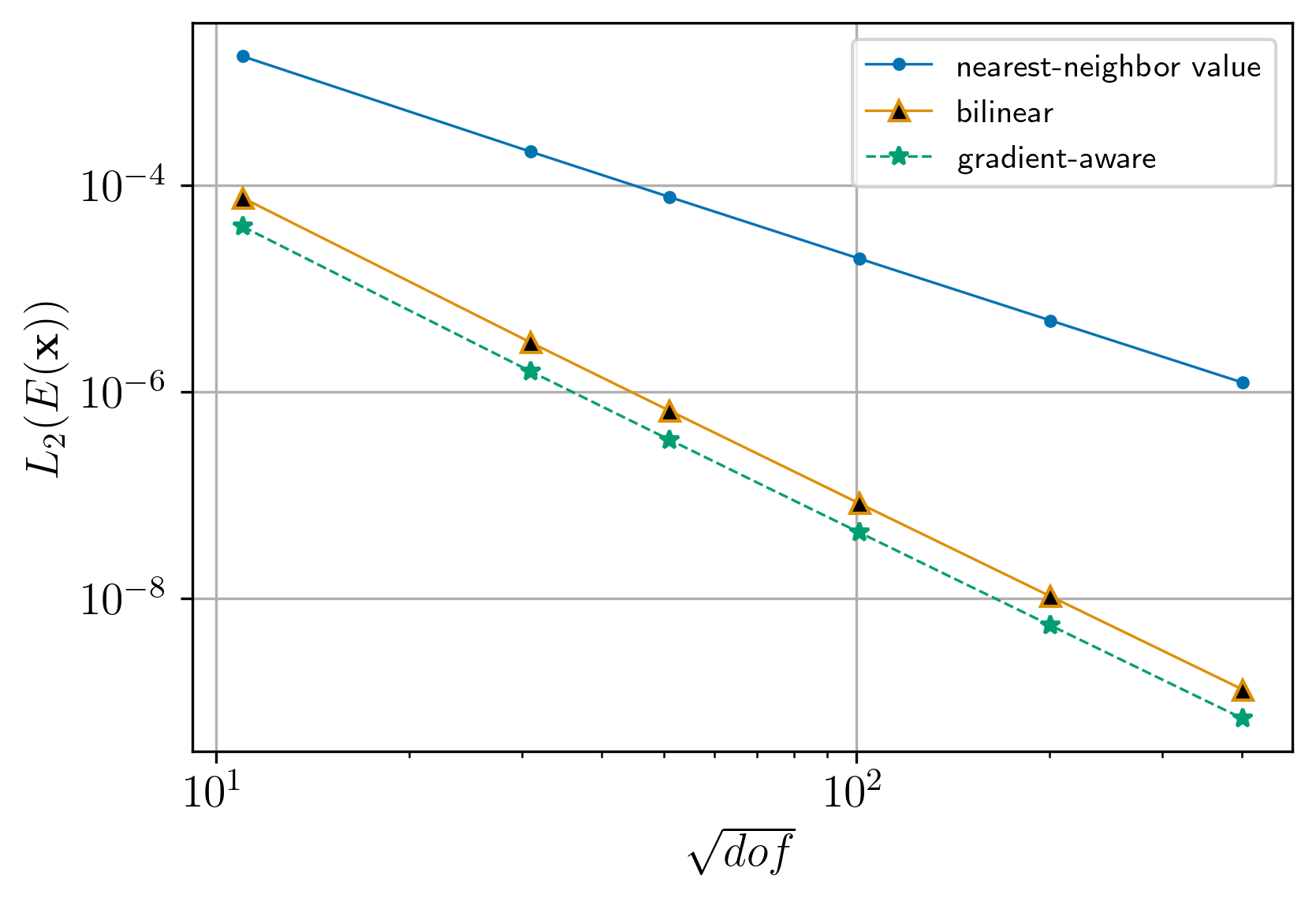}
    \end{subfigure}
    \caption{$L_{2}$ error over number of interpolation points per dimension $\sqrt{dof}$ for $f(\mathbf{x})$ (left) and $g(\mathbf{x})$ (right)}
    \label{fig:L2_loglog}
\end{figure}

\begin{figure}[H]
    \centering
    \begin{subfigure}{0.48\textwidth}
        \includegraphics[width=\textwidth]{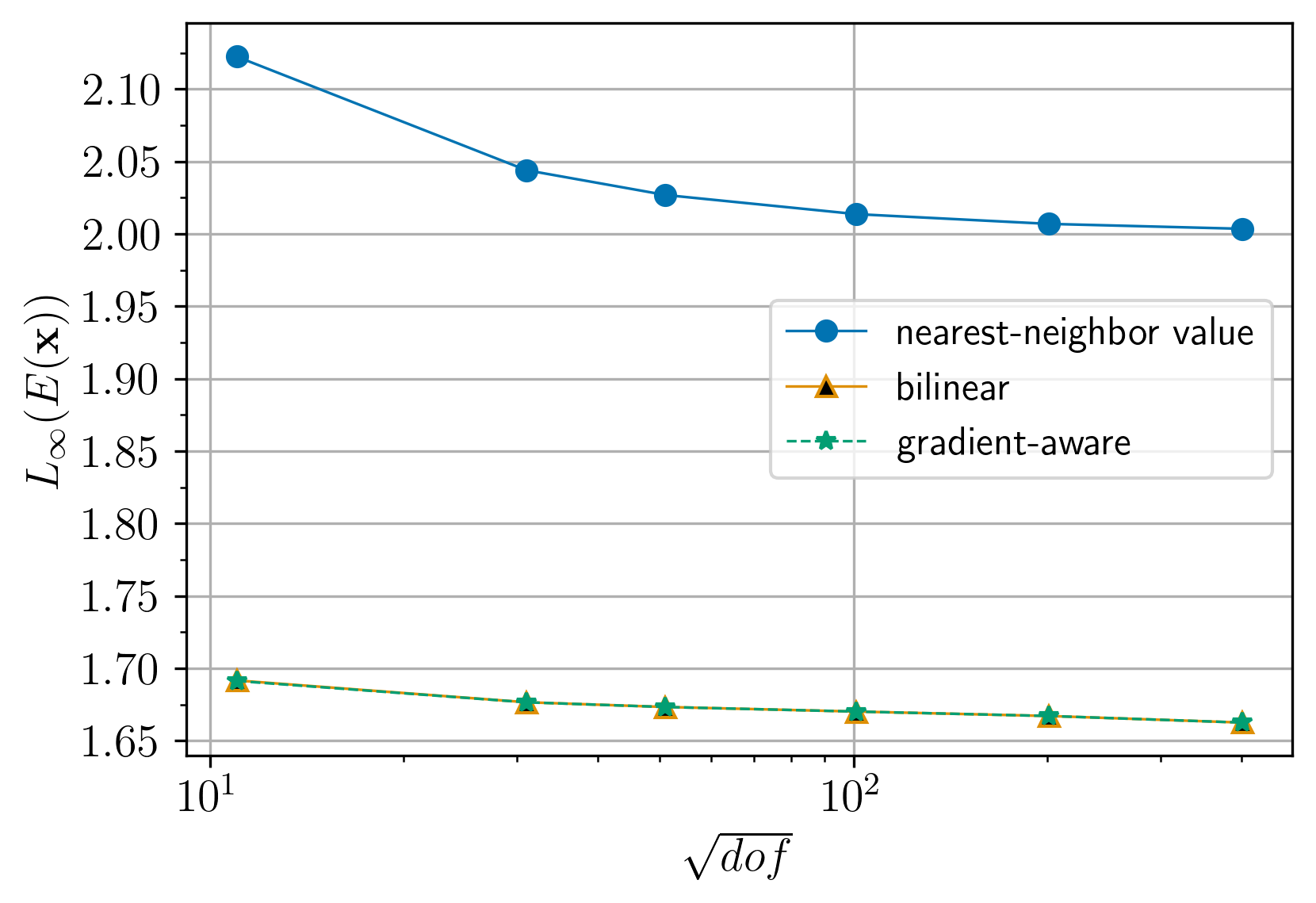}
    \end{subfigure}
    \begin{subfigure}{0.48\textwidth}
        \includegraphics[width=\textwidth]{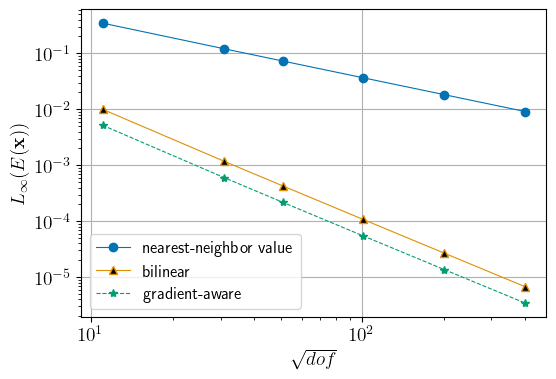}
    \end{subfigure}
    \caption{$L_{\infty}$ error over number of interpolation points per dimension $\sqrt{dof}$ for $f(\mathbf{x})$ (left) and $g(\mathbf{x})$ (right)}
    \label{fig:Linf_loglog}
\end{figure}

\begin{table}[H]
    \scalebox{0.72}{
    \centering
    \begin{tabular}{c|c|c|c|c|c|c}
        & \multicolumn{3}{|c|}{$f(x)$} & \multicolumn{3}{|c}{$g(x)$} \\ \hline
        $\sqrt{dof}$ & Nearest-Neighbor Value & Bilinear & Gradient-Aware & Nearest-Neighbor Value & Bilinear & Gradient-Aware\\ \hline \hline 
        11  & 7.52580E-03 & 5.66145E-03 & 6.21292E-03  &  1.78376E-03 & 7.59862E-05 & 4.00761E-05\\
        31  & 1.49797E-03 & 1.13244E-03 & 1.17074E-03  &  2.12410E-04 & 3.01991E-06 & 1.58321E-06\\
        51  & 7.00708E-04 & 5.30664E-04 & 5.41464E-04  &  7.75895E-05 & 6.61815E-07 & 3.46592E-07\\
        101 & 2.48863E-04 & 1.88714E-04 & 1.90564E-04  &  1.96115E-05 & 8.36385E-08 & 4.37671E-08\\
        201 & 8.81897E-05 & 6.68396E-05 & 6.71602E-05  &  4.93010E-06 & 1.05132E-08 & 5.49901E-09\\
        501 & 3.12153E-05 & 2.35964E-05 & 2.36521E-05  &  1.23568E-06 & 1.31792E-09 & 6.89148E-10
    \end{tabular}}
    \caption{$L_2$ error of interpolation of $f(\mathbf{x})$ and $g(\mathbf{x})$ for three candidate interpolants}
    \label{tab:L2_interp_f}
\end{table}

\begin{table}[H]
    \scalebox{0.72}{
    \centering
    \begin{tabular}{c|c|c|c|c|c|c}
        & \multicolumn{3}{|c|}{$f(x)$} & \multicolumn{3}{|c}{$g(x)$} \\ \hline
        $\sqrt{dof}$ & Nearest-Neighbor Value & Bilinear & Gradient-Aware & Nearest-Neighbor Value & Bilinear & Gradient-Aware\\ \hline \hline 
        11  & 2.12233e+00 & 1.69162e+00 & 1.69126e+00 &  3.45107e-01 & 9.97536e-03 & 5.17451e-03 \\
        31  & 2.04391e+00 & 1.67655e+00 & 1.67651e+00 &  1.20068e-01 & 1.17207e-03 & 5.98442e-04 \\
        51  & 2.02677e+00 & 1.67328e+00 & 1.67326e+00 &  7.25811e-02 & 4.26530e-04 & 2.17214e-04 \\
        101 & 2.01356e+00 & 1.67013e+00 & 1.67013e+00 &  3.64669e-02 & 1.07464e-04 & 5.46597e-05 \\
        201 & 2.00683e+00 & 1.66706e+00 & 1.66706e+00 &  1.82454e-02 & 2.69546e-05 & 1.37208e-05 \\
        501 & 2.00344e+00 & 1.66253e+00 & 1.66253e+00 &  9.09354e-03 & 6.74186e-06 & 3.44250e-06
    \end{tabular}}
    \caption{$L_{\infty}$ error of interpolation of $f(\mathbf{x})$ and $g(\mathbf{x})$ for three candidate interpolants}
    \label{tab:Linf_interp_f}
\end{table}

In \autoref{fig:L2_loglog}, \autoref{fig:Linf_loglog} and \autoref{tab:L2_interp_f}, \autoref{tab:Linf_interp_f}, $dof$ are the degrees of freedom for the interpolation problem, that is, the total number of interpolation points. With $\sqrt{dof}$, we refer to total number of interpolating points per dimension ($x$ or $y$).

Regarding error graphs and their respective tabular presentations, \autoref{fig:L2_loglog} and \autoref{tab:L2_interp_f} depict the error under the $L_2$ norm for the two testing functions. It is observed that the behavior of the proposed method together with the linear competitor implies a nearly equal rate of reduction. Additionally, in terms of error values, the proposed method can clearly outperform the nearest neighbor value scheme.

Moving to error under the $L_{\infty}$ norm, \autoref{fig:Linf_loglog} and \autoref{tab:Linf_interp_f} imply that $L_{\infty}$ is a decreasing function of $dof$ and in the case of $f$ specifically, it has values below $2$ which is the maximum height of discontinuity at point $(0.5,0.5)$. In other words, no overshoot is produced during the convergence. Apart from this, the new method converges with the same rate of a linear interpolator, and in smooth regions it outperforms other candidate methods. 


\subsection{Linear Wave Propagation}

In this validation case, a 2D water tank is constructed to simulate the propagation of 
a linear wave which coheres with the analytical solution extracted from the Airy wave 
theory. The goal in this experiment is to validate and assess the performance of the overset two-phase interpolation of a travelling discontinuity (volume fraction) over two 
stationary superimposed grids. The use of overset in such a simulation is trivial, but nonetheless, performance of the proposed method in terms of accuracy can be compared with the single grid solution. Moreover, this experiment serves as a first view of the overset procedure coupled with a solution approximated by FV.

In this case, we are using the two-phase Euler Equations to simulate a linear wave propagation in a 2D numerical water tank.

In order to assess accuracy, we will measure free-surface elevation along the direction of wave propagation before and after the effects of interpolation. The free-surface elevation in volume of fluid (VoF) simulations can be measured by finding the vertical position where volume fraction is equal to $0.5$. Of course, a cell with exact value $\alpha_l=0.5$ (see \autoref{appendix} for definition) does not exist, and therefore the height of the water column becomes is an implicit measurement of accuracy.

\begin{figure}[h]
    \centering
    \includegraphics[width=1\textwidth]{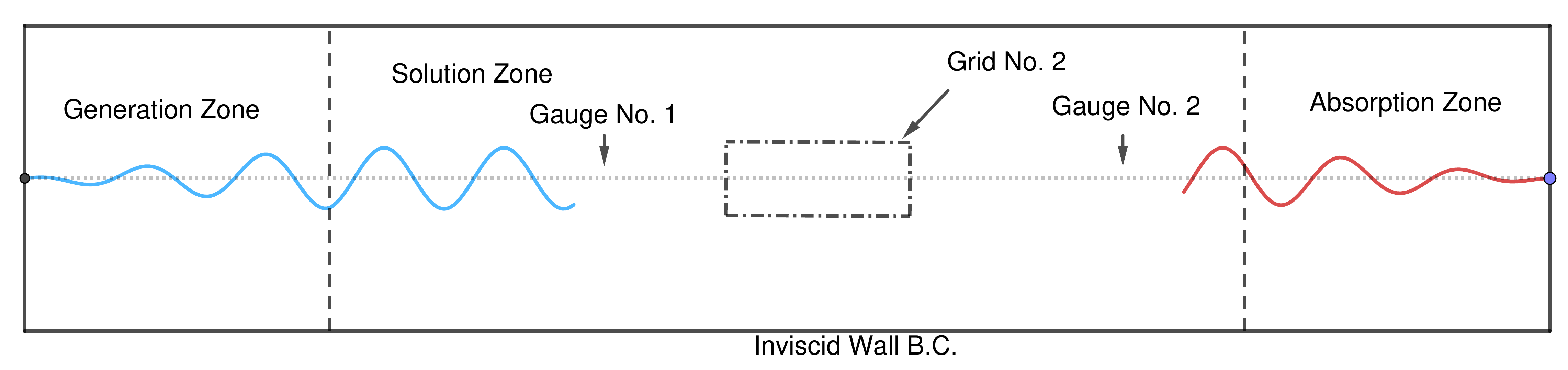}
    \caption{Schematic representation of the entire domain}
    \label{fig:2dwavetank}
\end{figure}

The background grid consists of three different zones along $x$-axis. Namely, the generation zone : $0-8\ m$, the solution zone : $8-24\ m$ and the damping (absorption) zone : $24-32\ m$. For a detailed description of the wave generation and absorption we refer to Ntouras et. al. \cite{ntouras2020coupled}. The generated wave properties are summarized in \autoref{tab:wave_char} while a schematic of the setup can be found in \autoref{fig:2dwavetank}. 

It is noted, that both grids have the same discretization in all directions. However, the inner grid (Grid No. 2 in  \autoref{fig:2dwavetank}) is slightly shifted both horizontally and vertically in order to avoid coincidence of cell coordinates.

\begin{table}[h]
    \centering
    \begin{tabular}{lrr}
        \multicolumn{3}{l}{Wave Details} \\\hline \hline 
        $H$  & $0.05$ & $m$ \\ \hline 
        $\lambda$ & $8$ & $m$ \\ \hline 
        $T$ & $2.243$ & $s$ \\ \hline 
        $g$ & $10.0$ & $m/s^2$ \\ \hline 
        $k$ & $0.7854$ & $1/m$ \\ \hline
        $\omega$ & $2.801$ & $rad/s$\\ \hline
    \end{tabular}
    \caption{Wave Details}
    \label{tab:wave_char}
\end{table}

Regarding mesh resolution, $150$ cells per wave length and $20$ per wave height were used according to \cite{ntouras2020coupled}. In order to extract wave signals, two stationary gauges are positioned (as shown again in \autoref{fig:2dwavetank} at a distance of $10\ m$ and $20\ m$ from the start of wave generation (from left)).

By positioning two stationary gauges before and after the overset grid in places where no interpolation takes place, we are trying to measure the loss of energy via amplitude reduction in the direction of wave propagation (left to right).


We first conduct a sensitivity study with one grid to validate that the results are convergent. Throughout this process, we increase the cell density by reducing cell dimensions by $10\ mm$ for every refinement.

\begin{figure}[H]
    \centering
    \begin{subfigure}[b]{0.48\textwidth}
        \includegraphics[width=0.98\textwidth]{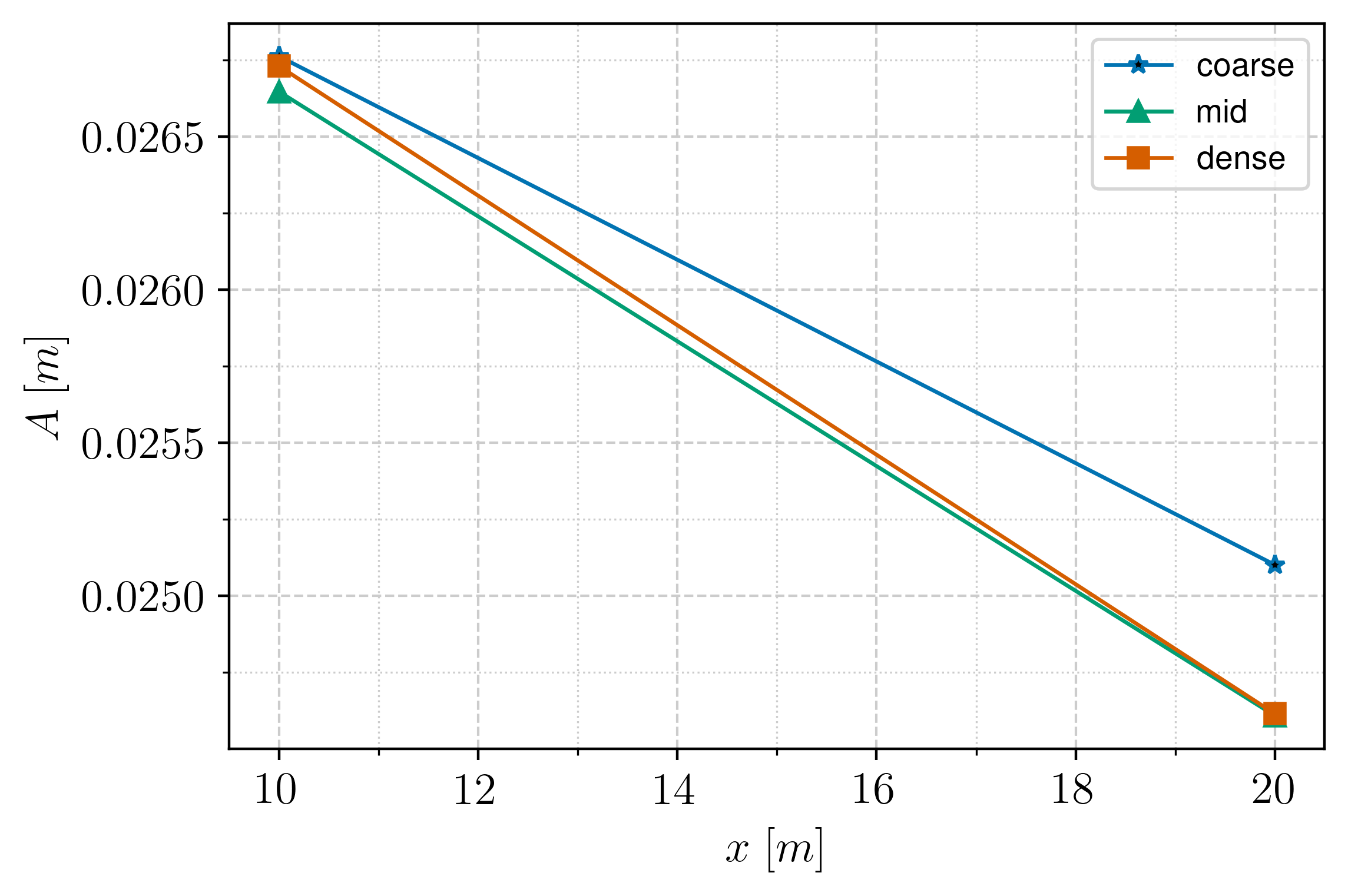}
    \end{subfigure}
    \begin{subfigure}[b]{0.48\textwidth}
        \includegraphics[width=0.98\textwidth]{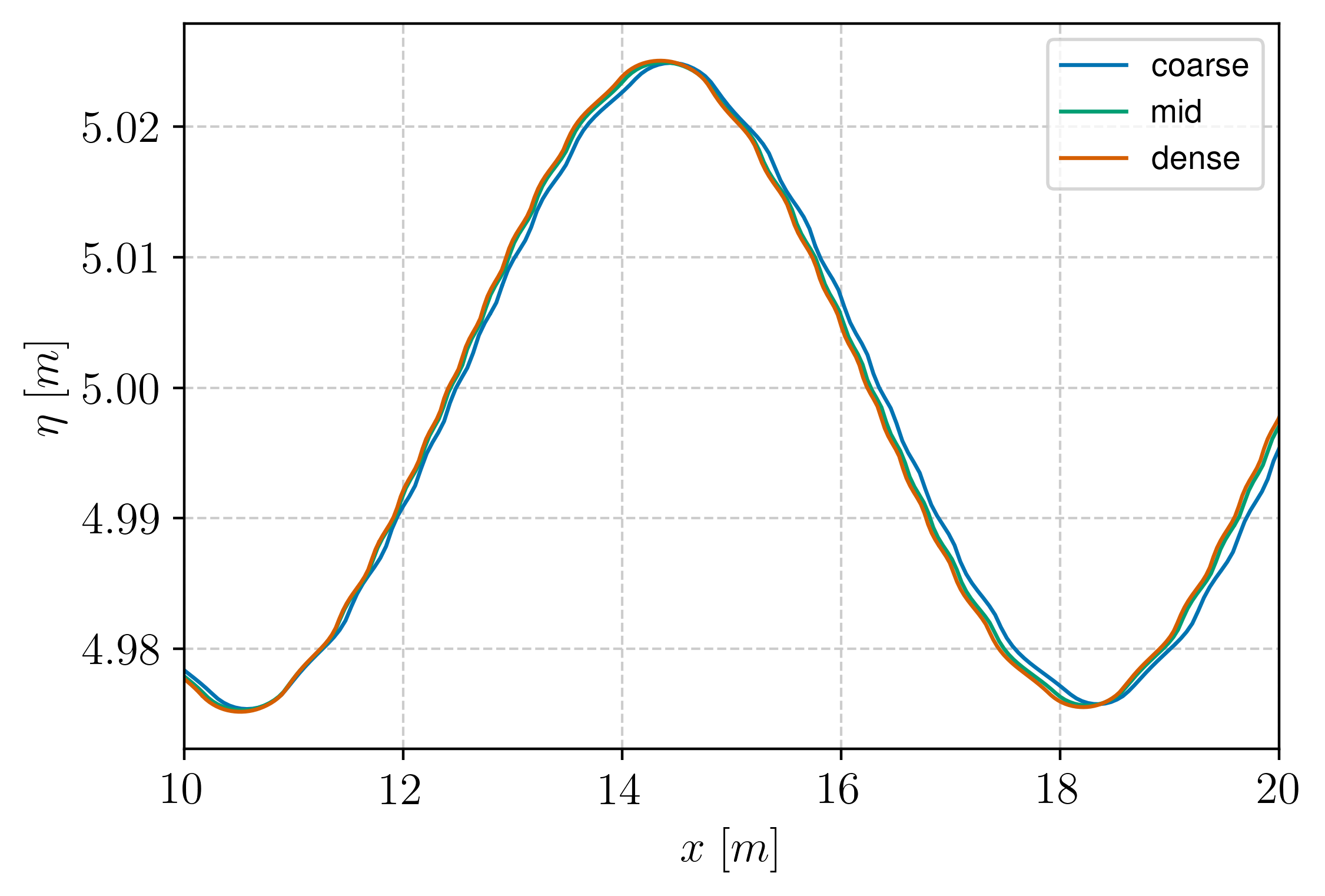}
    \end{subfigure}
    \caption{Sensitivity study with one grid. Left : the amplitude at the two gauges, Right : Free surface elevation over a portion of the solution domain}
    \label{fig:sens}
\end{figure}

By observing \autoref{fig:sens}, it is shown that the differences between middle and dense grids are reduced in comparison to the differences between the coarse and middle. As a result, the grids are properly designed. For the following comparison we use the middle grid.


Also, in the following comparison graphs, with NNV we refer to nearest-neighbor value and with GA we refer to gradient-aware algorithm. 

\begin{figure}[H]
    \centering
    \begin{subfigure}[b]{0.48\textwidth}
        \includegraphics[width=0.98\textwidth]{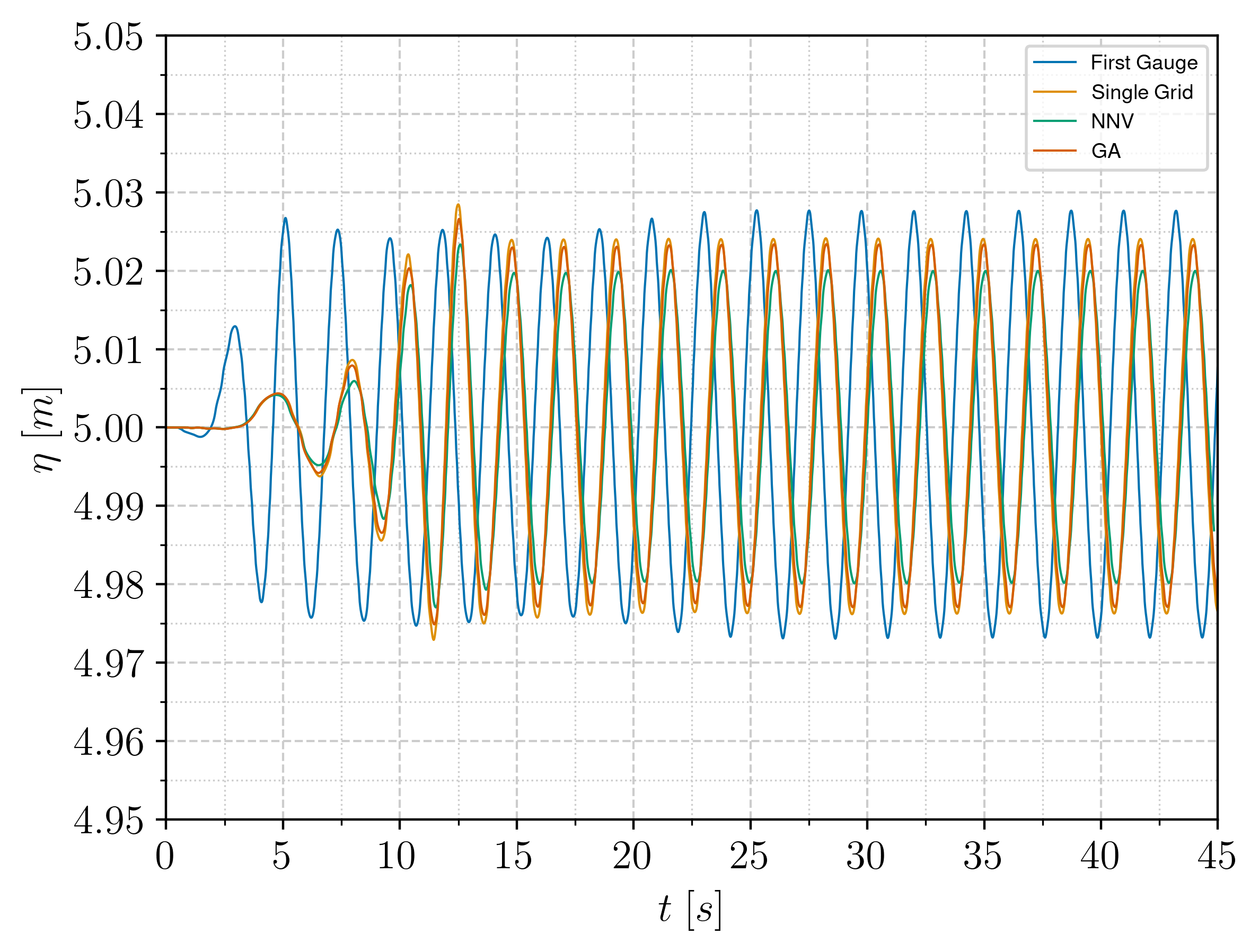}
    \end{subfigure}
    \begin{subfigure}[b]{0.48\textwidth}
        \includegraphics[width=0.98\textwidth]{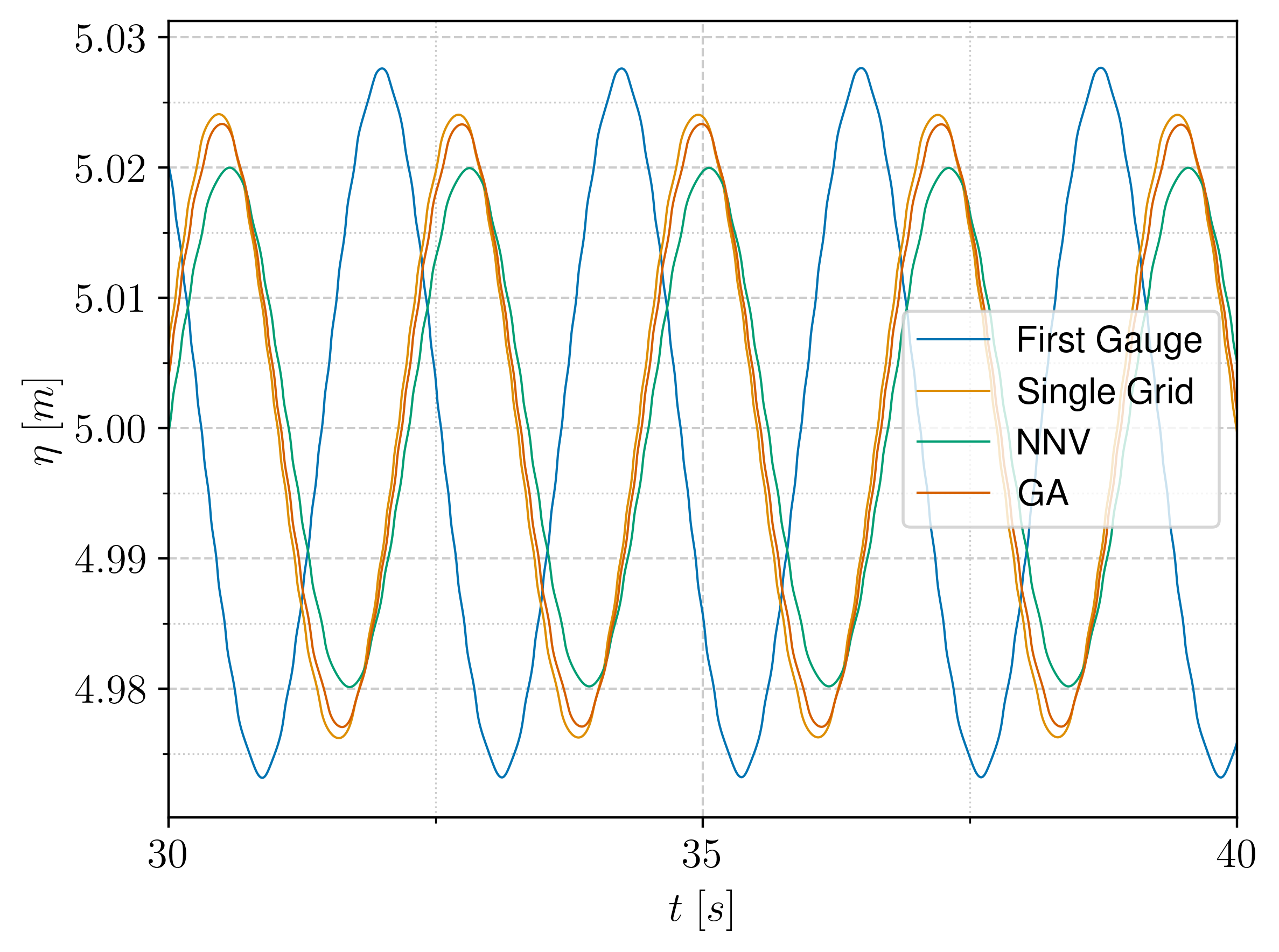}
    \end{subfigure} 
    \caption{Free surface elevation at two stations over time, with blue the signal from the first gauge at the single-grid simulation. The other three signals are the second gauge in three separate cases, each representing a one-grid simulation, a nearest neighbor value overset and a gradient-aware overset case, respectively. The right graph is a zoomed version of the left, focused on a smaller time interval.}
    \label{fig:signals_eta}
\end{figure}

In the two figures comprising \autoref{fig:signals_eta}, there are two remarks. First, the amplitude reduction between the single-grid and the GA simulation (measured both in the second gauge) has a very small difference compared to that of the single-grid and the NNV difference. 

Regarding the amplitude comparison, it is measured $35\ s$ into the simulation from the gauge signal and the normalized reduction is calculated
\begin{align}
    & \delta A_{norm} \% = \frac{A_{i,2nd} - A_{one-grid,1st}}{A_{one-grid,1st}}\cdot 100
\end{align}

where $i$ is one of three cases of simulation.

\begin{figure}[H]
    \centering
    \includegraphics[width=0.9\textwidth]{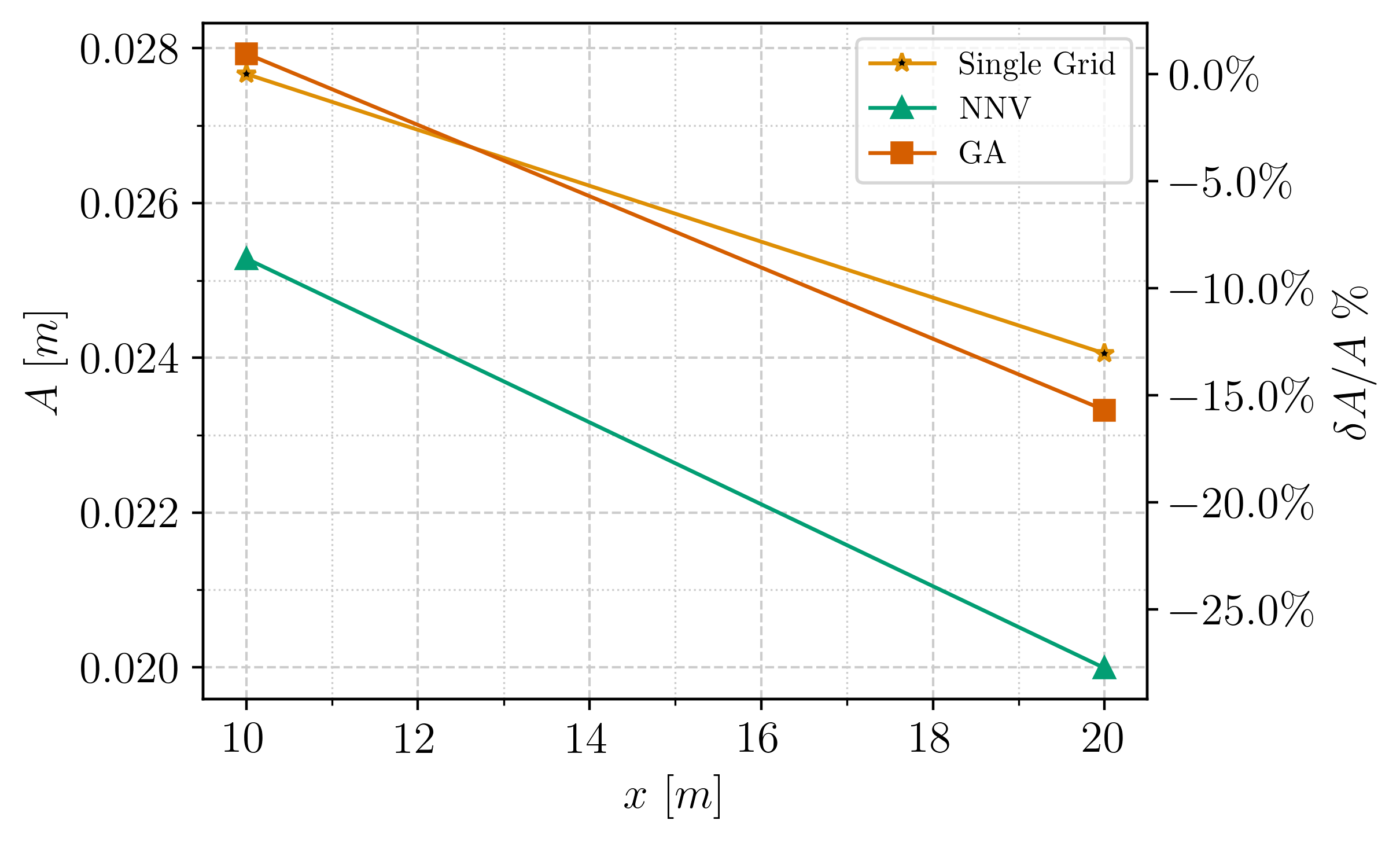}
    \caption{(left) Amplitude of wave over $x$-coordinate of two gauges indicating energy loss, (right) percentage of normalized amplitude reduction due to numerical diffusion and interpolation}
    \label{fig:ampl_red}
\end{figure}

\begin{table}[h]
    \centering
    \begin{tabular}{lcr}
        Simulations      & Wave Amplitude $[m]$ & Percentage of Reduction \\ \hline \hline 
        Single Grid, $1^{st}$ Gauge            & $0.0276$ & $-     \quad \%$ \\ \hline 
        Single Grid, $2^{nd}$ Gauge            & $0.0241$ & $13.054\quad \%$ \\ \hline
        Gradient-Aware, $2^{nd}$ Gauge         & $0.0233$ & $15.697\quad \%$ \\ \hline
        Nearest Neighbor Value, $2^{nd}$ Gauge & $0.0200$ & $27.747\quad \%$ \\ \hline
    \end{tabular}
    \caption{Wave amplitude values and reduction percentage with respect to 
    the $1^{st}$ station normalized.}
    \label{tab:height_loss}
\end{table}


As a first insight into \autoref{fig:ampl_red}, it is noted that amplitude reduction exists even with a single grid simulation. This is an expected result, given that numerical diffusion is an inherent flaw of low-order methods with the demand for a computational mesh.

It is observed that a large amount of the amplitude reduction is re-gained by the use of the new proposed scheme accounting for approximately $12\%$. Another observation at \autoref{fig:ampl_red} is that the first gauge signal also has differences for the three simulation setups (two overset with different interpolation and one single-grid setup), leading to the conclusion that the propagation of errors travels in both directions, affecting the flow in the opposite direction to the propagation vector (i.e., pointing left). More specifically, at the left gauge, the use of the NNV scheme accounts for approximately $7\%$ amplitude reduction with respect to the single grid measurement.

This concludes that in propagation problems, where general perturbations travel long distances through the whole computational domain, the preservation of accuracy allows for precise simulations with less diffusive results.

\subsection{Heave Decay of a floating sphere}\label{subsection:heave_decay}

In this last case, we seek to study the heave decay of a free-falling sphere of initially positive vertical distance from the undisturbed free surface. The purpose of this validating case is also twofold. Firstly, we would like to ensure that the entire overset algorithm gives valid results and is able to simulate surface-piercing (highly unsteady) phenomena on unstructured polyhedral grids. At the same time, results will be drawn from the performance of the gradient-aware interpolation algorithm in terms of scalability.

\subsection{Setup}

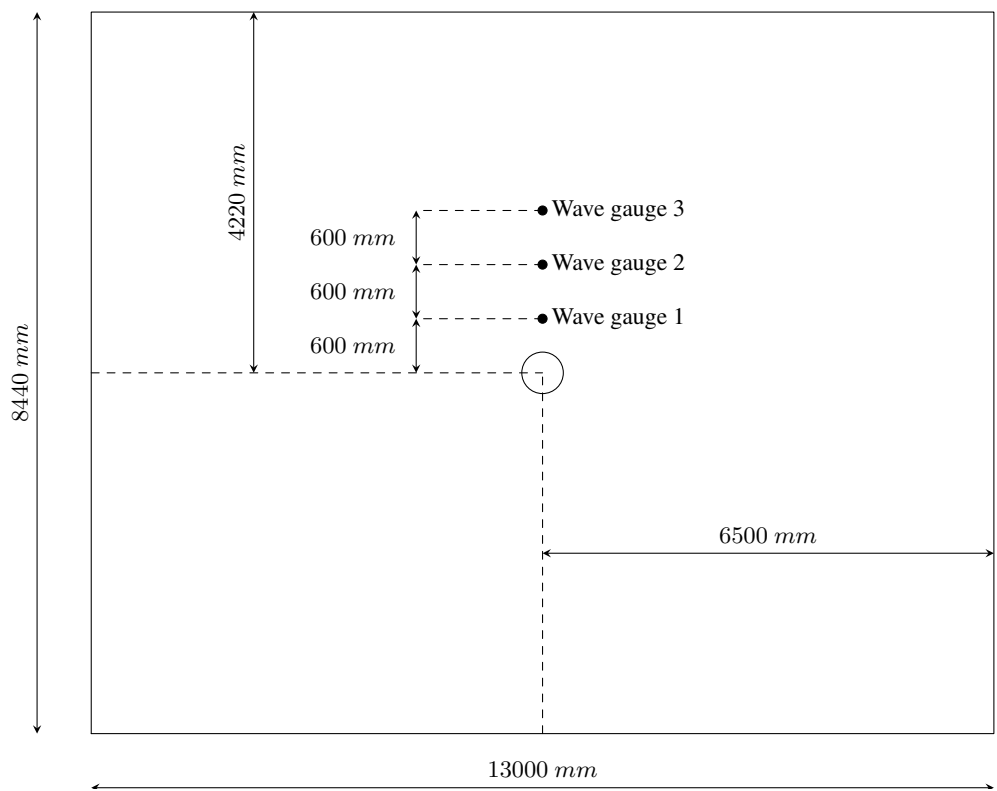
\begin{figure}[H]
    \centering
    \begin{tikzpicture}[line cap=round, line join=round, >=stealth, scale=1.2]
        \draw (0,0) -- (0,8) -- (10,8) -- (10,0) -- (0,0);
        \draw (5,4) circle (0.23);
        \draw [style=dashed] (5,0) -- (5,4);
        \draw [style=dashed] (0,4) --(5,4);
        \draw [style=solid, stealth-stealth] (5,2) -- (10,2);
        \draw [style=solid, stealth-stealth] (1.8,4) -- (1.8,8);
        \draw [style=solid, stealth-stealth] (-0.6,0) -- (-0.6,8);
        \draw [style=solid, stealth-stealth] (0,-0.6) -- (10,-0.6);
        \draw (5,-0.4) node {\small $13000\ mm$};
        \draw (-0.6,4) node[above,rotate=90] {\small $8440\ mm$};
        \draw (7.5,2.2) node {\small $6500\ mm$};
        \draw (1.8,6.) node[above,rotate=90] {\small $4220\ mm$};
        \foreach \i in {1,2,3} \draw[color=black, fill=] (5,4.+\i*0.60) circle (0.05) node[right] {\small Wave gauge \i};
        \foreach \i in {1,2,3} \draw[style=dashed] (5,4.+\i*0.60) -- (3.6,4.+\i*0.60);
        \foreach \i in {1,2,3} \draw (2.9,3.7+\i*0.6) node {\small $600\ mm$};
        \foreach \i in {1,2} 
        \draw [style=solid, stealth-stealth] (3.6,4.+\i*0.60+0.60) -- (3.6,4.+\i*0.60);
        \draw [style=solid, stealth-stealth] (3.6,4.60) -- (3.6,4);
    \end{tikzpicture}
    \caption{Experimental water tank representation}
    \label{tikz:watertank}
\end{figure}

This validating case was inspired by the study of heave decay from a consortium of laboratories, whose work can be found in \cite{kramer2021highly}. The experiments (numerical \& physical full-scale) were conducted for 3 initial heights, namely $0.1D,\ 0.3D,\ 0.5D$ and the solution domain is shown in the schematic representation of \autoref{tikz:watertank}.

The main features of the experiment are shown below.

\begin{table}[H]
    \centering
    \begin{tabular}{|clcr|}
        \hline
        Description & Properties & Values & Unit \\ \hline \hline 
        diameter & $D$ & $300$ & $mm$ \\
        mass & $m$ & $7056$& $g$ \\
        center of gravity & CoG  & $(0,0,-34.8)$ & $mm$ \\
        acceleration of gravity  & $g$ & $9.82$ & $m/s^2$ \\
        starting altitude & $H_0$& $150$ & $mm$ \\
        depth & $d$              & $900$ & $mm$ \\
        density of water & $\rho_w$ & $998.2$ & $kg/m^3$ \\
        density of air & $\rho_a$ & $1.2$ & $kg/m^3$ \\
        water kinematic viscosity & $\nu_w$ & $1.0\cdot 10^{-6}$ & $m^2/s$ \\
        air kinematic viscosity & $\nu_a$ & $15.1\cdot 10^{-6}$ & $m^2/s$ \\\hline
    \end{tabular}
    \caption{Heavy decay flow particulars. Definitions can be found in appendix.}
    \label{tab:heave_setup}
\end{table}

Here we will focus in one of the three cases, where the sphere is completely out of the water, i.e. $H_0 = 150\ mm$. The simulation setup can be found in \autoref{tab:heave_setup}. In the present work it is assumed that only one degree of freedom exists (heave), which is an adequately accurate assumption for a perfectly azimuth-symmetric numerical experiment.

In this case, the CFD solver is coupled to rigid body dynamics (RBD). Each time-step is divided into sub-steps in which the CFD and the RBD solver exchange forces and $CoG$ position and velocity, respectively. The time integration for the RBD solver is performed by the Newmark-beta method with coefficients $\gamma=0.5\ \& \ \beta=0.25$. In this numerical experiment, flow is considered to be turbulent and for this reason, a turbulence model is used. We proceed to use the modified $k-\omega$ SST model \cite{wilcox2008formulation} carefully modified for free-surface flows by B. Devolder \emph{et. al.} \cite{devolder2017application}. For more information on the coupling algorithm and the turbulence models the reader is referred to \cite{theodorakis2022investigation}.

Regarding the numerical parameters, the maximum value of $CFL$ for the set of simulations that where conducted was calculated to be $\approx 0.7$. The resulting time-step was set at a constant value of $dt = 5\cdot 10^{-4}\ s$. In all of the simulations we measured the $y^+$ value to be at approximately $1$ and no wall functions were employed.

\subsection{Sensitivity Results \& Meshing}

In order to provide a study over h-refinement for convergence purposes we proceed to decrease basic grid dimensions by $1\ mm$ at a time. This reduction is performed in horizontal and vertical background grid dimensions in the vicinity of the free surface as well as the overset cell size of the body-fitted grid. Lastly, we proceed with reduction of the wall surface mesh size by the same amount. As a result, the most dense grid has a free-surface cell height $h_c=3\ mm$ and length near the sphere $l_c = 5\ mm$. The total cell sizes of the three grids are approximately $6$, $10$ and $18$ million starting from the coarse to the most dense grid.

\begin{figure}[H]
    \centering
    \begin{subfigure}[b]{0.32\textwidth}
        \includegraphics[width=1.\textwidth]{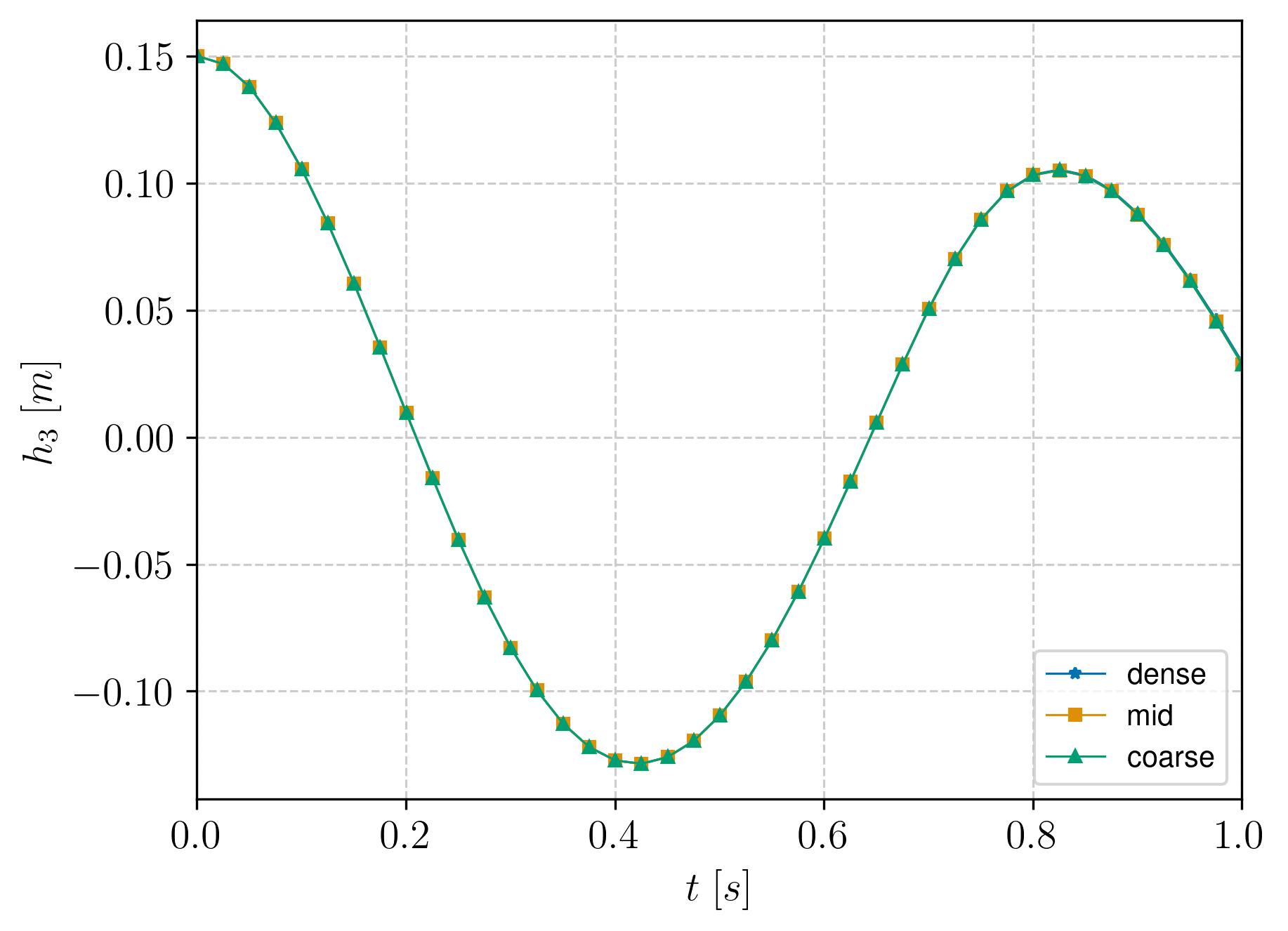}
    \end{subfigure}
    \begin{subfigure}[b]{0.32\textwidth}
        \includegraphics[width=1.\textwidth]{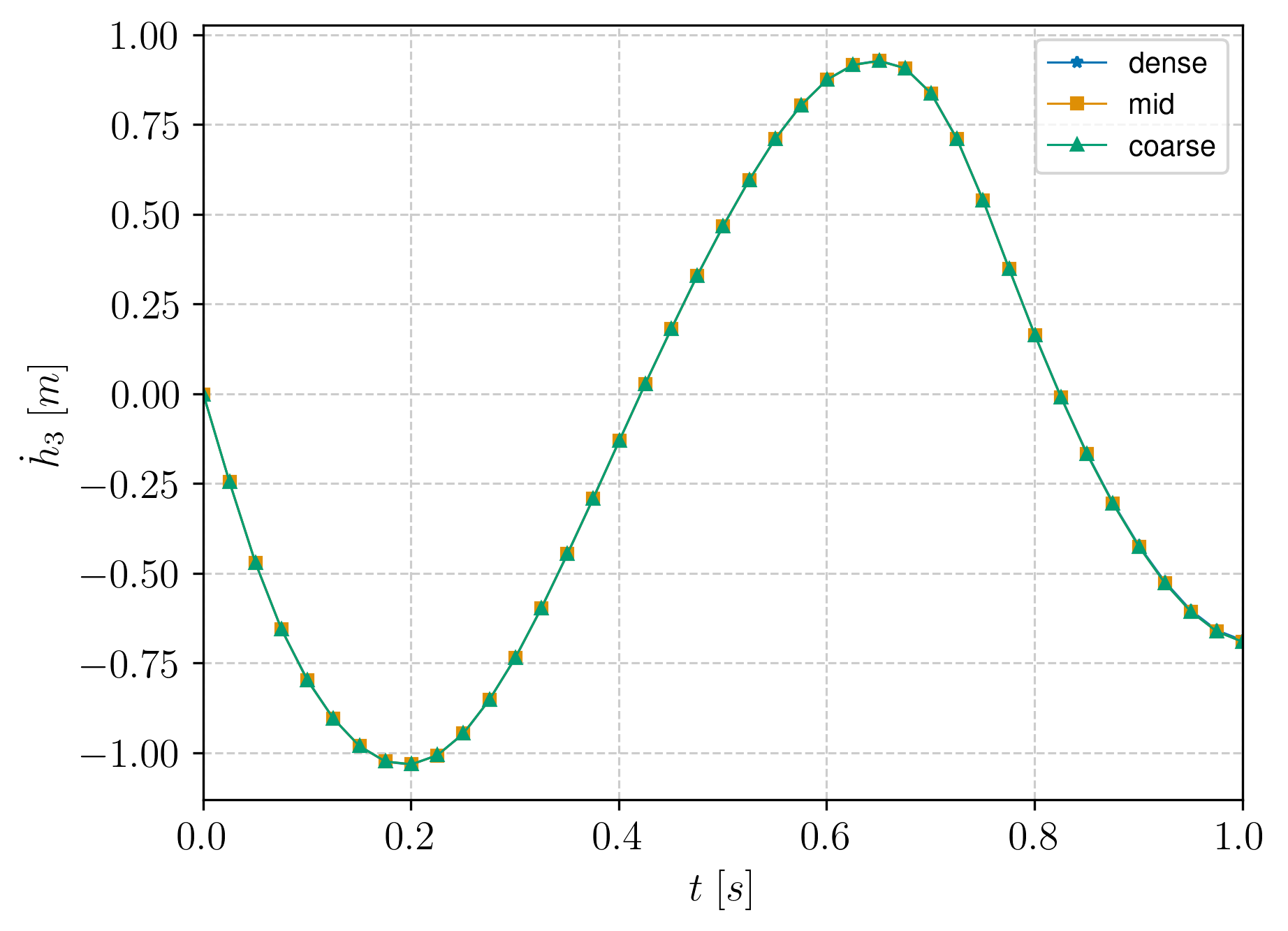}
    \end{subfigure}
    \begin{subfigure}[b]{0.32\textwidth}
        \includegraphics[width=1.\textwidth]{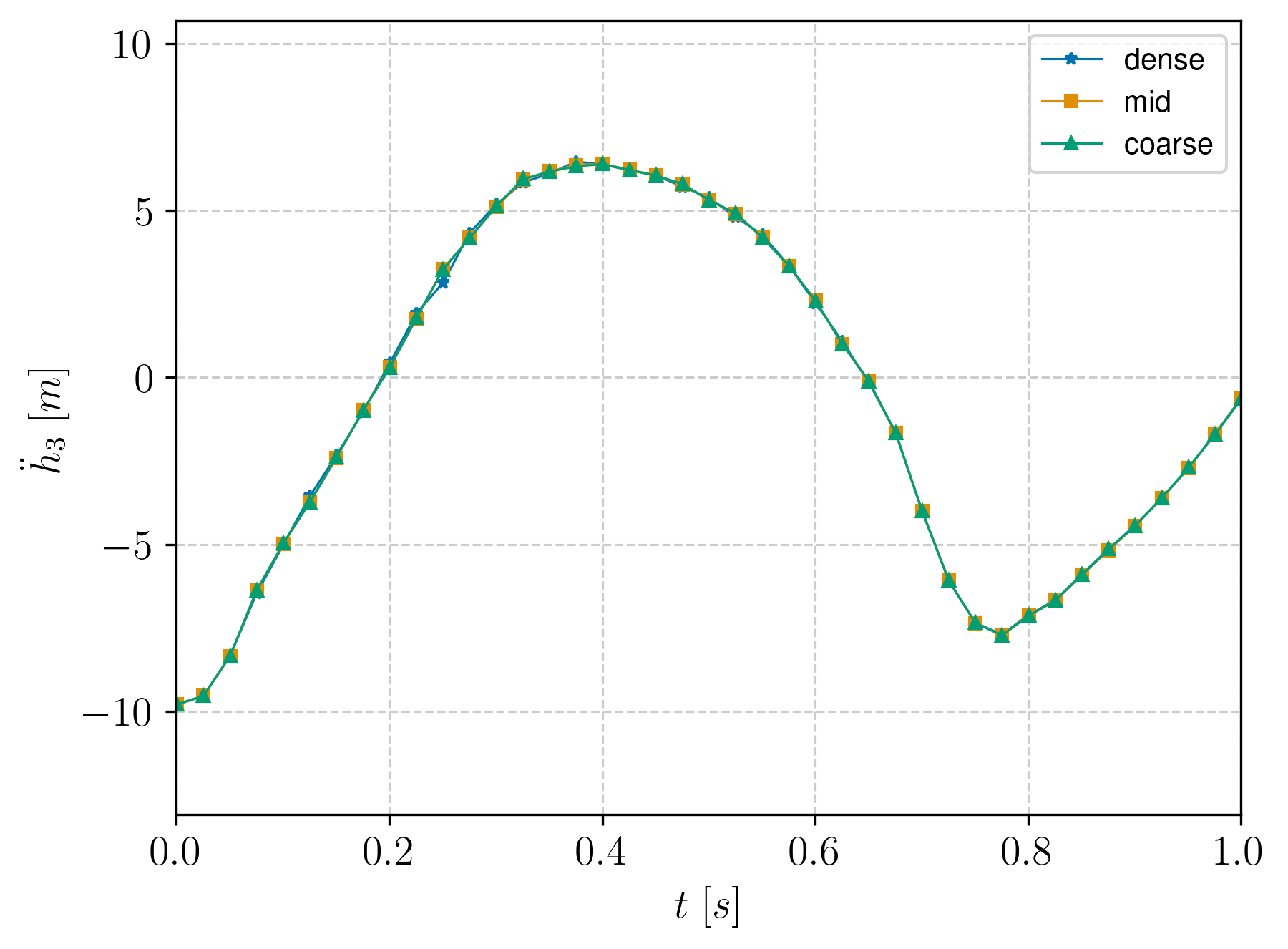}
    \end{subfigure}
    \caption{Sensitivity study for the heave displacement (left), heave velocity (middle), heave acceleration (right) signals.}
    \label{fig:sens_sphere}
\end{figure}

The response signals from vertical displacement, velocity and acceleration are shown in \autoref{fig:sens_sphere}.

In \autoref{fig:sens_sphere}, the signal sampling is at a rate of 50 time-steps in order to give a better visualization of the 3 curves. We observe that the response has very small changes with respect to grid refinement, and therefore the heave decay solution will not change relative to cell density. To proceed with the validation, we will use the dense grids in order to achieve the best possible description of the free surface elevation and to further examine the performance of the coupled overset solver in large data simulations.

\begin{figure}[h]
    \centering
    \begin{subfigure}[b]{0.32\textwidth}
        \includegraphics[width=1.0\textwidth]{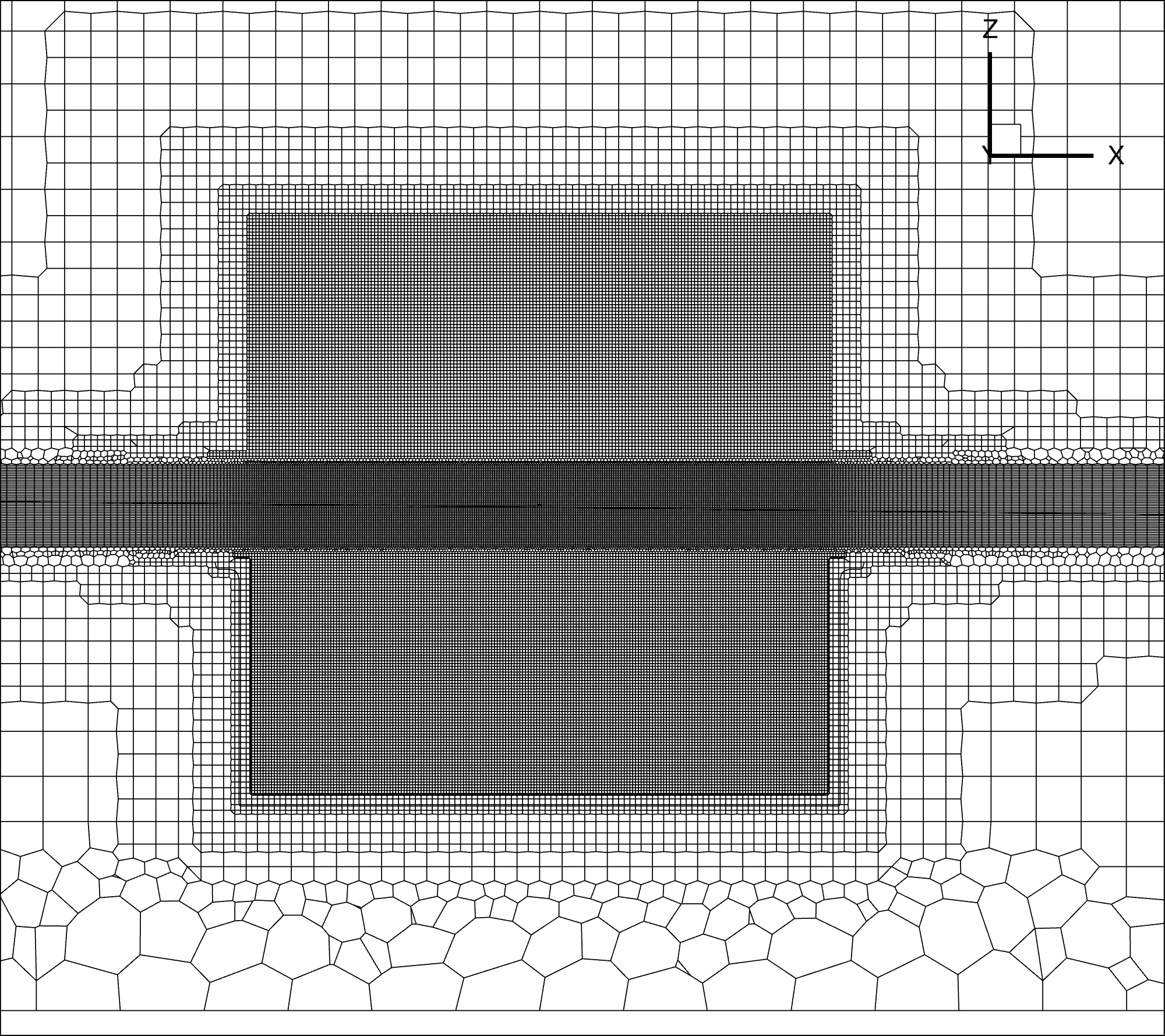}
    \end{subfigure}
    \begin{subfigure}[b]{0.32\textwidth}
        \includegraphics[width=1.0\textwidth]{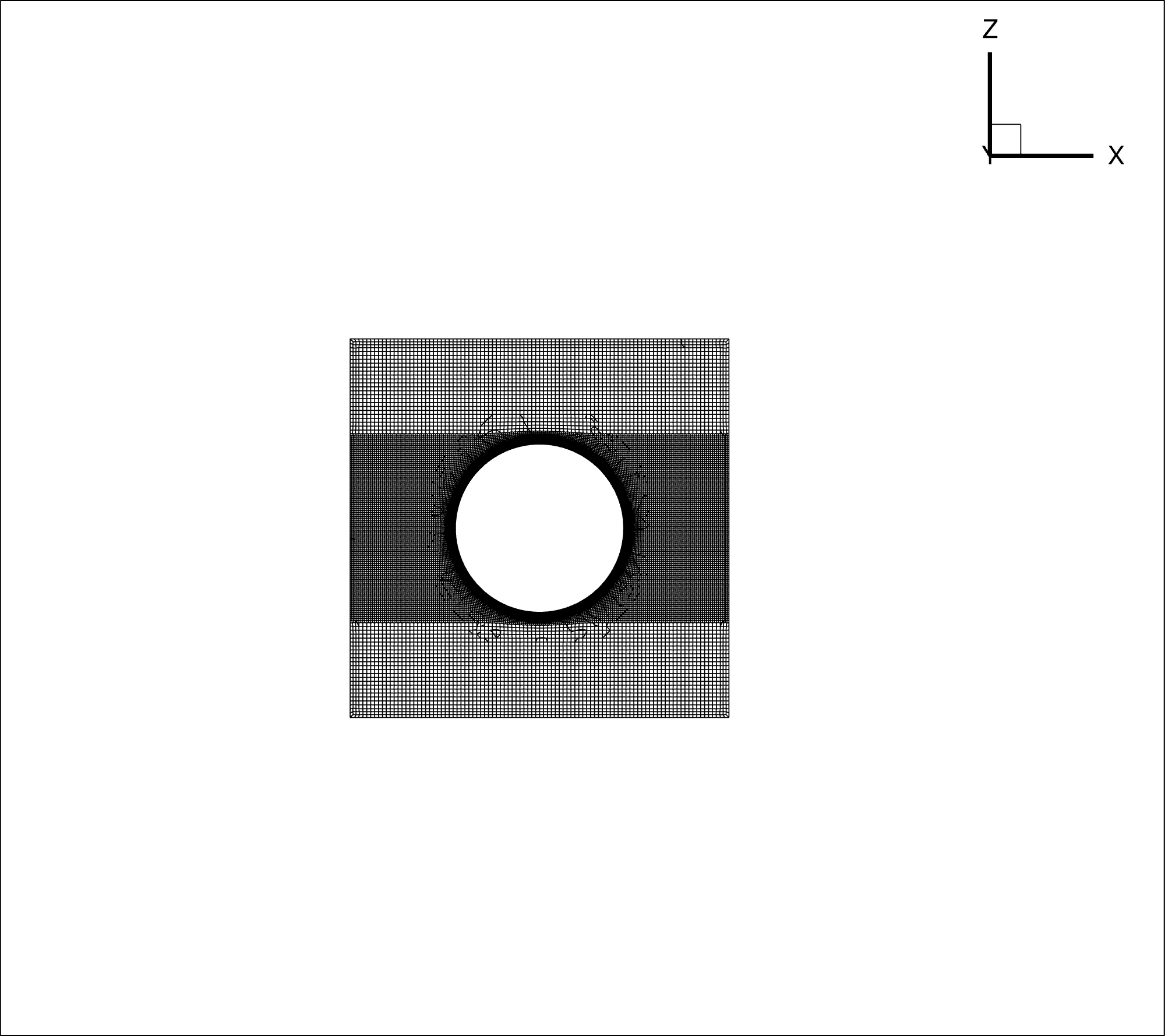}
    \end{subfigure}
    \begin{subfigure}[b]{0.32\textwidth}
        \includegraphics[width=1.0\textwidth]{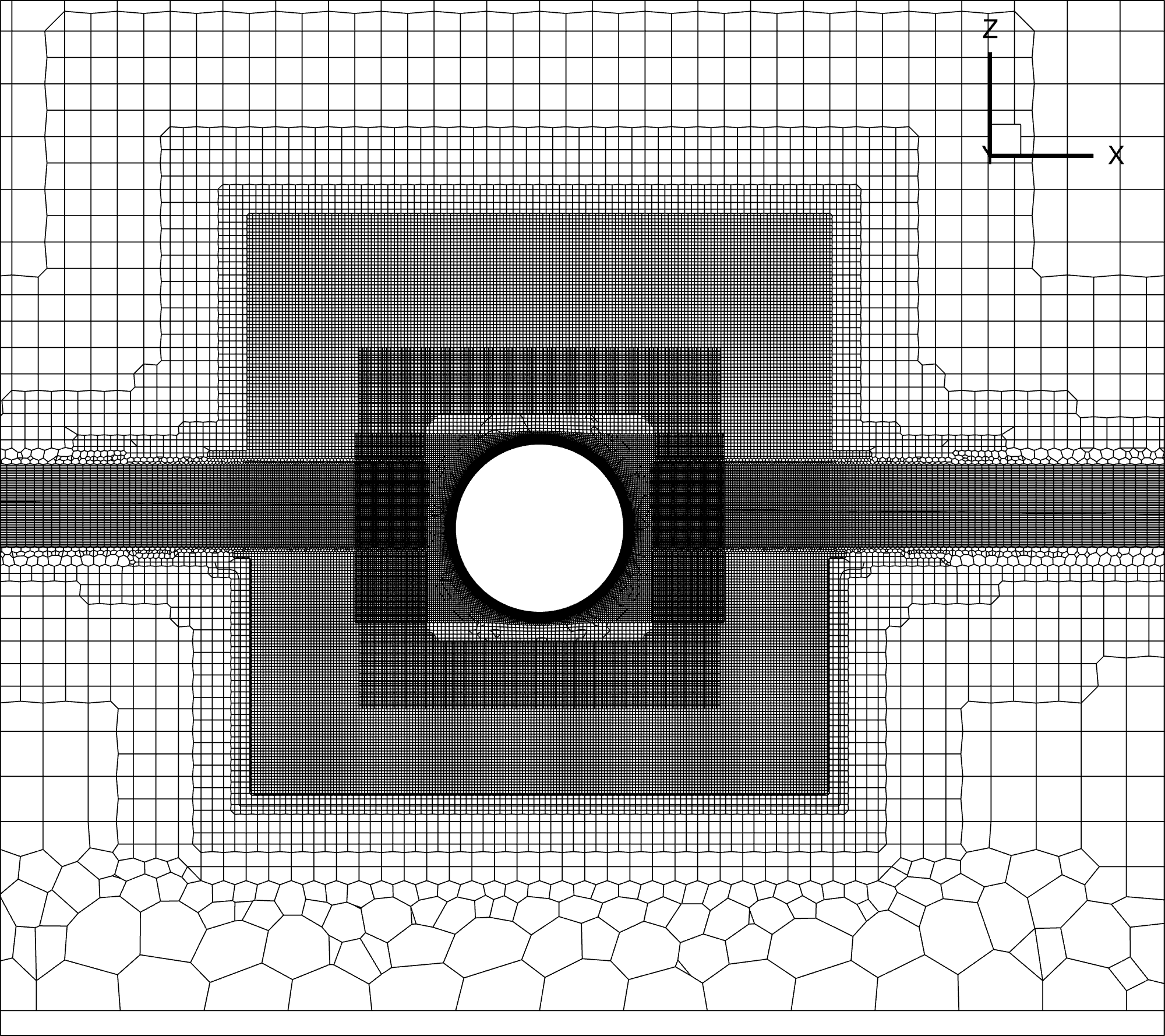}
    \end{subfigure}
    \caption{$y-z$ slice visualization of the mesh. (left to right) 1) Background mesh, 2) body-fitted mesh, 3) both meshes whilst showing only non-interpolated cells (visualization of solution cells)}
    \label{fig:mesh}
\end{figure}

\begin{figure}[H]
    \centering
    \includegraphics[width=0.6\textwidth]{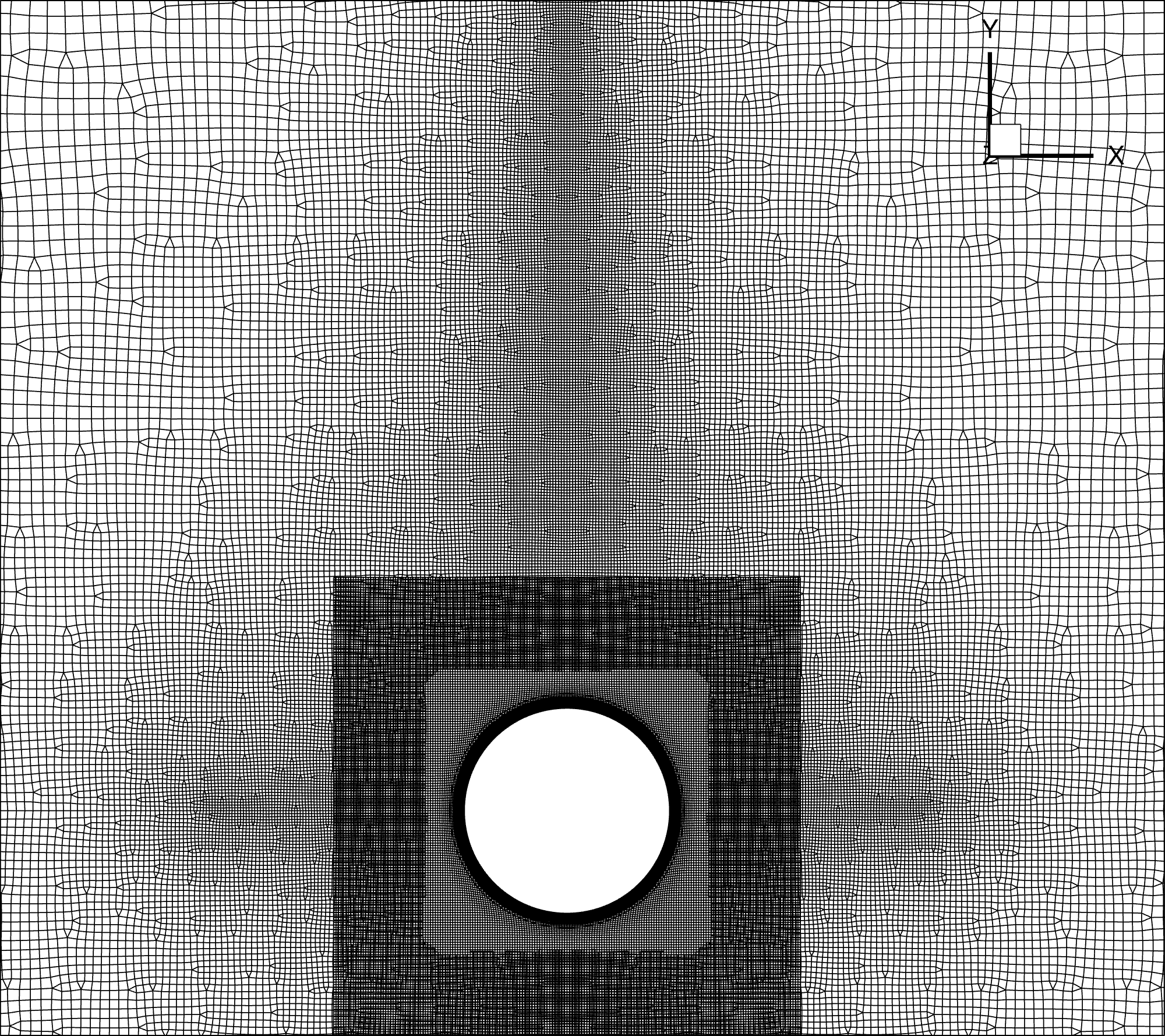}
    \caption{$x-y$ slice visualization of the to grids with the same value blanking as figure \autoref{fig:mesh}}
    \label{fig:horizontal_mesh}
\end{figure}

During the meshing procedure, it has to be ensured that in the vicinity of the free surface, adequate cell length is provided and preferably cells with horizontal faces. In the vertical direction $3\ mm$ constant cell height was used (as shown in \autoref{fig:mesh}). The goal is to resolve the first breaking waves of the phenomenon up to a desired point. Unfortunately, not all wave lengths are resolved due to mesh size restrictions. The last meshing constraint is the horizontal refinement near the free surface. This is imposed due to the measurement of the free surface elevation in three gauges (the decrease of cell size at $x=0,\ y>0$ is depicted in \autoref{fig:horizontal_mesh}). It is noted that both meshes in all simulations are general polygonal (some cells contain over $100$ nodes).

The solution in the finer grid is depicted below with iso-surface and contour slice snapshots. Interpolation was done with the gradient-aware algorithm.

\begin{figure}[H]
    \centering
    \begin{subfigure}[b]{0.23\textwidth}
        \includegraphics[width=1.\textwidth]{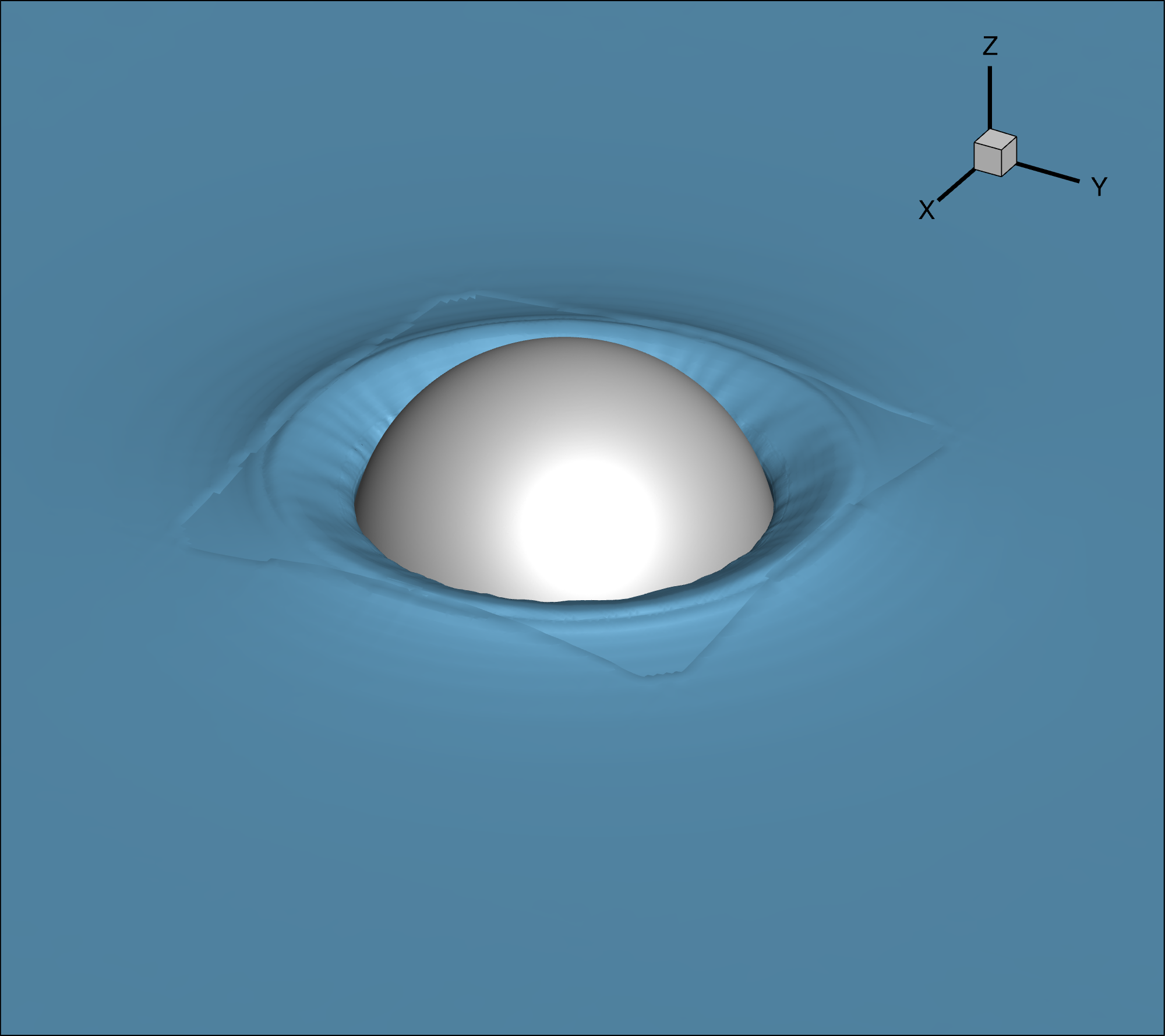}
    \end{subfigure}
    \begin{subfigure}[b]{0.23\textwidth}
        \includegraphics[width=1.\textwidth]{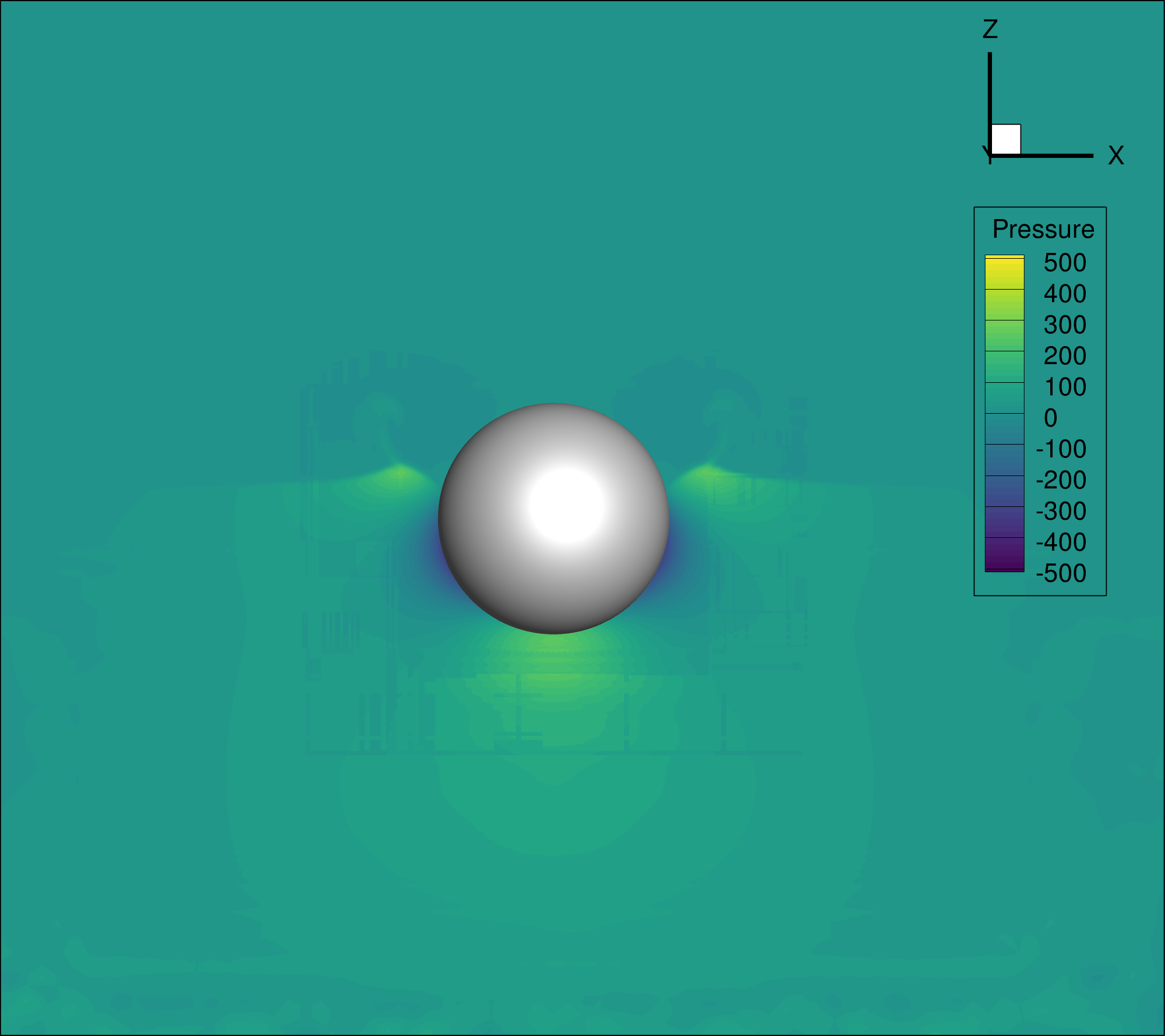}
    \end{subfigure}
    \begin{subfigure}[b]{0.23\textwidth}
        \includegraphics[width=1.\textwidth]{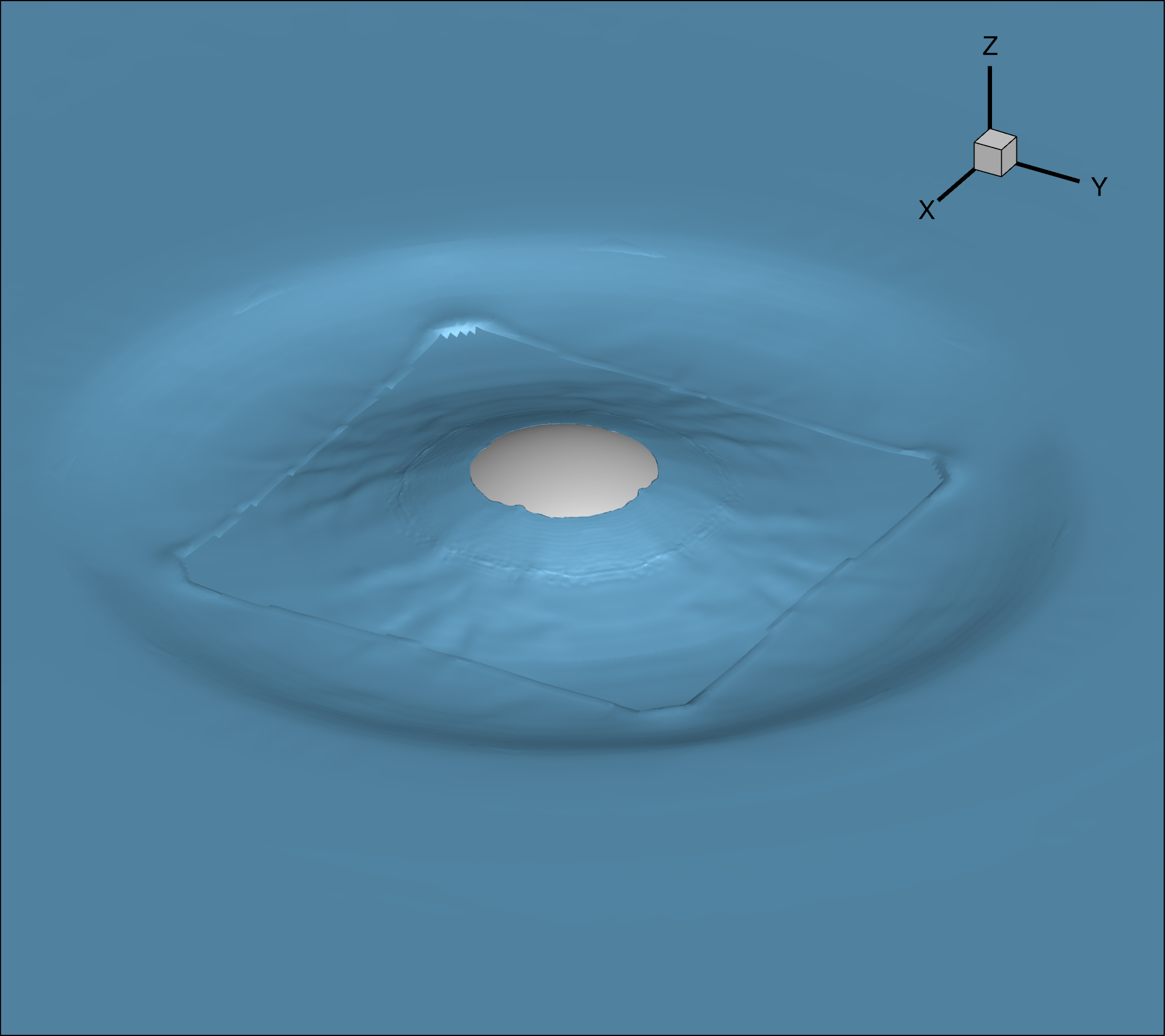}
    \end{subfigure}
    \begin{subfigure}[b]{0.23\textwidth}
        \includegraphics[width=1.\textwidth]{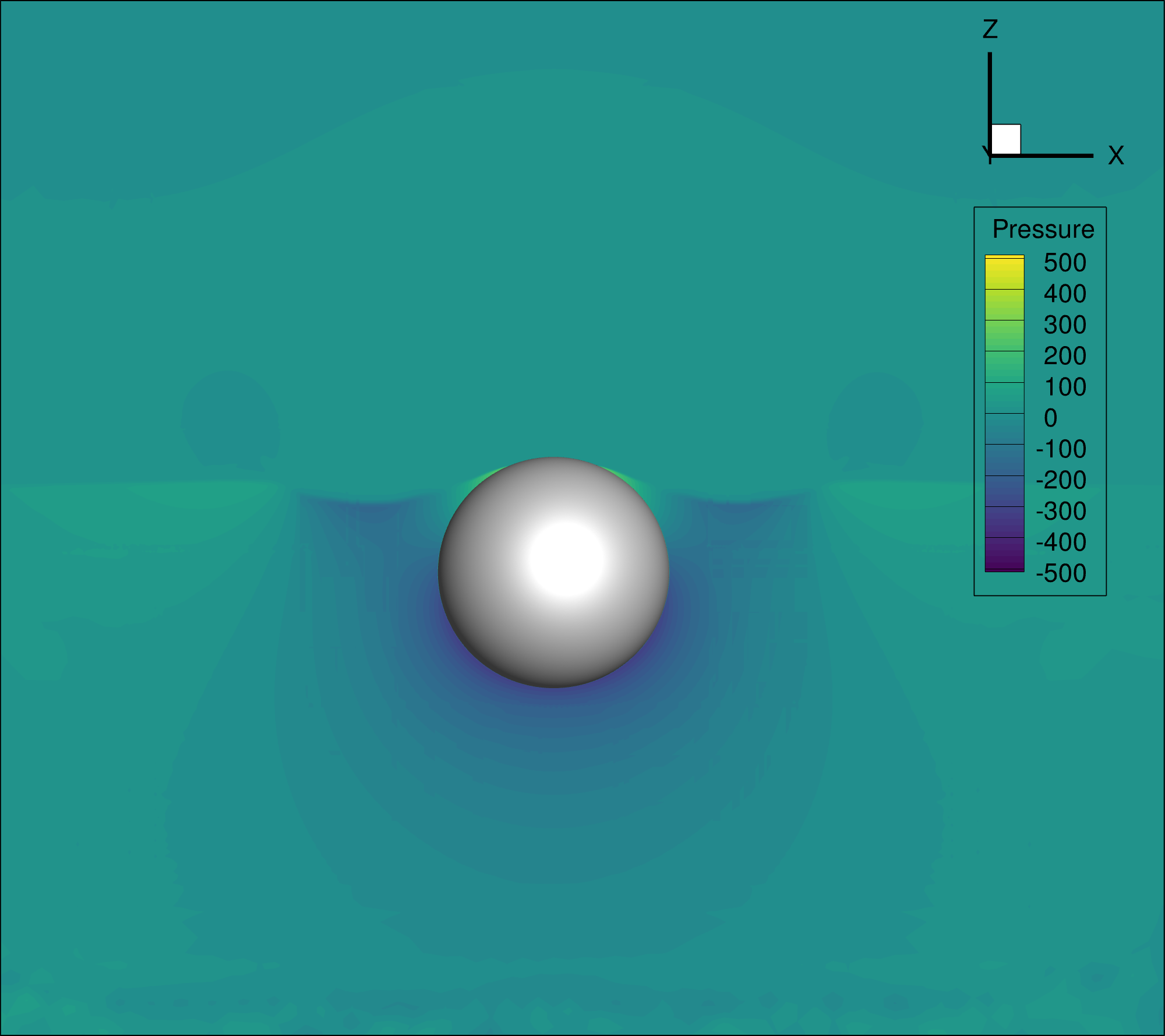}
    \end{subfigure}
    \begin{subfigure}[b]{0.23\textwidth}
        \includegraphics[width=1.\textwidth]{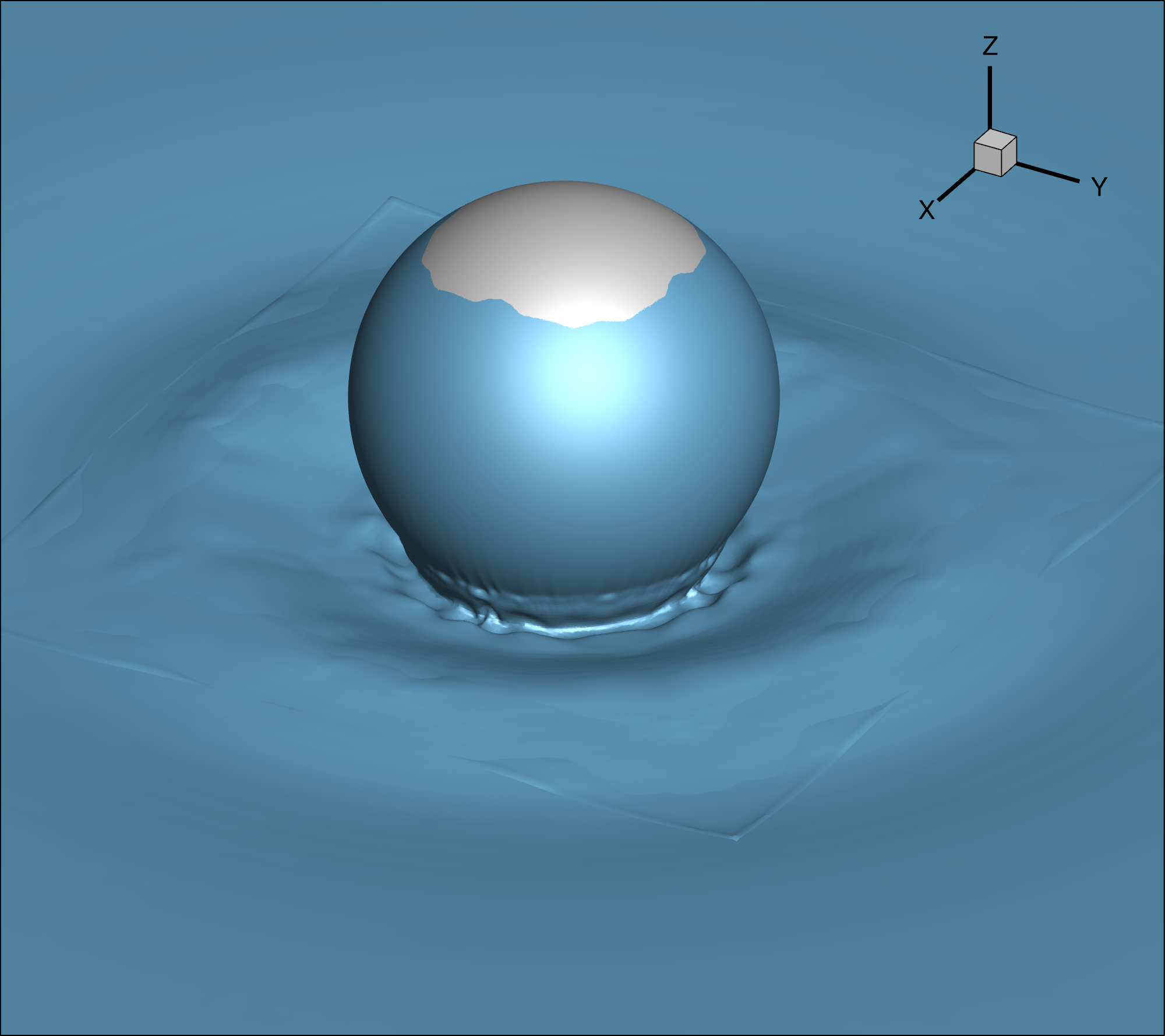}
    \end{subfigure}
    \begin{subfigure}[b]{0.23\textwidth}
        \includegraphics[width=1.\textwidth]{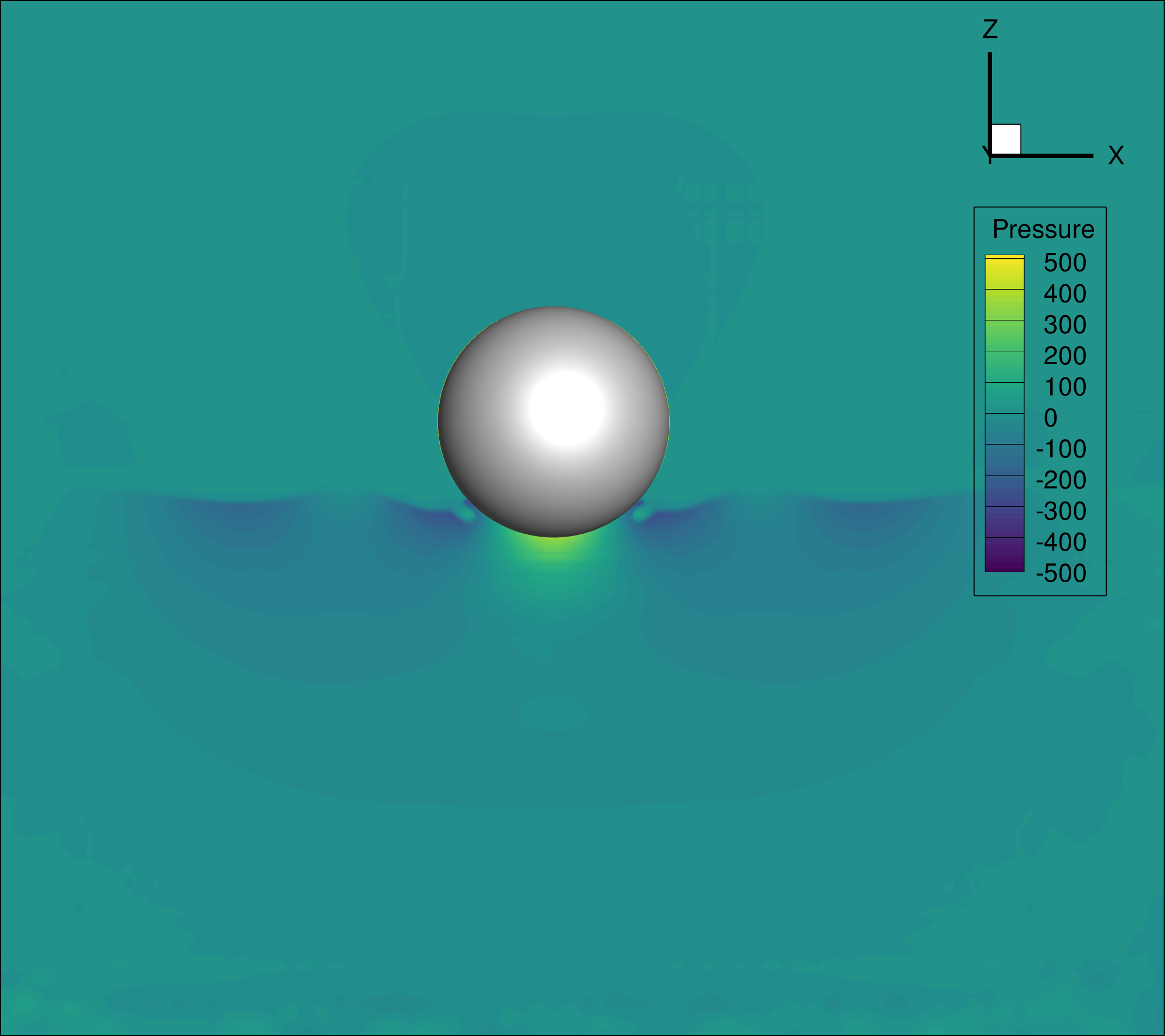}
    \end{subfigure}
    \begin{subfigure}[b]{0.23\textwidth}
        \includegraphics[width=1.\textwidth]{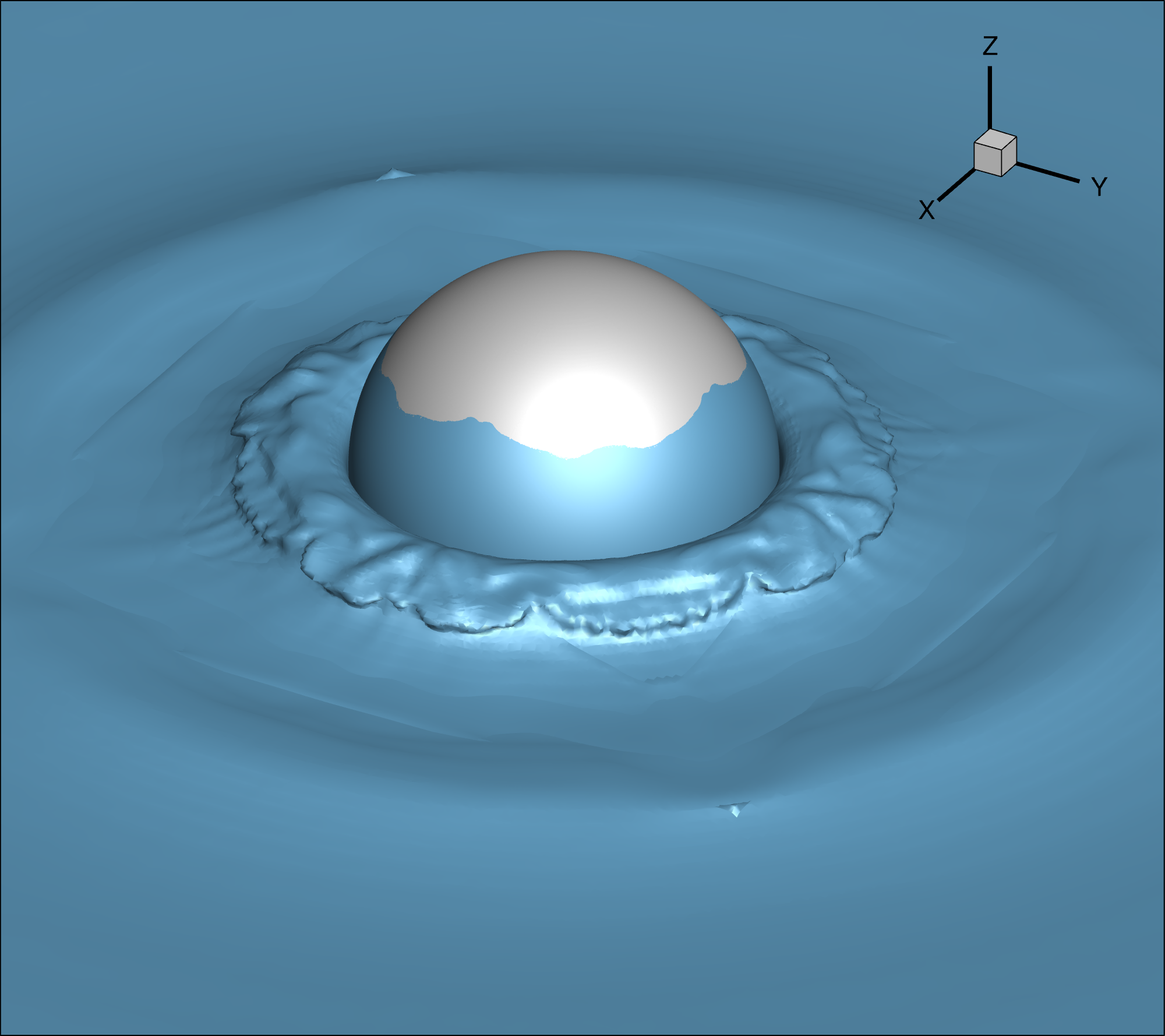}
    \end{subfigure}
    \begin{subfigure}[b]{0.23\textwidth}
        \includegraphics[width=1.\textwidth]{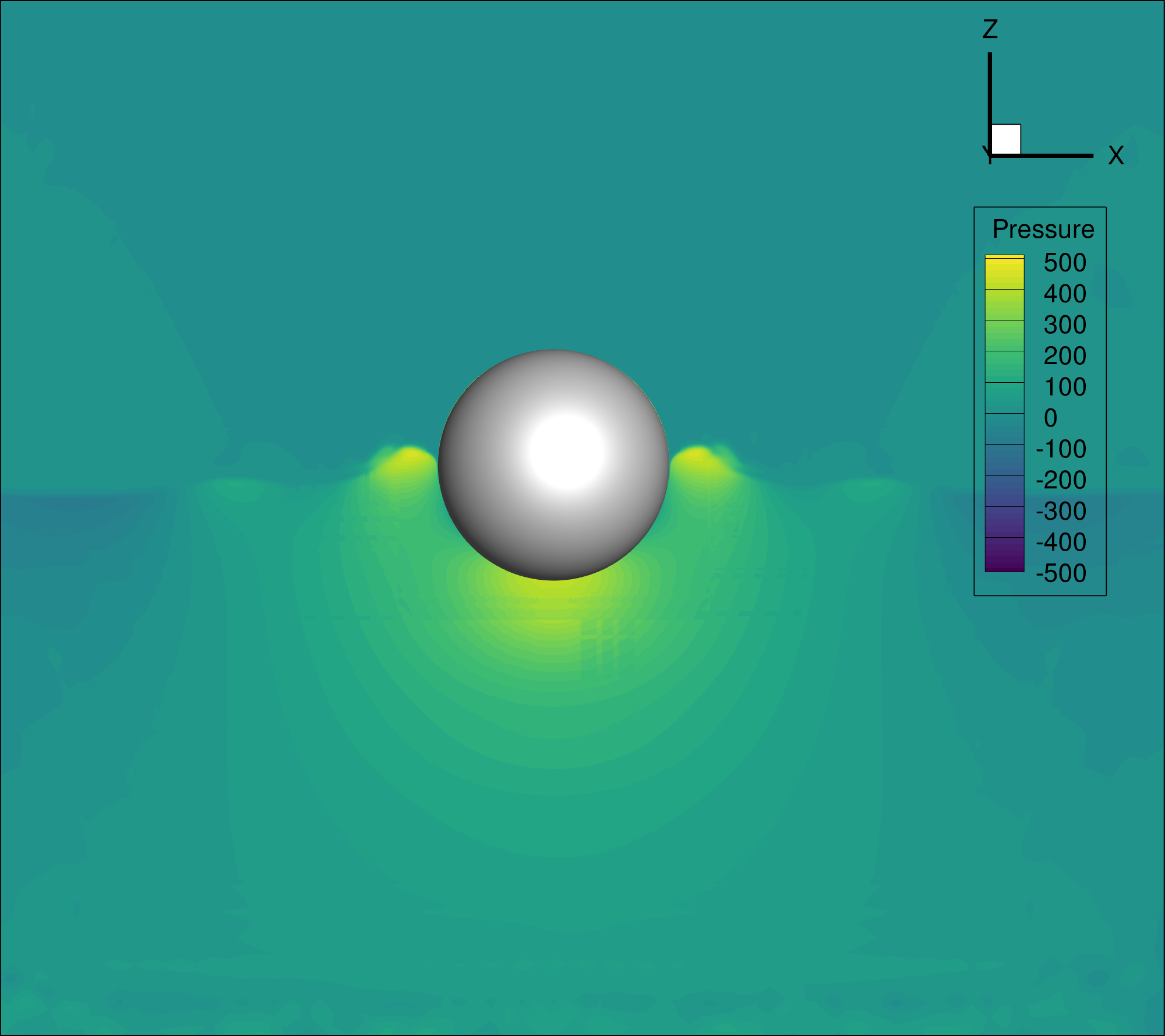}
    \end{subfigure}
    \caption{(left to right \& top to bottom in pairs): Density iso-surface at the average value of air \& water (left subplot) and dynamic pressure contour (right subplot) at times $t=0.25,\ 0.50,\ 0.75,\ 1.0\ s$ respectively}
    \label{fig:press_fs}
\end{figure}

In \autoref{fig:press_fs} it is worth noticing that the coupled overset solver manages to handle the propagation of the waves in relation with the marching of time. Additionally, in the last two snapshots (below \& right) the breaking of the near-wall radiating wave stands out as well as the cylindrical propagation of the previous perturbation. 

\subsection{Comparison}

The comparison is performed against the experimental data shown in \cite{kramer2021highly}. We first compare the heave displacement for the whole simulation and we present the free surface elevation signals from the three gauges.

\begin{figure}[H]
    \centering
    \begin{subfigure}[b]{0.48\textwidth}
        \includegraphics[width=1.\textwidth]{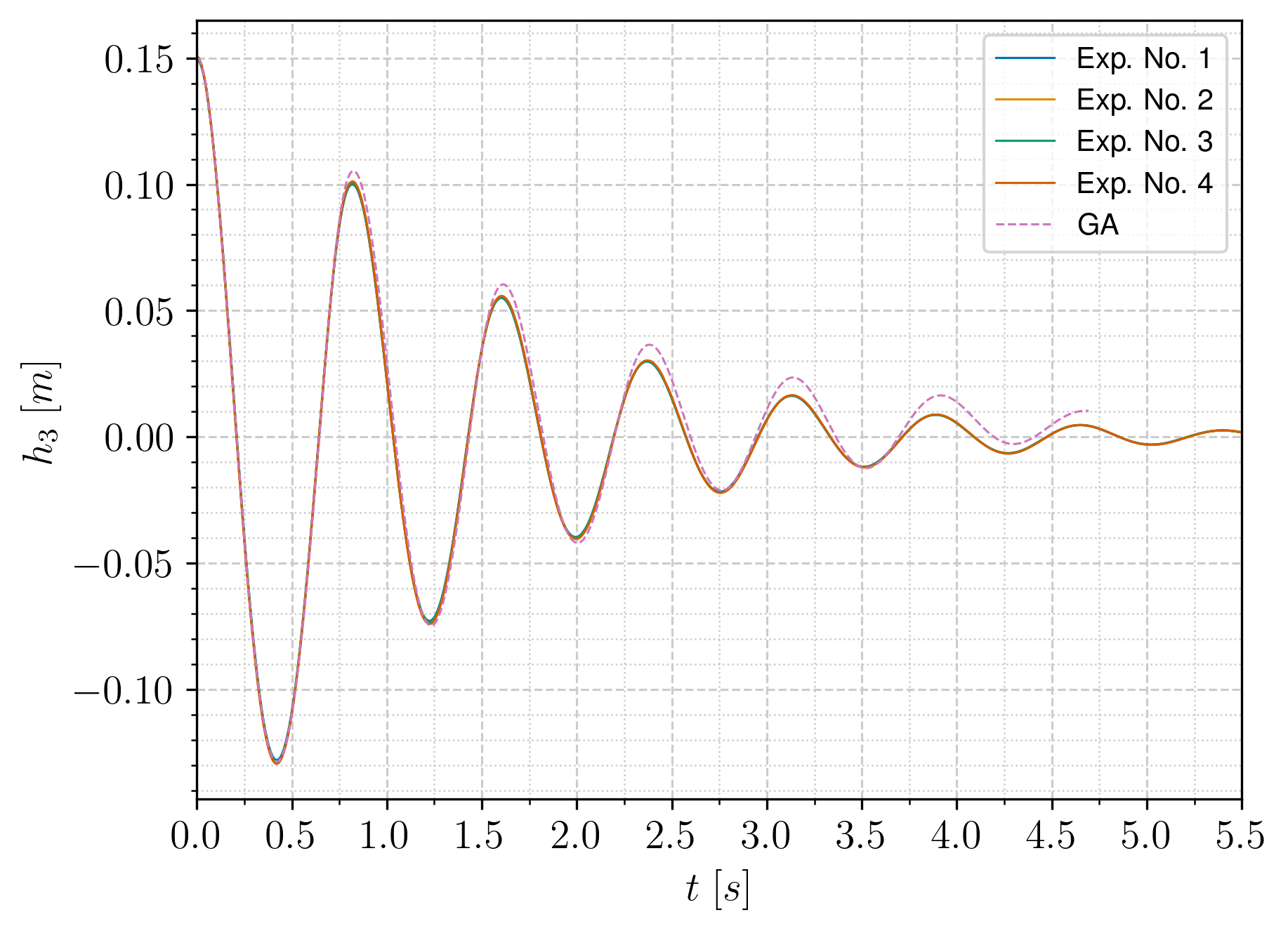}
    \end{subfigure}
    \begin{subfigure}[b]{0.48\textwidth}
        \includegraphics[width=1.\textwidth]{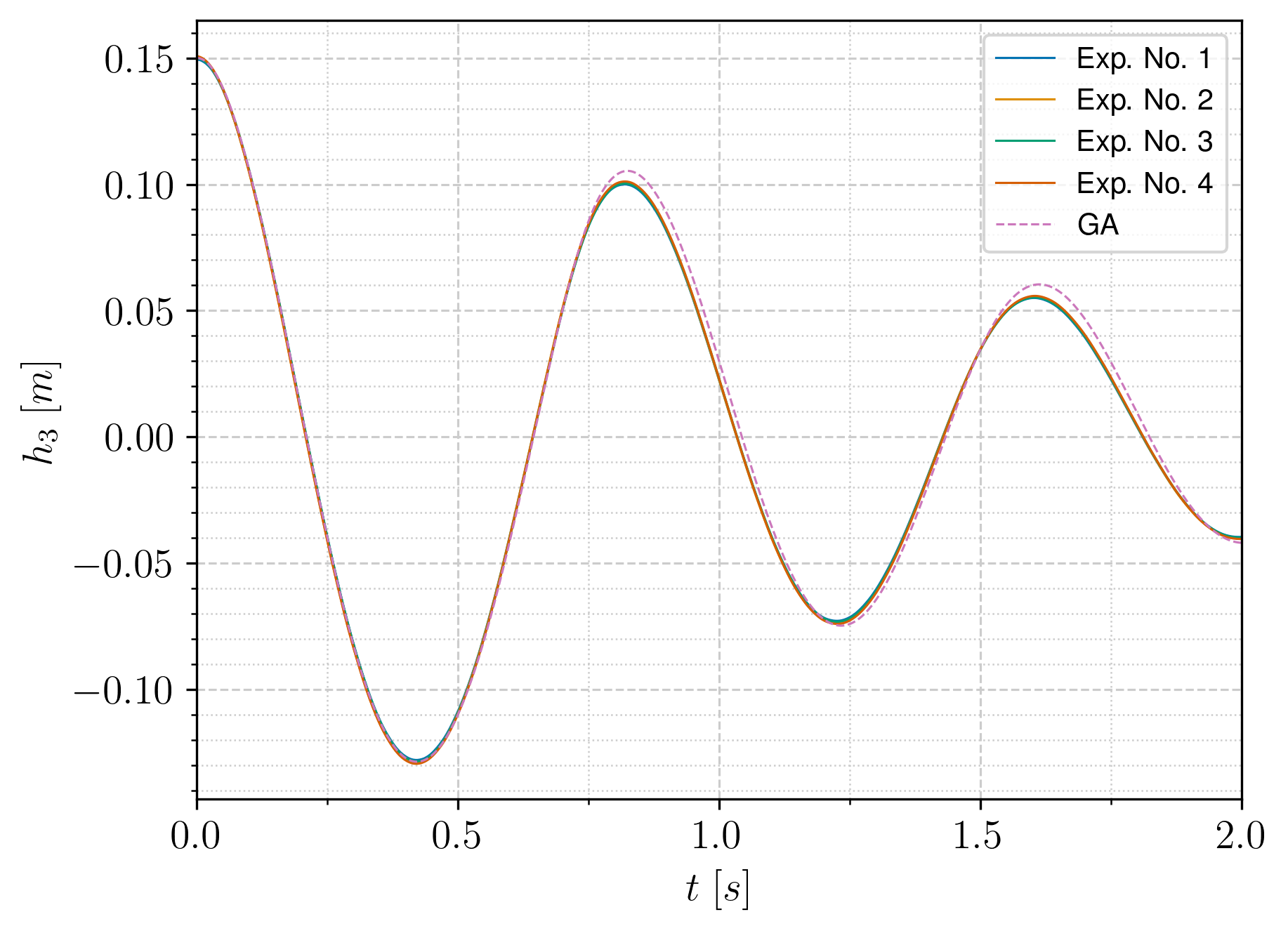}
    \end{subfigure}
    \caption{(Left) Heave displacement of the floating sphere over time of simulation and experiments. (Right) Heave displacement over the first $2\ s$ time interval.}
    \label{fig:heave_decay}
\end{figure}

\begin{figure}[H]
    \centering
    \includegraphics[width=1.\textwidth]{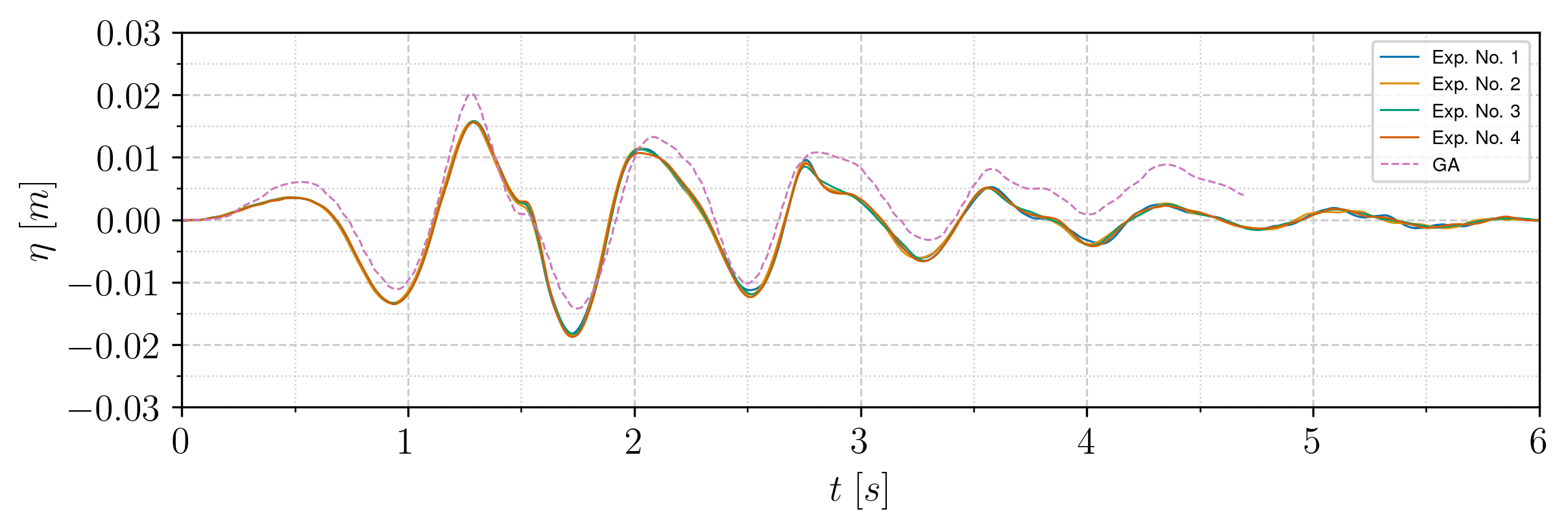}
    \caption{Free surface elevation over time at gauge No. 1 positioned at $(x,y) = (0, 0.6)\ m$ for overset simulation and experiments.}
    \label{fig:hgauge1}
\end{figure}

\begin{figure}[H]
    \centering
    \includegraphics[width=1.\textwidth]{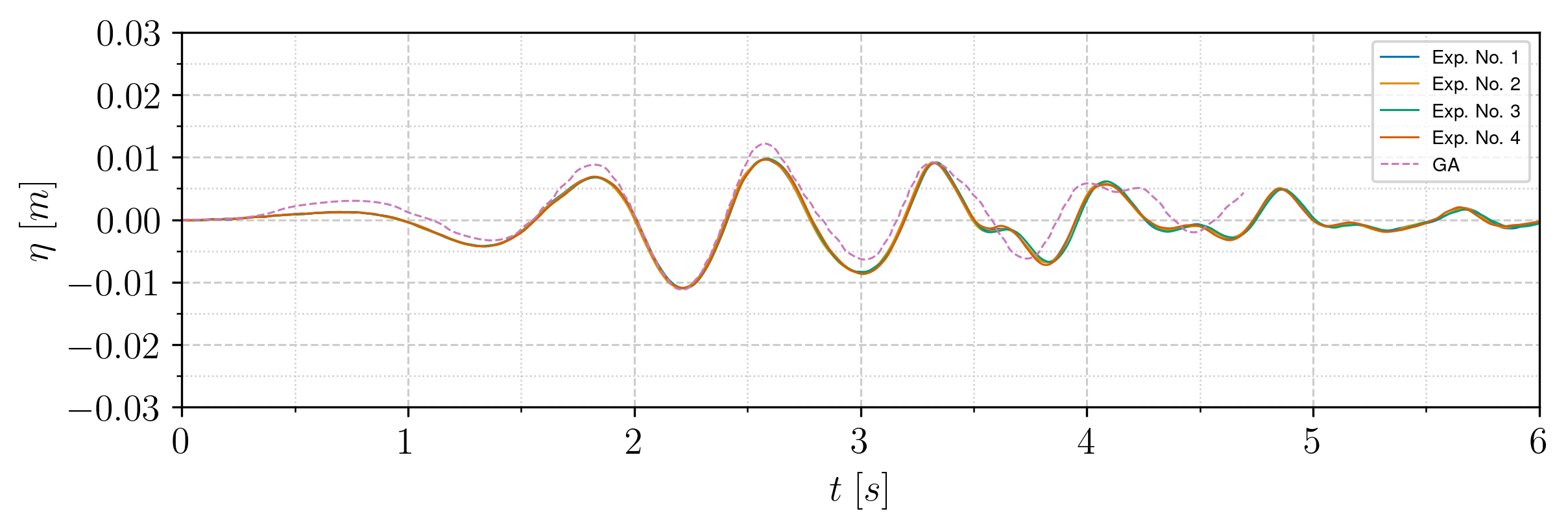}
    \caption{Free surface elevation over time at gauge No. 2 positioned at $(x,y) = (0, 1.2)\ m$ for overset simulation and experiments.}
    \label{fig:hgauge2}
\end{figure}

\begin{figure}[H]
    \centering
    \includegraphics[width=1.\textwidth]{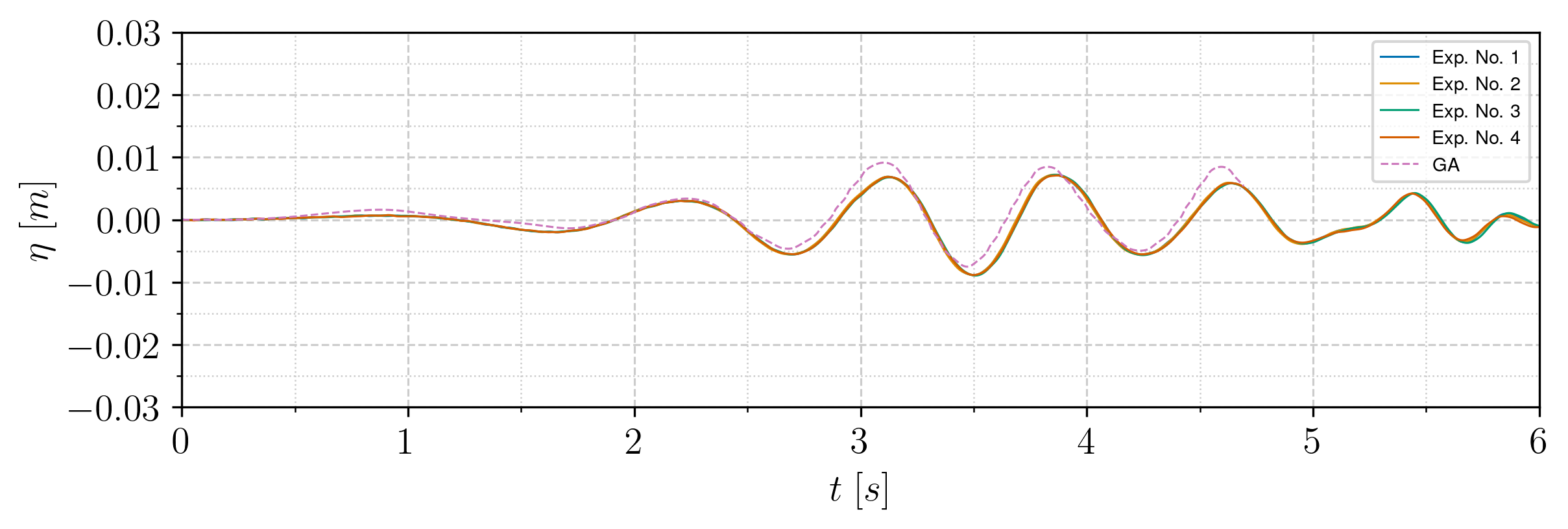}
    \caption{Free surface elevation over time at gauge No. 3 positioned at $(x,y) = (0, 1.8)\ m$ for overset simulation and experiments.}
    \label{fig:hgauge3}
\end{figure}

In \autoref{fig:heave_decay} we come to the result that the numerical simulation is in good agreement with experimental data. Moreover, in \autoref{fig:hgauge1}, \autoref{fig:hgauge2} and \autoref{fig:hgauge3} we can distinguish that the coupled overset solver manages to recreate the radiating wave structures with details.

We can also notice that after the $4th$ heave decay period the heave displacement and the radiating wave height drop at a total value of approximately $8-10\ mm$. We recall that the vertical cell length during the meshing is $h_c = 3\ mm$, that is $2-3$ cells per oscillation amplitude. This, of course, is not an adequate amount of cells in order to track the behavior of the free surface and the force contributions on the falling sphere. Therefore, we do not expect any agreement with the experimental data after four periods of heave decay.

\subsection{Scalability}\label{sub:scalability}

The last part of investigating the proposed overset algorithm is to validate the scaling in parallel architectures. This examination will concern the strong scaling.

These measurements will be performed for the validation case examined in \autoref{subsection:heave_decay}, which is a fairly dense grid ($18$ million cells). In order to measure the performance in parallel execution, we have to synchronize the CPUs before obtaining the clock time. That way, we make sure that for every possible computing in-balance among CPUs, we will measure the worst CPU candidate. The \emph{MaPFlow} solver uses the concept of pseudo-time to perform internal iterations. This means that for every (real) timestep, the overset algorithms (boundary cutting \& donor searching) are performed only once, whereas the donor data sending and the interpolation are performed at every internal iteration.

After obtaining the clock time for a series of parallel executions with an increasing number of processors, we will measure the speedup in the following way.

\begin{align}\label{eq:scale}
    & S = \frac{\Delta t_{0}}{\Delta t_{i}}
\end{align}

In \autoref{eq:scale} we divide the total time needed to perform a single (real) time-step for execution No. $i$ from the time-step of the first execution.

The number of CPUs for the first execution is $60$ and the last execution is performed with $600$ CPUs.

\begin{figure}[H]
    \centering
    \includegraphics[width=0.75\textwidth]{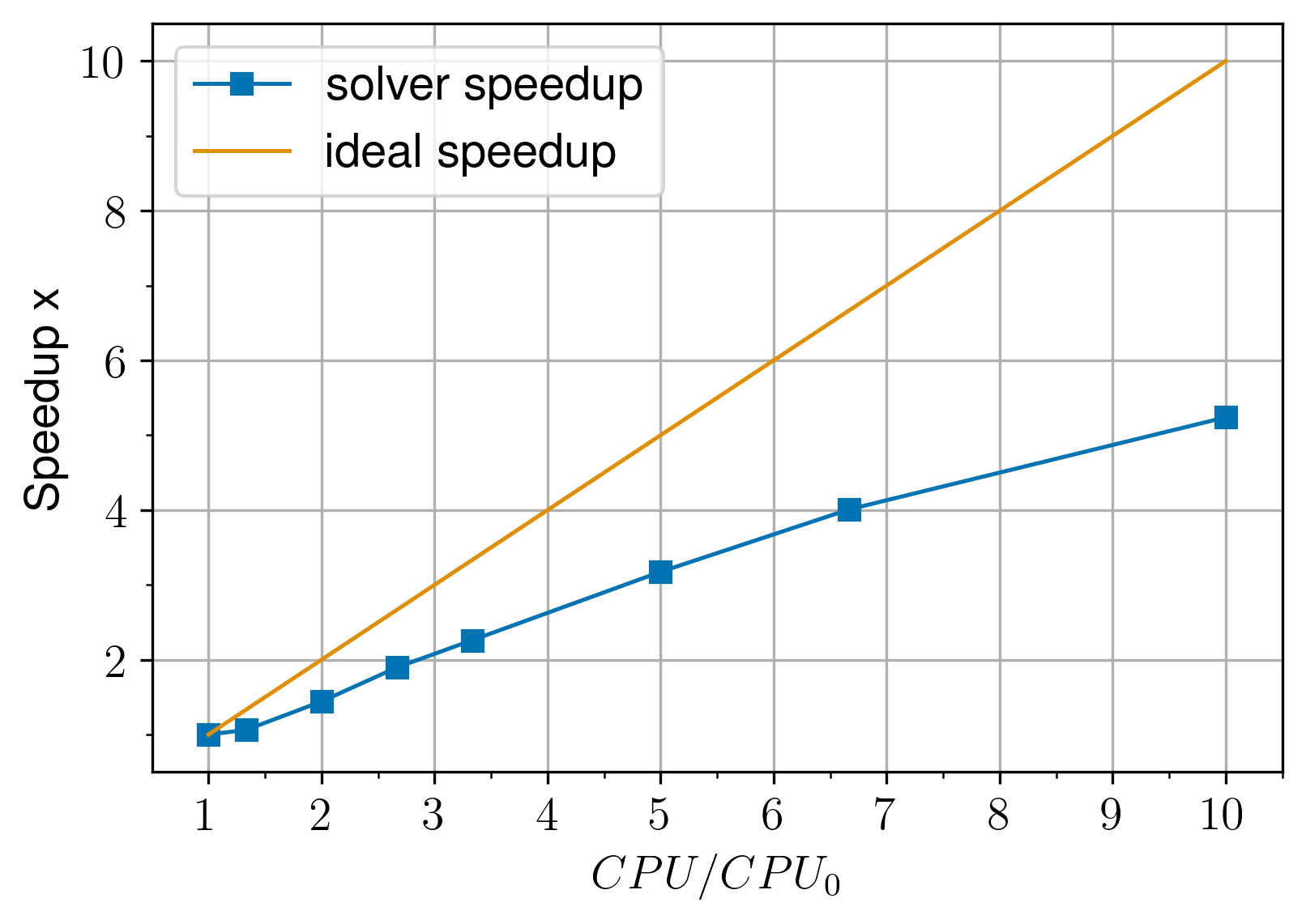}
    \caption{Parallel speedup of execution time over the fraction of execution CPUs from initial execution CPUs.}
    \label{fig:speedup}
\end{figure}

\begin{table}[H]
    \centering
    \begin{tabular}{ccccccccc}
        $CPUs$ & 60 & 80 & 120 & 160 & 200 & 300 & 400 & 600 \\\hline
        $CPUs/CPU_0$ & 1 & 1.33 & 2 & 2.667 & 3.333 & 5 & 6.667 & 10 \\\hline
        Speedup & 1 & 1.062 & 1.442 & 1.897 & 2.261 & 3.175 & 4.008 & 5.240\\ \hline
    \end{tabular}
    \caption{Measured values of parallel speedup of execution time with scaling of CPUs and number of CPUs of execution.}
    \label{tab:speedup}
\end{table}

In \autoref{fig:speedup} we compare the rate of increase of execution speed with the rate of increase of CPUs. The first simulation, which provides the numerator in the fraction of \autoref{eq:scale}, is ran with $60$ CPUs. Regarding the results of \autoref{fig:speedup} and the \autoref{tab:speedup}, it stands out that the coupled overset solver has a speedup approximately half of the increase of CPUs for every parallel execution. Lastly, when examining the slope of every line segment in \autoref{fig:speedup}, we observe that the coupled-solver manages to scale by a fair amount even in massively parallel executions (3-digit amount of CPUs).

Especially for the \say{serd/recv} task, there is a great dependence on the number of donors. This can be caused especially in small number of CPUs, due to the fact that cartesian boxes will possibly occupy a large amount of space relatively to the volume of the convex hull of donors. The use of a box aligned to the eigenvectors of the covariance matrix of donor coordinates would be a more suitable solution with a small amount of extra calculations. The definition and application of these types of boxes can be found in TIOGA \cite{roget2014robust}. Their use would be a good future addition to the current algorithm, but for the current applications, we observed that we are able to achieve robustness with significant algorithmic simplicity of implementation. 

\section{Conclusions}\label{section:conclusions}

The current study is devoted to the design of an overset algorithm that is able to handle the most generic types of grids in a parallel fashion and target general problems of (practical) CFD simulations with discontinuous solutions. The topology algorithm incorporates elements from various previous works on topology algorithms and introduces novelties (handling of multi-block general polyhedral meshes, which are at the forefront of meshing technology) that make it attractive for use in high-performance simulations. The scaling performance of the topology algorithm was presented in \autoref{sub:scalability} where a steady scaling up to several hundreds of processors was observed.

Additionally, this work addresses the problem of unstructured interpolation for fields with discontinuities by introducing a novel interpolation algorithm, especially friendly to Finite Volume solvers. This scheme was first validated with an analytical solution, where promising results were drawn in terms of boundness and accuracy. The first test of the coupled scheme with the FV solver on a 2D propagation problem was then presented, in which a significant amount of wave energy was regained in contrast to existing numerical diffusion and compared with the competitor of the nearest neighbor value. The last validating case of the heave decay served as a test of performance in surface-piercing problems, and the solution came in good agreement with experimental data. Finally, from the scalability test, we came to the conclusion that the interpolation algorithm poses no computing threshold in the strong scaling of the parallel execution.


\section*{Acknowledgments}
This work was supported by computational time granted from the Greek Research \& Technology Network (GRNET) in the National HPC facility – ARIS – under project ”DYNASEA” with ID pr012019.The polyhedral mesh production was performed with the help of the ANSA pre-processor, a product of BETA-CAE Systems.

\section*{Declaration of Competing Interest}

The authors declare that they have no known competing financial interests or personal relationships that could have
appeared to influence the work reported in this paper.


\appendix\label{appendix}

\section{Multi-Phase Incompressible Navier-Stokes}


In this work, we consider free surface flows and consequently the incompressible form of the Navier-Stokes equations is employed. The two immiscible fluid phases are resolved via the Volume of fluid (VoF) method. The governing equations of the problem are expressed in a vector notation below:

\begin{align}\label{eq:cont2}
    & \nabla \cdot \mathbf{u} = 0
\end{align}
\begin{align}\label{eq:ns2}
    & \frac{\partial \mathbf{u}}{\partial t} + \left(\mathbf{u}\cdot \nabla\right) \mathbf{u} = -\frac{1}{\rho}\nabla p + \frac{1}{\rho}\mathbf{g} + \nu \Delta \mathbf{u} + \frac{1}{\rho}\mathbf{F}
\end{align}
\begin{align}\label{eq:vof2}
    & \frac{\partial \alpha_l}{\partial t} + \left( \mathbf{u}\cdot \nabla \right) \alpha_l = 0 
\end{align}
\begin{align}\label{eq:partition_of_unity}
    & \alpha_l + \alpha_a = 1
\end{align}

Considering \autoref{eq:cont2} up to \autoref{eq:partition_of_unity}, $\mathbf{u} = \mathbf{u}\left(\Vec{x},t\right)$ is the velocity vector in $x,y$ and $z$ directions, $p = p\left(\mathbf{x},t\right)$ is the pressure, $\rho_m=\left(1-\alpha_l\right)\rho_a + \alpha_l \rho_l$ is the mixture density of the control volume and $\alpha_l$ and $\alpha_a$ are the volume fraction of mixture for liquid and air respectively. \autoref{eq:cont2}, \autoref{eq:ns2}, \autoref{eq:vof2} and \autoref{eq:partition_of_unity} represent the continuity condition, the momentum equation, the transport equation for the volume fraction and the partition of unity respectively. \autoref{eq:partition_of_unity} indicates that for $N$-phases, we solve $N-1$ transport equations of the \autoref{eq:vof2} type.

For the pressure-velocity coupling, the artificial compressibility method (AC) is employed firstly introduced by A. J. Chorin \cite{chorin1997numerical} and therefore further manipulations have to be done, to cast a conservative expression.

Following the work of Ntouras \cite{ntouras2020coupled} where the discretized total system of equations is constructed, we express the system in the following brief representation

\begin{align}\label{eq:system2}
    & \Gamma \frac{\partial}{\partial \tau} \int_{D_k}\overrightarrow{Q}\; \mathrm{d}D +  \Gamma_e \frac{\partial}{\partial t}\int_{D_k} \overrightarrow{Q} \; \mathrm{d}D_k + \int_{\partial D_k} \Big(\overrightarrow{F}_c - \overrightarrow{F}_v \Big) \; \mathrm{d}S_k = \int_{D_k} \overrightarrow{S}_q \; \mathrm{d}D_k
\end{align}

\begin{align}
    & \overrightarrow{U} = \left[\ 0\quad \rho \vec{u}\quad  \alpha_l\ \right]^T, && \overrightarrow{Q} = \left[\ p\quad \vec{u}\quad \alpha_l\ \right]^T
\end{align}
\begin{align}
    & \frac{\partial \overrightarrow{U}}{\partial t} =\frac{\partial U}{\partial Q}\frac{\partial \overrightarrow{Q}}{\partial t} =\Gamma_e \frac{\partial \overrightarrow{Q}}{\partial t}
\end{align}

where
\begin{align}
    \Gamma_e = \frac{\partial U}{\partial Q} = 
    \begin{bmatrix}
       0 & 0 & 0           \\[0.3em]
       0 & \rho I_{3\times 3} & \vec{u}\delta \rho \\[0.3em]
       0 & 0 & 1
    \end{bmatrix}, && 
    \Gamma = 
    \begin{bmatrix}
        \frac{1}{\beta \rho_m} & 0 & 0           \\[0.3em]
        0 & \rho_m I_{3\times 3} & \vec{u}\delta \rho \\[0.3em]
        \frac{\alpha_l}{\beta \rho_m} & 0 & 1
     \end{bmatrix}
\end{align}
is the Jacobian matrix and the preconditioning matrix of Kunz \cite{kunz2000preconditioned}, $\overrightarrow{F}_c$ and $\overrightarrow{F}_v$ are the convective and viscous fluxes respectively, $\overrightarrow{S}_q$ are the source terms, $\beta$ and $\tau$ are the artificial compressibility factor and the artificial time, respectively, associated with the term $\frac{1}{\beta}\frac{\partial \overrightarrow{Q}}{\partial \tau}$ and $\delta \rho=\rho_l - \rho_a$ is the difference between the two densities.

\bibliographystyle{abbrv}
\bibliography{references}  






\end{document}